\documentclass[11pt,a4paper]{article}
\usepackage{amsmath}
\usepackage{graphicx}
\usepackage{xcolor}
\usepackage[caption=false]{subfig}
\usepackage{mathrsfs,mathtools}
\usepackage{physics,amssymb}
\usepackage{siunitx}
\usepackage{bm}
\usepackage{float}
\usepackage{braket}
\usepackage{listings}

\usepackage{cases}
\usepackage{comment}
\usepackage{soul}
\usepackage{cancel}
\usepackage{cases}
\usepackage[utf8]{inputenc}
\usepackage{url}
\usepackage{float}
\usepackage{longtable}
\usepackage[normalem]{ulem}
\usepackage{xspace}
\usepackage{aas_macros}
\setstcolor{red}
\usepackage{jcappub}
\hypersetup{colorlinks=true
,urlcolor=DARKBLUE
,anchorcolor=DARKBLUE
,citecolor=DARKBLUE
,filecolor=DARKBLUE
,linkcolor=DARKBLUE
,menucolor=DARKBLUE
,linktocpage=true
,pdfproducer=medialab
}

\usepackage{graphicx,epsfig,color,xcolor}

\usepackage[english]{babel}
\usepackage{amssymb,amsfonts,amsmath,physics}
\usepackage{mathtools}
\usepackage{siunitx}
\usepackage{verbatim}
\usepackage{ulem}
\usepackage{soul}
\usepackage{lipsum, babel}
\usepackage{acronym}
\usepackage{aas_macros}

\usepackage{hyperref}

\newcommand{\be}{\begin{equation}} \newcommand{\ee}{\end{equation}}
\newcommand{\bea}{\begin{eqnarray}} \newcommand{\eea}{\end{eqnarray}}

\usepackage{fancyhdr}
\usepackage{amsmath}
\usepackage{mathrsfs}
\usepackage{float}
\usepackage{xcolor}
\usepackage{graphicx, epsfig, amssymb} 
\usepackage{amsmath, amsfonts}
\usepackage{bm} 
\usepackage{tensor}
\usepackage{soul}
\usepackage{hyperref}
\usepackage{physics}

\definecolor{MONZA}{HTML}{CF000F}
\definecolor{DARKBLUE}{HTML}{00008b}
\definecolor{DARKMAGENTA}{HTML}{8b008b}
\definecolor{DARKCYAN}{HTML}{008B8B}
\definecolor{DARKORANGE}{HTML}{FF8C00}
\definecolor{OBSERVATORY}{HTML}{049372}
\definecolor{GREENBAMBOO}{HTML}{006442}
\definecolor{TURQUOISE}{HTML}{36D7B7}
\definecolor{JUNGLEGREEN}{HTML}{26C281}

\begin{document}

\title{A new approach for simulating PBH formation from generic curvature fluctuations with the Misner-Sharp formalism}

\author[a,b]{Albert Escriv\`a}

\emailAdd{escriva.manas.alberto.k0@f.mail.nagoya-u.ac.jp}
\affiliation[a]{Institute for Advanced Research, Nagoya University, \\
Furo-cho Chikusa-ku, Nagoya 464-8601, Japan}
\affiliation[b]{Department of Physics, Nagoya University, \\
Furo-cho Chikusa-ku, Nagoya 464-8602, Japan}

\date{\today}
\abstract{Primordial Black Holes (PBHs) may have formed in the early Universe due to the collapse of super-horizon curvature fluctuations. Simulations of PBH formation have been essential for inferring the initial conditions that lead to black hole formation and for studying their properties and impact on our Universe. The Misner-Sharp formalism is commonly used as a standard approach for these simulations. Recently, type-II fluctuations, characterized by a non-monotonic areal radius, have gained interest. In the standard Misner-Sharp approach for simulating PBH formation with these fluctuations, the evolution equations exhibit divergent terms ($0/0$), which complicate and prevent the simulations. We formulate a new approach to overcome this issue in a simple manner by using the trace of the extrinsic curvature as an auxiliary variable, allowing simulations of type-II fluctuations within the Misner-Sharp formalism. Using a set of standard exponential-shaped curvature profiles, we apply and test our new approach and numerical code based on pseudospectral methods to study the time evolution of the gravitational collapse, threshold values of type A/B PBHs and PBH mass. Interestingly, we identify cases of type-II fluctuations that do not necessarily result in PBH formation.}
\maketitle
\flushbottom

\acresetall

\acrodef{GW}{gravitational wave}
\acrodef{CMB}{cosmic microwave background}
\acrodef{PBH}{primordial black hole}
\acrodef{DM}{Dark Matter}
\acrodef{MS}{Misner-Sharp}
\acrodef{FLRW}{Friedmann‐-Lema\^itre--Robertson--Walker}
\section{Introduction}
Primordial Black Holes (PBHs) are black holes that could have formed in the early Universe through various mechanisms \cite{Zeldovich:1967lct,Hawking:1971ei,Carr:1974nx,Carr:1975qj} (see \cite{Escriva:2022duf} for a comprehensive review covering various perspectives and a detailed list of different PBH formation mechanisms). These PBHs could constitute a significant fraction of dark matter, particularly in the so-called asteroid mass range \cite{Chapline:1975ojl}. Additionally, they may help resolve certain cosmic mysteries \cite{Carr:2019kxo,Carr:2023tpt}. Recent advancements and future prospects in gravitational wave studies, along with observational efforts \cite{2016PhRvX...6d1015A,2023PhRvX..13d1039A}, are crucial in this context, as they provide opportunities to test both the direct and indirect detection of PBHs and a quantification of their role in the dark matter.

One of the standard mechanisms for PBH production is the collapse of super-horizon curvature fluctuations generated during inflation. These fluctuations may collapse, forming black holes during the radiation epoch after re-entering the cosmological horizon. From now on, we will restrict ourselves to this scenario for PBH production. For this scenario, significant efforts have been made to explore it using numerical relativity \cite{1978SvA....22..129N,1980AZh....57..250N,Niemeyer:1999ak,Musco:2008hv,Musco:2012au,Musco:2018rwt,Escriva:2019nsa,Escriva:2021pmf,Yoo:2021fxs,deJong:2021bbo,Escriva:2022yaf,Escriva:2023qnq}. The reason is straightforward: the statistical estimation of PBH production is highly sensitive (in particular, exponentially sensitive \cite{Carr:1975qj}) to the initial conditions. Therefore, accurately determining the initial conditions that lead to black hole formation is essential for precisely predicting the abundance of PBHs and their implications in our Universe. In general, only numerical relativity studies can provide an answer to this (see \cite{Aurrekoetxea:2024mdy} for a review of numerical relativity in the broader context of cosmology). Typically, the assumption of spherical symmetry has been employed in numerical simulations (see \cite{Escriva:2021aeh} for a review), based on the consideration that large peaks (which are necessary for PBHs to constitute a significant fraction of dark matter) are roughly spherical \cite{1986ApJ...304...15B}. Therefore, the gravitational collapse can be considered spherical (see \cite{Escriva:2024aeo} for a recent study that tests this assumption using non-spherical simulations, specifically in the case of a monochromatic power spectrum with Gaussian statistics). However, some works have investigated PBH formation in non-spherical settings \cite{Yoo:2020lmg,deJong:2023gsx,Yoo:2024lhp,Escriva:2024lmm}, which may have some impact on specific scenarios where non-sphericities become important (for instance in a matter-dominated era \cite{Khlopov:1980mg}). While this is an interesting avenue that warrants further exploration, for the purposes of this study, which focuses on a radiation-dominated Universe, we will assume PBH formation in spherical symmetry.

As introduced in Ref.~\cite{PhysRevD.83.124025}, curvature fluctuations $\zeta$ can be classified into two types: type-I and type-II. Type-I fluctuations correspond to cases where the areal radius $R = \sqrt{A/4\pi}$ is a monotonically increasing function, where $A$ is the area of the $2$-sphere as a function of the radial coordinate $r$ at constant $t$. In contrast, type-II fluctuations correspond to cases with a non-monotonic $R$ associated with sufficiently large fluctuations, satisfying the condition that exist a region where $\partial_{r}R<0$. This results in a characteristic throat structure, which also appears in other scenarios like baby Universes, domain wall formation and inflating monopoles \cite{PhysRevD.53.655,Cho:1997rb,Deng:2016vzb,Deng:2017uwc,Deng:2020mds}.

Type-I fluctuations have traditionally been considered the standard in the literature. However, type-II fluctuations have recently gained significance, motivating their numerical exploration. Recent studies have specifically highlighted their importance in scenarios related to non-Gaussianities \cite{Gow:2022jfb} and non-linear statistics \cite{Fumagalli:2024kxe}. Crucially, type-II fluctuations have been found to surround vacuum bubbles that form during inflation in specific models \cite{escriva2023formation}, providing a strong motivation for their detailed analysis trhough simulations. While these fluctuations were once commonly overlooked due to the assumed rarity of large amplitudes, their importance in the PBH scenario has become a key focus of current research.

Remarkably, reference \cite{Uehara:2024yyp} presented numerical simulations of type-II fluctuations in a radiation-dominated Universe, following the formalism of \cite{Shibata:1999zs} with the BSSN formalism \cite{PhysRevD.52.5428,PhysRevD.59.024007} and the numerical code of \cite{Yoo:2014boa,Okawa:2014nda} (COSMOS code, which originally follows SACRA code \cite{Yamamoto:2008js}). From the numerical results obtained, a new classification of PBHs based on the trapping horizon configuration was proposed, with type A PBHs corresponding to the standard formation of a marginally trapped surface during the numerical evolution of over-threshold fluctuations, and type B PBHs featuring bifurcated trapping horizons for sufficiently large curvature fluctuations above the threshold for type A PBH formation. Recently, references \cite{Shimada:2024eec, Inui:2024fgk} documented cases where specific fluctuation shapes (from large negative non-Gaussianity models) exhibit a formation threshold in the type-II region, meaning that some type-II fluctuations do not lead to PBH formation. This contrasts with the common understanding that type-II fluctuations always collapse to form black holes. These counter-intuitive results underscore the necessity of detailed numerical investigations into type-II fluctuations, especially regarding the effect of profile dependence, which is known to be significant.

A standard and successful numerical approach that has been employed to study PBH formation under the assumption of spherical symmetry is the so-called Misner-Sharp formalism \cite{PhysRev.136.B571} (see for some examples \cite{Niemeyer:1999ak,Musco:2004ak,Nakama:2013ica,Bloomfield:2015ila,Escriva:2019nsa,Milligan:2025zbu,Ning:2025ogq}), which essentially corresponds to Einstein's equations in the comoving gauge. However, in the case of simulating type-II fluctuations, the standard formulation suffers from divergent terms of the form $(0/0)$, which are associated with the existence of a neck structure in the initial conditions, where the areal radius satisfies $\partial_{r}R = 0$. This makes the simulation challenging and potentially causes it to crash immediately after starting or when the initial condition approaches the boundary in the parameter space that separates the type-I and type-II regions. The Misner–Sharp formalism is highly valued for its simplicity and efficiency when simulating type-I fluctuations. Therefore, resolving the issues associated with the neck structure is essential to fully exploit the Misner-Sharp formalism for studying type-II fluctuations in various contexts and scenarios.

In this work, we address this issue with the goal of developing a new approach for numerical simulations of PBH formation using the Misner-Sharp formalism that can effectively handle generic curvature fluctuations, type-I and type-II. Our objective is to successfully simulate type-II fluctuations using the Misner–Sharp formalism and to explore key features such as the collapse dynamics, trapping horizon configurations, formation thresholds, and PBH masses. We demonstrate that using auxiliary equations enables us to absorb the divergent terms in the Misner-Sharp equations in a straightforward manner. We follow and update the numerical methodology as in \cite{Escriva:2019nsa} and make a basic version of our new code publicly available here \cite{codigo_albert}. Throughout the paper, we use geometrized units with $G = c = 1$.

\section{Misner-Sharp equations and approach for type-II fluctuations}
The Misner-Sharp equations \cite{1964PhRv..136..571M} describe the motion of a relativistic fluid with spherical symmetry. This corresponds to the Einstein field equations written in the comoving slicing and comoving threading, which both define the comoving gauge. To obtain them, first of all, we need to consider a perfect fluid with the energy-momentum tensor,
\begin{equation}
T^{\mu \nu} = (p+\rho)u^{\mu}u^{\nu}+pg^{\mu\nu},
 \label{eq:tensor_energy}
\end{equation}
and with the following spacetime metric in spherical symmetry:
\begin{equation}
\label{2_metricsharp}
ds^2 = -A(r,t)^2 dt^2+B(r,t)^2 dr^2 + R(r,t)^2 d\Omega^2,
\end{equation}
where $R(r,t)$ is the areal radius, $A(r,t)$ is the lapse function, $d\Omega^{2} = d\theta^2+\sin^2(\theta) d\phi^2$ is the line element of a two-sphere and we have chosen zero-shift vector $\beta^{i}=0$ (comoving threading). The components of the four-velocity $u^{\mu}$ are given by $u^{t}=1/A$ and $u^{i}=0$ for $i=r,\theta,\phi$, since we are considering comoving coordinates (comoving slicing). 

Solving the Einstein field equations, the following quantities appear:
\begin{align}
\frac{1}{A(r,t)}\frac{\partial R(r,t)}{\partial t} &\equiv D_{t}R \equiv U(r,t),\nonumber \\
\frac{1}{B(r,t)}\frac{\partial R(r,t)}{\partial r} &\equiv D_{r}R \equiv \Gamma(r,t),
\label{2_covariantR}
\end{align}
where $D_{t}$ and $D_{r}$ are the proper time and distance derivatives, respectively. From now on, we will use the notation $R'\equiv \partial_{r} R$ (partial radial-derivative) and $\dot{R}\equiv \partial_{t} R$ (partial time-derivative). $U$ is the radial component of the four-velocity associated with an Eulerian frame (so not comoving), which measures the radial velocity of the fluid with respect to the origin of the coordinates. The Misner-Sharp mass $M(r,t)$ is introduced as:
\begin{equation}
M(R) \equiv \int_{0}^{R} 4\pi \tilde{R}^{2} \rho \, d\tilde{R}\, ,
\label{eq:misner_sharp}
\end{equation}
which is related with $\Gamma$, $U$ and $R$ though the constraint:
\begin{equation}
\Gamma = \sqrt{1+U^2-\frac{2 M}{R}},
\label{eq:gamma_constraint}
\end{equation}
where $\Gamma$ is called the generalised Lorentz factor, which includes the gravitational potential energy and kinetic energy per unit mass. Finally, the differential equations governing the evolution of a spherically symmetric collapse of a perfect fluid in general relativity are:
\begin{align}
\label{eq:U_time}
D_{t}U &= -\left[\frac{\Gamma}{(\rho+p)}D_{r}p+\frac{M}{R^{2}}+4\pi R p \right], \\
\label{eq:rho_time}
D_{t}\rho &= -\frac{(\rho+p)}{\Gamma R^{2}}D_{r} (U R^{2}), \\
\label{eq:R_time}
D_{t} R &= U, \\
D_{t}M &= -4 \pi R^{2} U p, \\
\label{GG}
D_{t}\Gamma &= \frac{A'}{A}\frac{U}{B},\\
\label{BB}
D_{t}B &= \frac{U'}{\Gamma},\\
\label{2_eqconstraint}
D_{r}M &= 4\pi \Gamma \rho R^{2}, \\
\label{2_lapse}
D_{r} A &= \frac{-A}{\rho+p}D_{r}p\, .
\end{align}
We refer the reader to \cite{1964PhRv..136..571M} for the details of the derivation. This previous set of equations is typically used to numerically study PBH formation \cite{Niemeyer:1999ak,Musco:2004ak,Nakama:2013ica,Bloomfield:2015ila,Escriva:2019nsa,Milligan:2025zbu,Ning:2025ogq}. Let's now consider a linear equation of state defined by $p= w \rho$ with constant parameter $w$, being $w=1/3$ for a radiation-dominated Universe. The lapse equation in Eq.~\eqref{2_lapse} can be solved analytically considering $A(r \rightarrow \infty,t) = 1$ to match with the \ac{FLRW} background,
\begin{equation}
\label{eq:lapse}
A(r,t) = \left(\frac{\rho_{b}(t)}{\rho(r,t)}\right)^{\frac{w}{w+1}},
\end{equation}
where $\rho_{b}(t) = \rho_{0}(t_{0}/t)^{2}$ is the energy density of the \ac{FLRW} background and $\rho_{0} = 3 H_{0}^{2}/8\pi$ is the initial value at time $t = t_0$, with $H_0$ being the initial Hubble factor defined as $H(t) = \dot{a}/a$ and $a$ is the scale factor (see \cite{Escriva:2019nsa} for the analytical solution using the same notation we use in this work). From Eqs.\eqref{eq:U_time}-\eqref{BB}, not all time-evolution equations are strictly necessary for simulating PBH formation. For instance, in \cite{Escriva:2019nsa} only $U,R,\rho,M$ are necessary and $B,\Gamma$ are obtained from their definitions using the other variables. However, the divergence term associated with type-II fluctuations is essentially caused by the time evolution of the energy density Eq.\eqref{eq:rho_time} and the function $B$ Eq.\eqref{BB} due to the existence of the term $U'/R'$. This is responsible for disrupting the numerical evolution from the beginning, since for type-II fluctuations, it is true that $R' = 0$ at least at one point in $r$. To address this issue, we define the term involved in $\dot{\rho}$, as an auxiliary variable to absorb the divergent term, i.e.,
\begin{equation}
K \equiv -\left(\frac{U'}{R'}+ 2 \frac{U}{R}\right).
\label{eq:K}
\end{equation}
This is indeed the trace of the extrinsic curvature $K = \gamma^{ij} K_{ij}$ with $K_{ij}=-\mathcal{L}_{n} \gamma_{ij}/2=(-\partial_t \gamma_{ij}+\mathcal{L}_{\beta}\gamma_{ij})/(2A)$ where $\mathcal{L}$ is the Lie-derivative, $\gamma_{ij}$ is the spatial metric of Eq.\eqref{2_metricsharp} and $n$ is the unit normal vector orthogonal to the hypersurfaces of constant $t$ ($\Sigma_t$), i.e., $n^{\mu} = (1/A, -\beta^{i}/A)$. In the comoving threading $\beta^{i} =0$ we simply have $K_{ij}=-\partial_t \gamma_{ij}/(2A)$. In the case of the Misner–Sharp formalism with the metric of Eq.\eqref{2_metricsharp}, we can easily verify, using the definition of the extrinsic curvature, that $K = -\left( \frac{\dot{B}}{B} + \frac{2 \dot{R}}{R} \right) / A$, which matches Eq.\eqref{eq:K} when the time-evolution equations Eq.\eqref{BB} and Eq.\eqref{eq:R_time} are used. In the flat \ac{FLRW} background, $K$ is proportional to the Hubble factor, $K_{b} = -3 H$. Therefore, we can consider Eq.~\eqref{eq:K} as a local quantity of the Hubble factor that incorporates the modification of the \ac{FLRW} background due to the cosmological fluctuation during the time evolution\footnote{In contrast to the comoving gauge, $K$ is forced to a constant value in the CMC (constant mean curvature) gauge, see for instance \cite{Harada:2015yda}.}. 

Additionally, we must carefully arrange the other terms to avoid $R'$ in the denominators, using the relation $R' = B \Gamma$, since the neck structure associated with type-II arises from a zero in the $\Gamma$ factor, rather than from the function $B$. From the definition in Eq.~\eqref{eq:K}, we can rewrite the time evolution of $B$ in terms of $K$ as $D_t B = -B  (K + 2U/R)$. Next, we need a time-evolution equation for $K$. The time-evolution equation for $K$ is given by (see for instance \cite{2016nure.book.....S,Baumgarte_Shapiro_2021})
\begin{equation}
\dot{K} = -\nabla^i \nabla_{i} A+A \left( K_{ij}K^{ij}+4 \pi (\rho+S)+\beta^{i} \nabla_{i} K\right),
\end{equation}
where $S=\gamma^{ij}T_{ij}$ and $\nabla_{i}$ is the covariant derivative in the spatial index "i". After some computations (we refer the reader to the appendix \ref{apendix_K} for details) we can obtain Eq.\eqref{eq:K_simply} which, along with the previous equations written in a convenient form for the numerical simulation as in \cite{Escriva:2019nsa}, gives us:
\begin{align}
\label{eq:u_simply}
\dot{U} &= -A\left[  \frac{M}{R^2}+4 \pi R w \rho \right]+\frac{A' \Gamma}{B}, \\
\label{eq:rho_simply}
\dot{\rho} &= A\rho (1+w) K, \\
\label{eq:r_simply}
\dot{R} &= A U, \\
\label{eq:M_simply}
\dot{M} &= -4  \pi A w \rho U R^2  \\
\label{eq:G_simply}
\dot{\Gamma} &= A' \frac{U}{B}, \\
\label{eq:B_simply}
\dot{B} &= -A B\left(K+2\frac{ U}{R}\right),\\
\label{eq:K_simply}
\dot{K} &= A\Biggl[\left(K+2\frac{ U}{R}\right)^2+2\left( \frac{U}{R}\right)^2 +4\pi \rho (1+3w) \Biggr] \\ \nonumber
&- \frac{1}{B^2} \left( A'' + A' \left(2 \frac{R'}{R} - \frac{B'}{B} \right)\right).\\ \nonumber
\end{align}
Notice that when considering the FLRW solution with $A_b=1, K_{b}=-3 H$, we recover the second \ac{FLRW} equation, $\dot{H} = -3(1+w)H^{2}/2$, from Eq.\eqref{eq:K_simply}. Instead of solving the time-evolution for the Misner-Sharp mass $M$ Eq.\eqref{eq:M_simply} it can be obtained at each time step during the numerical evolution through the constraint equation, Eq.\eqref{eq:gamma_constraint}, using
\begin{equation}
    M = \frac{R}{2} \left( 1+U^2-\Gamma^2\right).
    \label{eq:misner_sharp_mass}
\end{equation}

For type-II fluctuations, solving $\Gamma$ with the constraint Eq.\eqref{eq:gamma_constraint} would not be convenient, since $\Gamma$ takes negative values in the throat structure and therefore we solve it with the time-evolution equation Eq.\eqref{eq:G_simply}. The use of the trace of the extrinsic curvature $K$ as an auxiliary variable Eq.\eqref{eq:K_simply} in the Misner-Sharp equations allows for a formulation free from divergent terms associated with type-II fluctuations, and this corresponds to the new approach we use in this work. Finally, we will use the Hamiltonian constraint equations to check the accuracy of our numerical solution,
\begin{equation}
\mathcal{H} \equiv \frac{M'}{B}-4 \pi R^{2} \rho \Gamma.
\label{eq:H_constraint}
\end{equation}
In the next section, we introduce the initial conditions.
\section{Initial conditions, compaction function and trapping horizons}

At early times after inflation, cosmological perturbations have a physical wavelength $L$ much larger than the Hubble radius $H^{-1}$ \cite{Shibata:1999zs}. Consequently, we adopt the long-wavelength approximation \cite{Lyth:2004gb, Tanaka:2007gh} to determine the initial form of the metric and hydrodynamical variables. This approach involves expanding the exact solutions in a power series of the parameter

\begin{equation}
\epsilon(t) \equiv \frac{1}{H(t) L(t)},    
\end{equation}

and retaining only the lowest non-vanishing order in the regime $\epsilon(t) \ll 1 $ where $L(t) \equiv R_m(t) = a(t)r_m e^{\zeta(r_m)}$ and $r_m$ the comoving lenghtscale of the fluctuation. In the limit $\epsilon \rightarrow 0$, the spacetime metric in the zeroth-order of the long-wavelength solution, under the assumption of spherical symmetry, takes the form

\begin{equation}
ds^2 = -dt^2 + a^2(t) e^{2\zeta(r)} \left( dr^2 + r^2 d\Omega^2 \right),
\label{eq:spacetime_metric}
\end{equation}

where we have chosen a zero shift, $\beta^{i} = 0$ (comoving threading), and $\zeta(r)$ is the comoving super-horizon curvature fluctuation. We can now define two types of fluctuations for $\zeta$. Type-I fluctuations correspond to cases where the areal radius $R_b = a r e^{\zeta}$ is a monotonic function, whereas type-II corresponds to fluctuations in $\zeta$ that lead to non-monotonic behavior in $R$ ($R'< 0$) \cite{PhysRevD.83.124025}, which originates a throat-like structure in the initial condition. In particular, the points where $R' = 0$ satisfy $1 + r_{\textrm{II}} \zeta'(r_{\textrm{II}}) = 0$. The initial conditions for the system of equations in Eqs. \eqref{eq:u_simply} - \eqref{eq:K_simply} are then obtained by employing the quasi-homogeneous solution of the Misner-Sharp equation \cite{Polnarev:2006aa,Polnarev:2012bi}. This involves expanding in terms of the parameter $\epsilon$ to isolate the growing mode solution up to the leading order in the expansion,

\begin{align}
    U &= H R (1+\epsilon^2 \tilde{U}),\nonumber\\
    \rho &= \rho_b(1+\epsilon^2 \tilde{\rho}),\nonumber\\
    R &= a r e^{\zeta} (1+\epsilon^2 \tilde{R}),\nonumber.\\
    M &= \frac{4 \pi}{3} \rho_b R^3 (1+\epsilon^2 \tilde{M}),\nonumber\\
    B &= a e^{\zeta} (1+\epsilon^2 \tilde{B}),\nonumber\\
    A &= 1 + \epsilon^2 \tilde{A},\nonumber\\
    K &= -3 H (1+\epsilon^2 \tilde{K}).\nonumber\\
    \label{eq_ms}
\end{align}
Introducing this expansion into the Misner-Sharp equations and taking the leading order term in $\epsilon^{2}$ (we refer the reader to Appendix \ref{sec:derivation_QHS} for details of the computations), we obtain:
\begin{align}
\label{eq:U_tilde}
\tilde{U} &= \frac{1}{5+3w} e^{-2 (\zeta-\zeta_m)} \zeta' \left( \frac{2}{r}+\zeta' \right) r^2_m, \\ 
\label{eq:rho_tilde}
\tilde{\rho} &= \frac{-2(1+w)}{5+3w}e^{-2 (\zeta-\zeta_m)} \left[ \zeta''+\zeta'\left(\frac{2}{r}+\frac{\zeta'}{2}\right) \right]r^2_m,  \\ 
\label{eq:Rtilde}
\tilde{R} &= \frac{1}{1+3w}\left( -\frac{w}{w+1}\tilde{\rho}+\tilde{U}\right), \\ 
\tilde{M}&= -3(1+w) \tilde{U} , \\ 
\label{eq:Btilde}
\tilde{B} &= \frac{-1}{1+3w} (\tilde{{\rho}}+2 \tilde{U}) , \\ 
\label{eq:tilde_A}
\tilde{A} &= -\frac{w}{w+1}\tilde{\rho}, \\ 
\label{eq:tilde_K}
\tilde{K} &= -\frac{\tilde{\rho}}{3(1+w)}, 
\end{align}
where $\zeta_m \equiv \zeta(r_m)$.

Once the initial curvature $\zeta$ and its derivative are introduced in Eqs.\eqref{eq:U_tilde}-\eqref{eq:tilde_K}, the initial conditions for the numerical evolution of the cosmological fluctuation are set. 

The initial conditions from Eqs.~\eqref{eq:U_tilde}–\eqref{eq:tilde_A}, at first order in $\epsilon^{2}$, coincide with those in \cite{Musco:2018rwt} (see also \cite{Polnarev:2006aa,Polnarev:2012bi,Harada:2015yda}), but differ for $\tilde{B}$ [Eq.~\eqref{eq:Btilde}] (see Appendix~\ref{sec:derivation_QHS} for details). In that reference, the expansion of $B$ is written in terms of a transformation between the areal radial coordinate $\hat{r} = re^{\zeta(r)}$ (which makes the spacetime metric of Eq.\eqref{eq:spacetime_metric} resemble the flat FLRW metric with a non-homogeneous curvature $\chi(\hat{r})$)\footnote{With this coordinate, the spatial metric reads as
\begin{equation}
    d\hat{\Sigma}^2(\epsilon \rightarrow 0) = a^2 \left(\frac{d\hat{r}^2}{1 - \chi(\hat{r})\hat{r}^2} + \hat{r}^2 d{\Omega}^2\right)
    \label{eq:mtric_K}
\end{equation}
} to the conformally flat coordinate $r$ (see \cite{Harada:2015yda} for the transformation between the coordinates $r, \hat{r}$ and the curvatures $\zeta(r), \chi(\hat{r})$) using the result of \cite{Polnarev:2006aa}. However, the metric written in the $\hat{r}$ coordinate cannot cover an extremal surface, corresponding to the coordinate singularity $\chi(\hat{r}) \hat{r}^2 = 1$ \cite{PhysRevD.83.124025,Harada:2024jxl}, and therefore a coordinate singularity associated with the throat structure $R'(r_{\textrm{II}}) = 0$ for type-II fluctuations will be present. For this reason, we make the expansion directly from the asymptotic value $B(\epsilon \rightarrow 0) = a e^{\zeta}$ in the $r$ coordinate to avoid this issue and obtain a new initial condition free from divergent terms related to type-II fluctuations.

Although we provide the term $\tilde{K}$ at first order in $\epsilon^{2}$, we have found that setting up the initial condition from the exact definition in Eq.\eqref{eq:K} makes the numerical evolution much more stable for the cases tested, with the Hamiltonian constraint decreasing from the beginning of the simulation rather than growing. The definition given in Eq.\eqref{eq:K} contains the divergent term associated with the type-II fluctuations; however, using an alternative definition for the time-derivative of the function $B$, obtained from the other Misner-Sharp equations (see for instance \cite{Deng:2017uwc} in the context of PBH formation from vacuum bubbles), we have derived a new equation,

\begin{equation}
\label{eq:alternative}
    K = \frac{-1}{A}\left(\frac{\dot{B}}{B} +2 \frac{\dot{R}}{R}\right)=  \frac{-1}{U}\left(4 \pi \rho R-\frac{M}{R^2}+\frac{\Gamma'}{B} \right) -\frac{2 U}{R},
\end{equation}
which does not have a divergent term at $r_{\rm II}$. Notice that this expression exhibits divergent behavior when $U=0$, which is expected to occur during the non-linear regime of the collapse, when the Eulerian velocity will likely become negative. However, we are only using Eq.\eqref{eq:alternative} to set up the initial conditions, where the background $U_{b}=HR_{b}$ is positive definite (see Eq.\eqref{eq_ms}), and therefore the behavior at $r\rightarrow 0$ of $K(r \rightarrow 0,t_0)$ is regular if the initial conditions in $\zeta$ are also regular. Then, the initial condition for $\Gamma$ is directly determined from the other variables with $\Gamma = R' / B$. The initial condition for the Misner-Sharp mass $M$ is obtained using the constraint equation Eq.\eqref{eq:misner_sharp_mass}, once the other variables have been fixed.

On the other side, a relevant quantity in the context of PBH formation is the compaction function (first introduced in \cite{Shibata:1999zs}), which is defined in the comoving gauge (see \cite{Harada:2023ffo,Harada:2024trx} for a recent discussion in other gauges and about its definition) as twice the mass excess over the areal radius
\begin{equation}
    \mathcal{C} = 2 \frac{M-M_b}{R},
\end{equation}
where $M_b = 4\pi \rho_b R^3/3$. At leading order in the gradient expansion, the compaction function corresponds to a non-linear relation in terms of the curvature fluctuation $\zeta$ \cite{Harada:2015yda}\footnote{Note that the factor $f(w)$ depends on the gauge choice; see \cite{Harada:2023ffo}},
\begin{equation}
\mathcal{C}(r) = f(w)\left( 1-(1+r \zeta')^2\right)\, \,  , \,\, f(w) = \frac{3(1+w)}{5+3w} ,
\end{equation}
which is a time-independent quantity. This is essential for setting up time-independent initial conditions for PBH formation, as long as the fluctuations remain in the super-horizon scale regime. In particular, the peak value of the compaction function $\mathcal{C}(r_m)$, where $r_m$ is the radial coordinate value at which the maximum is found, satisfies $\zeta'(r_m) + r_m \zeta''(r_m) = 0$, and has been found useful for characterize the initial conditions and defining the criteria for black hole formation in the case of type-I fluctuations \cite{Musco:2018rwt,Escriva:2019phb}. For type-II fluctuations, the compaction function is characterized by a minimum at the scale $r_m$, surrounded by two maxima, both of which satisfy $1 + r_{\rm II}  \zeta'(r_{\rm II}) = 0$ (see for instance \cite{Uehara:2024yyp}). The maximum value of $\mathcal{C}$ in the comoving gauge is given by $f(w)$, which corresponds to the points $r_{\rm II}$ where $R'(r_{\rm II})=0$.

In this work, as a source for the initial curvature fluctuation $\zeta$, we use the following standard exponential-shaped curvature profiles modulated by the exponent $\beta$, with $\mu$ being the peak value of $\zeta$,
\begin{equation}
    \zeta(r)= \mu \, e^{ -\left(r/r_m\right)^{2\beta} }.
    \label{eq:profile_exp}
\end{equation}

The parameter $\beta$ modulates the stiffness of the decay of the shape of $\zeta$. With this curvature profile (which we assume follows Gaussian statistics), we aim to explore the effect of profile dependence by varying the parameter $\beta$. Typically, in our simulations, we need $\epsilon \lesssim 10^{-1}$ to satisfy the quasi-homogeneous solution and ensure a small initial violation of the Hamiltonian constraint, which we adjust by modifying the initial length-scale of the fluctuation $r_m$ in terms of the Hubble $H(t_0)$, $r_m = r_{N}R_{H}(t_0)$ (since we take $a(t_0)=1$), where $r_N$ represents the number of initial cosmological horizons $R_{H}(t_0)$. However, for profiles with large derivatives in $\zeta$ (large $\beta$), the initial Hamiltonian constraint may be violated more significantly. For this reason, for such cases we increase the length-scale of the fluctuation with respect to the initial cosmological horizon, reducing the initial value of $\epsilon$.

\begin{figure}[t]
\centering
\includegraphics[width=2.6 in]{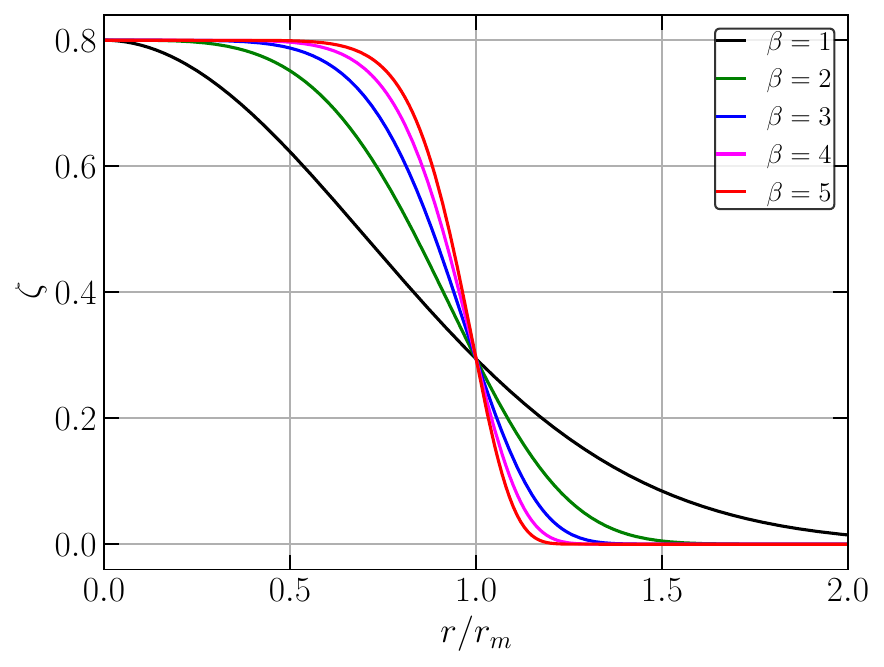}
\hspace*{-0.3cm}
\includegraphics[width=2.6 in]{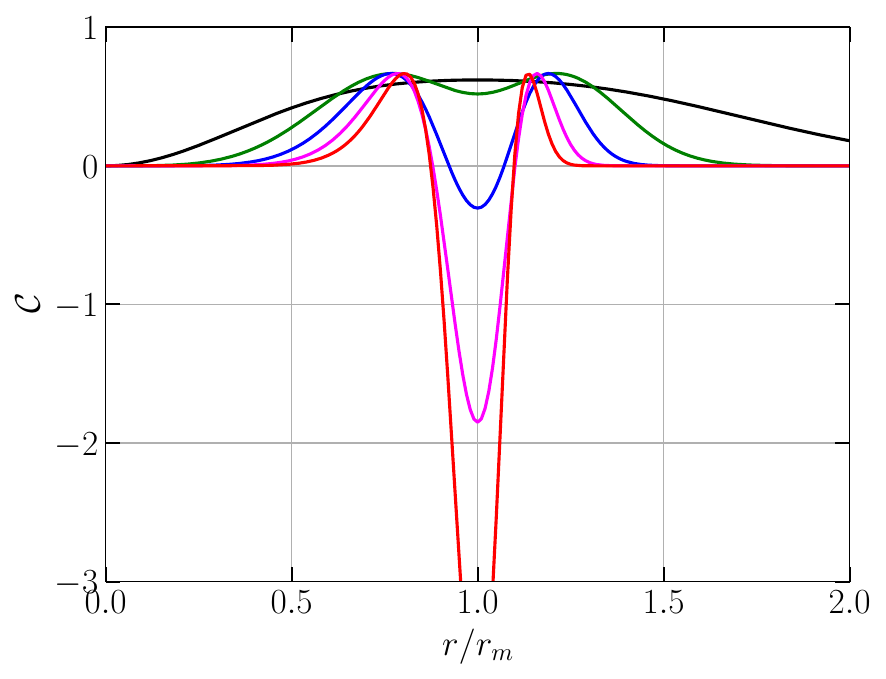}
\hspace*{-0.3cm}
\caption{Profiles of the curvature $\zeta(r)$ (left-panel) and compaction function (right panel) for different values of $\beta$ and fixing $\mu=1.0$}
\label{fig:initial_zeta}
\end{figure}

To obtain the critical value of $\mu_{c,\rm II}$ that gives us the boundary between the type-I and type-II regions (type-II fluctuations occur when $\mu >\mu_{c,\rm II}$), we simply need to solve $R'(r_m) = 0 \Rightarrow 1 + r_m \zeta'(r_m) = 0$, from which we find $\mu_{c,\rm II} = e / (2\beta)$. The case $\mu=\mu_{c,\rm II}$ corresponds to the marginal case with a single point $r_{\rm II}$ where $R'(r_{\rm II})=0$, which results in a compaction function with $\mathcal{C}''(r_m)=0$. Plots for the different initial conditions can be found in Fig.\ref{fig:initial_zeta}. It can be seen that the type-II configuration, with two characteristic peaks in the compaction function, appears when the amplitude $\mu$ exceeds the critical value $\mu_{c,\rm II}$.

For the purpose of studying the dynamics of the fluctuations, we define two reference quantities: the time of horizon crossing $t_H$ when the fluctuation reenters the cosmological horizon neglecting the curvature perturbation $\zeta$\footnote{The horizon crossing time defined here is not the true one, which should be obtained through the numerical simulation to account for the full non-linear evolution of the fluctuation. However, for simplicity, we define a time-scale $t_H$, neglecting the effects of the fluctuation $\zeta$ on the FLRW background and its dynamics.} and the mass of the cosmological horizon at horizon reentry $M_H$,
\begin{equation}
\label{eq:background_tH_MH}
t_H = t_0 \left( a_0 r_m H_0 \right)^{\frac{1}{1 - \alpha}} ,   \,\,  M_H = \frac{1}{2 H(t_H)} = \frac{1}{2}\left( a_0 r_m H_0^\alpha \right)^{\frac{1}{1 - \alpha}} \,\, ,\alpha = \frac{2}{3(1 + w)}.
\end{equation}


On the other side, to identify trapped surfaces and characterize the trapping horizons of type A/B PBH configurations introduced in \cite{Uehara:2024yyp} following the nomenclature of \cite{PhysRevD.49.6467,PhysRevD.53.1938}, we consider the expansion $\Theta^{\pm} \equiv h^{\mu\nu} \nabla_{\mu}k_{\nu}^{\pm}$ of null geodesic congruences $k^{\pm}_{\mu}$ orthogonal to a spherical surface $\Sigma$. Here $h_{\mu\nu}$ is  the metric induced on $\Sigma$. There are two such congruences, which we may call inward ($k_{\mu}^{-}$) and outward-directed ($k_{\mu}^{+}$), with components $k_{\mu}^{\pm} = (-A,\pm B,0,0)$, such that $k^+\cdot k^- = -2$. 

Surfaces $\Sigma$ with $\Theta^{-}<0$, while $\Theta^{+}>0$ are called ``untrapped" which corresponds to a flat spacetime. If both expansions are negative, the surface is called ``trapped", while if both are positive, the surface is ``anti-trapped". In terms of the Misner-Sharp variables (see for instance \cite{Faraoni:2016xgy,Helou:2016xyu}), we have
\begin{gather}
    \Theta^{\pm}=\dfrac{2}{R}(U\pm \Gamma).
    \label{eq:thetas}
\end{gather}

In a spherically symmetric spacetime, any point in the $(r,t)$ plane can be thought of as a closed surface $\Sigma$ with proper radius $R(r,t)$. Trapping horizons at the point $r_*$ then satisfy $\Theta^{+} \Theta^{-} = 0$, which, using the constraint Eq.\eqref{eq:gamma_constraint} and Eq.\eqref{eq:thetas}, gives $2M(r_*,t) = R(r_*,t)$. The trapping horizons can then be either inner ($\mathcal{L}_{-} \Theta^{+} > 0$), outer ($\mathcal{L}_{-} \Theta^{+} < 0$) or degenerate ($\mathcal{L}_{-} \Theta^{+} = 0$), where the symbol $\mathcal{L}_{-}$ denotes the Lie derivative along the congruence $k^{-}$. In general, we can define the Lie derivative of $\Theta^{q} =(\Theta^{+},\Theta^{-})$ along the congruence $k_{p}^{\mu} = (k^{\mu(+)},k^{\mu(-)})$ as $\mathcal{L}_{p} \Theta^{q} = k_{p}^{\mu} \partial_{\mu} \Theta^{q}= (D_{t}+p D_{r})\Theta^{q}$, where $q,p$ take the values $(+1,-1)$. In terms of the Misner-Sharp variables, we can write the following general expression,
\begin{align}
\label{eq:lie_congruences}
\nonumber
\mathcal{L}_{p} \Theta^{q} &= \frac{2}{R} \biggl[ \frac{M}{R^{2}}(q \,p-1) -\frac{w}{w+1}\frac{\rho'}{\rho B}(\Gamma+q \, U)-p\left(K+\frac{2 U}{R}\right)(\Gamma+q U) \\
&-4 \pi \rho R(q \, p+w)-\frac{(U+q \Gamma)(U+p \Gamma)}{R} \biggr],
\end{align}

where we made use of Eq.\eqref{eq:K} to eliminate the divergences associated with type-II fluctuations. We refer the reader to the appendix \ref{apendix_L} for details about the derivation.

In this work, we are mainly interested on the identification of type A/B PBH (we refer the reader to \cite{Uehara:2024yyp} for a more detailed discussion and analysis about the horizon structure) during the numerical evolution, and therefore for practical purposes we simply classify the horizons as: the apparent horizon of a spherically symmetric black hole corresponds to a marginally trapped surface which fulfills $\Theta^{+}(r_*,t)=0, \, \Theta^{-}(r_*,t)<0$ (which satisfies $U(r_*,t) = -\Gamma(r_*,t)$), the cosmological horizon corresponds to a marginally anti-trapped surface where $\Theta^{-}(r_*,t)=0, \, \Theta^{+}(r_*,t)>0$ (which satisfies $U(r_*,t) = \Gamma(r_*,t)$) and bifurcated trapping horizons corresponds to $\Theta^{+}(r_*,t) = \Theta^{-}(r_*,t) = 0$, which implies that $U(r_*,t) = \Gamma(r_*,t) = 0$ and its formation denotes type B PBH. This means that at the location $r_{*}$ of the bifurcated trapping horizon, the Eulerian velocity $U$ vanishes at the same $r_*$ where $R'=0$ is found. In section \ref{sec:num_results}, we will focus on studying the congruence expansion $\Theta^{\pm}$ in the numerical simulations to identify type A/B PBH.

\section{Numerical methodology}
\label{sec:method}
To implement the new approach, we develop a numerical code based on, and updating, the methodology of \cite{Escriva:2019nsa}, named SPriBHoS-II (extending the original SPriBHoS\footnote{SPriBHoS was the first publicly available code for spherically symmetric simulations of PBH formation from the collapse of adiabatic fluctuations with a perfect fluid.}). A basic version is available at \cite{codigo_albert}. We use the Pseudo-Spectral Chebyshev Collocation Method for computing radial derivatives in Eqs.~\eqref{eq:u_simply}–\eqref{eq:K_simply}, and evolve the system in time using a fourth-order Runge-Kutta (RK4) scheme. Spectral methods offer exponential error decay with $N_{\rm cheb}$, as they compute derivatives globally using all grid points.

To introduce spectral methods, we briefly summarize the main ideas following \cite{Escriva:2019nsa}; for a detailed treatment, see \cite{doi:10.1137/1.9780898719598}. Consider approximating a function $f(x)$ using $N_{\rm cheb}$ Chebyshev polynomials $f_{N_{\rm cheb}}(x) = \sum_{k=0}^{N_{\rm cheb}} c_{k} T_{k}(x)$ where $T_k(x)$ are Chebyshev polynomials. The coefficients $c_k$ are determined by imposing $f_{N_{\rm cheb}}(x_k) = f(x_k)$ at the Chebyshev collocation points $x_k = \cos(k\pi / N_{\rm cheb})$, which satisfy $T'_k(x_k)=0$. The interpolated function can be written as
\begin{align}
f_{N_{\rm cheb}}(x) &= \sum_{k=0}^{N_{\rm cheb}} L_{k}(x) f(x_k),\
\, \, L_k(x) = \frac{(-1)^{k+1}(1 - x^2) T'{N_{\rm cheb}}(x)}{\bar{c}_k N_{\rm cheb}^2 (x - x_k)},
\label{eq:spec2}
\end{align}
with $\bar{c}_k = 2$ for $k = 0, N{\rm cheb}$ and $\bar{c}_k = 1$ otherwise. The $L_k(x)$ are Lagrange interpolation polynomials. The $n$-th derivative at $x_i$ is then given by $f^{(n)}_{N{\rm cheb}}(x_i) = \sum_{k=0}^{N_{\rm cheb}} L^{(n)}_k(x_i) f(x_k)$. The Chebyshev differentiation matrix is then defined as $D^{(n)}=\{L^{(n)}_{k}(x_{i})\}$, we referee the reader to \cite{Escriva:2019nsa} for details about the exact numerial coefficients.

The $n$-th derivative is computed via repeated multiplication of the Chebyshev differentiation matrix: $D^{(n)} = (D^{(1)})^n$. Numerical stability is governed by the CFL condition \cite{1967IBMJ...11..215C}, requiring smaller time steps $dt$ as the spatial resolution ($N_{\rm cheb}$) increases. Following \cite{Escriva:2019nsa}, we use a time-dependent conformal step size: $dt = dt_0 (t/t_0)^\alpha$, with $dt_0 = 10^{-3}$. The radial domain is defined as $\Omega = [r_{\rm min}, r_{\rm max}]$, with $r_{\rm min} = 0$ and $r_{\rm max} = N_H R_H(t_0)$, where $N_H \approx 10^2$. A linear mapping is applied to transform the spectral domain $[-1, 1]$ to the physical domain:

\begin{equation}
\tilde{x}_{k} = \frac{r_{\rm max}+r_{\rm min}}{2}+\frac{r_{\rm max}-r_{\rm min}}{2}x_{k} \, ;\,\,\, \tilde{D} = \frac{2}{r_{\rm max}-r_{\rm min}} D.
\end{equation}

$\tilde{D}$ and $\tilde{x}_{k}$ are the new Chebyshev matrix and grid points rescaled to our domain $\Omega$. To implement a Dirichlet boundary condition at a given $x_{k}$, such that $f(\tilde{x}=\tilde{x}_{k})=u_{D,bc}$, we just need to replace the value at the grid point: $f_{N_{\rm cheb}}(\tilde{x}=\tilde{x}_{k})=u_{D,bc}$ directly. In contrast, for a Neumann boundary condition such that $f^{(1)}(\tilde{x}=\tilde{x}_{k})=u_{N, \rm bc}$, we have $(\tilde{D}_{(k)} \cdot f_{N_{\rm cheb}})(\tilde{x}=\tilde{x}_{k})=u_{N, \rm bc}$, where $\tilde{D}_{(k)}$ is the $k$-th row of the matrix $\tilde{D}$.

The boundary conditions we have implemented are: (I) Dirichlet condition, with  $U(r=0,t) = R(r=0,t) = 0$ and $\Gamma(r=0,t) = 1$. (II) Neumann condition with the pressure gradient vanishing at the outer boundary of the grid, $\rho'(r_{\rm max},t) = 0$, to approach the FLRW background solution and help prevent reflections from density waves. Additionally, $ K'(r=0,t) = 0 $ ensures the regularity of the evolution at the center.

This choice of boundary conditions has worked well for the purposes of our study. To impose the boundary conditions for (I) at the points $k=0, N_{\rm cheb}$, we directly modify the values of the fields at the boundary points. For the case (II) we simply do:  

\begin{equation}
    u_{\rm k, new} = u_{k}+\frac{1}{\tilde{D}_{k,k}}(u_{\rm N,bc}- \tilde{D}_{(k)} \cdot f_{N_{\rm cheb}}),
\end{equation}
where $u_{\rm k,new}$ represents the new value of the field at $f_{N_{\rm cheb}}(\tilde{x} =\tilde{x}_k)$ such that fulfills the Neumann condition.

On the other hand, for initial data close to its critical threshold or for sharp profiles, large gradients will develop during the numerical evolution, for which more refinement in the specific domain where the gradients are developed would be convinient. To manage these cases, we have implemented a fixed mesh refinement procedure updating the one used in \cite{Escriva:2020tak}. This procedure consists of using overlapping Chebyshev grids to increase the number of points in regions where large gradients develop. For this purpose, we split the domain into several grids that we denote with index $l$: $\tilde{x}_{k,l}$, $\tilde{D}_l$, with $N^{(l)}_{\rm cheb}$ points in each $l$ grid. This requires imposing boundary conditions across the domain to propagate the information during the numerical evolution. To achieve this, in this work, we enforce continuity of the field values and their derivatives: $ f(\tilde{x}_{N^{(l)}{\rm cheb},l}) = f(\tilde{x}_{0,l+1}) $ and $ f^{(1)}(\tilde{x}_{N^{(l)}{\rm cheb},l}) = f^{(1)}(\tilde{x}_{0,l+1}) $ across the domains.

\begin{align}
\nonumber
u_{N^{(l)}_{\rm cheb},l}=u_{0,l+1} &= \frac{1}{\tilde{D}_{(N^{(l)}_{\rm cheb},N^{(l)}_{\rm cheb}),l}-\tilde{D}_{(0,0),l+1}}\bigl[ \tilde{D}_{(0),l+1} \cdot f_{N^{(l+1)}_{\rm cheb}} -\tilde{D}_{(N^{(l)}_{\rm cheb}),l} \cdot f_{N^{(l)}_{\rm cheb}} \, \\ 
&+ \tilde{D}_{(N^{(l)}_{\rm cheb},N^{(l)}_{\rm cheb}),l}\, f_{N^{(l)}_{\rm cheb}}(\tilde{x}_{N^{(l)}_{\rm cheb},l})- \tilde{D}_{(0,0),l+1} f_{N^{(l+1)}_{\rm cheb}}(\tilde{x}_{0,l+1}) \bigr].
\end{align}
This worked well for our purposes. In addition, for certain cases (mainly the computation of the PBH mass for fluctuation amplitudes very close to its threshold), we implemented an automatic grid refinement by monitoring the numerical accuracy of the simulation using the Hamiltonian constraint (Eq. \ref{eq:H_constraint}). Specifically, once the averaged Hamiltonian constraint increases by a factor $\sim 3$, we refine the existing grid in the central region near the horizon, increasing the number of points by a factor of approximately $\sim 1.2$. The data is interpolated in the new refinment grid by using Eq.\eqref{eq:spec2}.


\section{Numerical results}
\label{sec:num_results}
In the following subsections, we show the numerical results of our study using the curvature profiles $\zeta$ from Eq.\eqref{eq:profile_exp}. While we usually consider a perfect fluid with a constant equation of state $w$ in the analytical equations, our numerical study focuses on the case of a radiation-dominated Universe with $w=1/3$. The aim of this section is twofold: on one hand, to test and demonstrate the accuracy and robustness of the new approach when simulating type-II fluctuations; on the other hand, to explore various aspects of type-II fluctuations using the specific shape template defined in Eq.~\eqref{eq:profile_exp}, by varying the parameter $\beta$.

\subsection{Time-evolution of the gravitational collapse}
We study the dynamics of different hydrodynamic variables, in particular $R$, $U$, $\rho$, $M$, and $\Gamma$, as well as the time evolution of the compaction function $\mathcal{C}$, the compactness $2M/R$, and the expansion of the congruences $\Theta^{\pm}$ with the new Misner-Sharp approach. We consider three different scenarios focused on type-II fluctuations: PBH formation from type-II fluctuations without the formation of a bifurcated trapping horizon (leading to type-II A PBHs), type-II fluctuations with the formation of a bifurcated trapping horizon (leading to type-II B PBHs), and type-II fluctuations that do not form PBHs.

\begin{figure}[t]
\centering
\includegraphics[width=1.9 in]{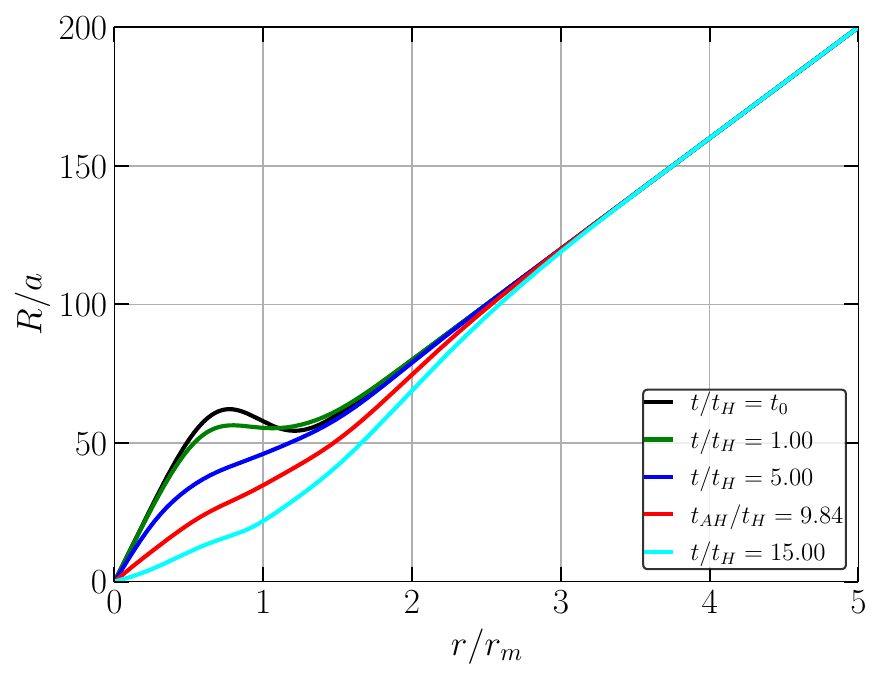}
\hspace*{-0.3cm}
\includegraphics[width=1.9 in]{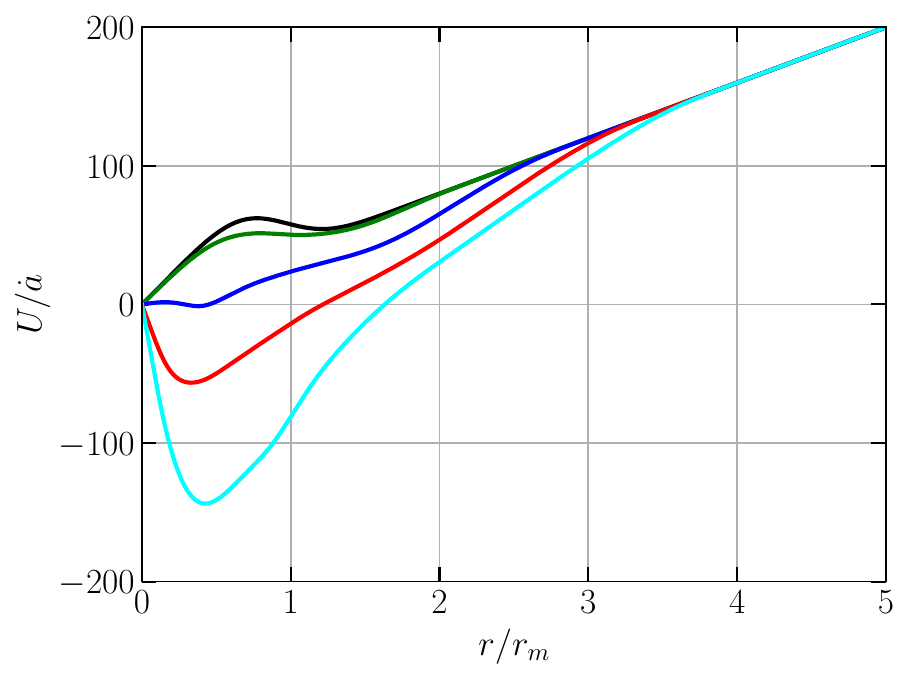}
\hspace*{-0.3cm}
\includegraphics[width=1.9 in]{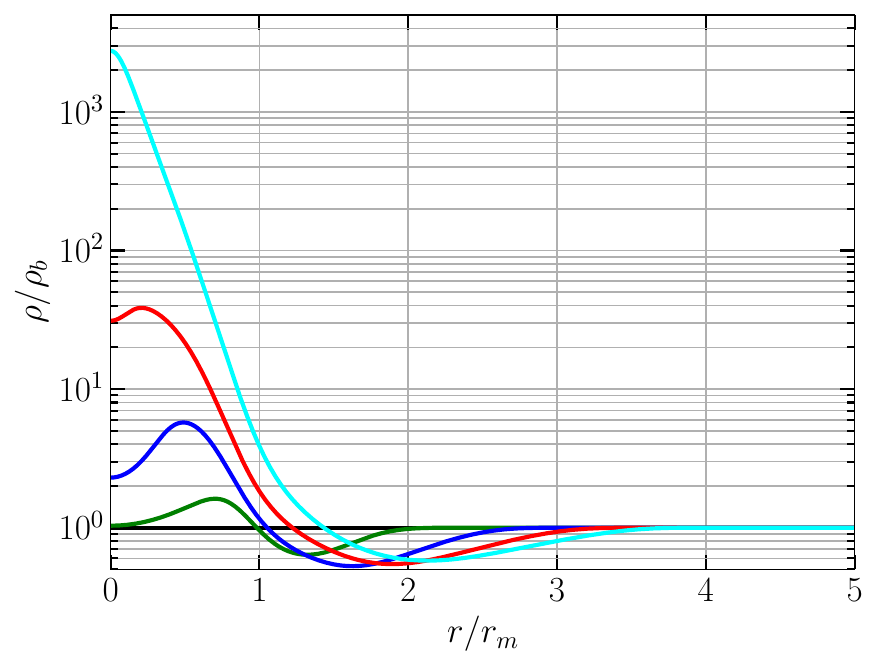}
\hspace*{-0.3cm}
\includegraphics[width=1.9 in]{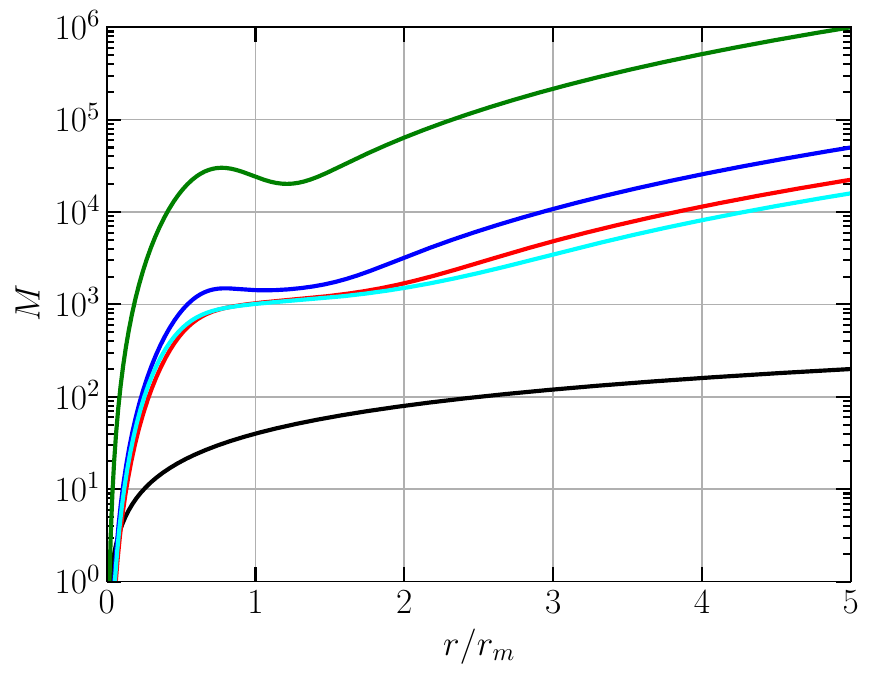}
\hspace*{-0.3cm}
\includegraphics[width=1.9 in]{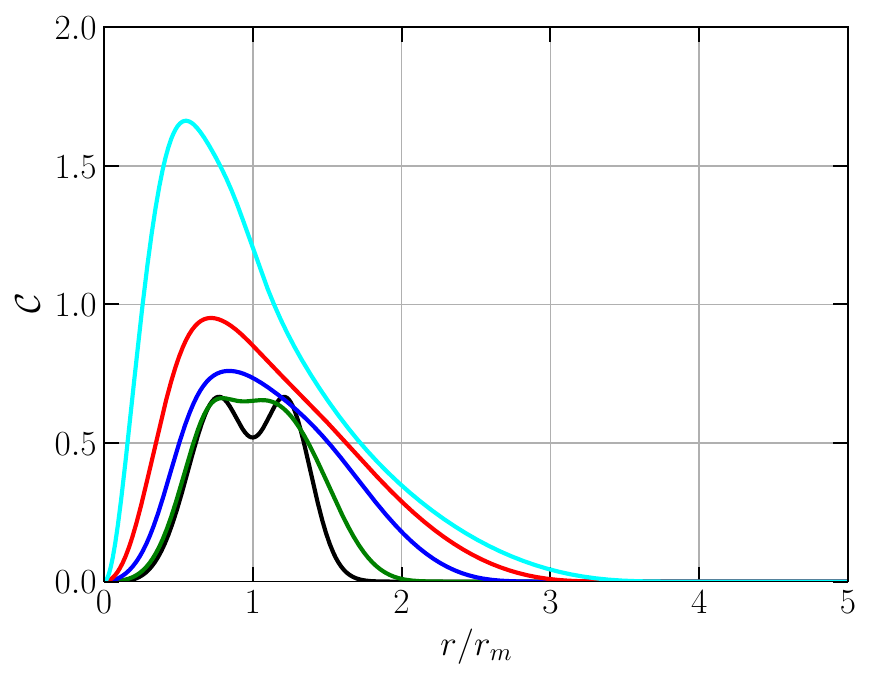}
\hspace*{-0.3cm}
\includegraphics[width=1.9 in]{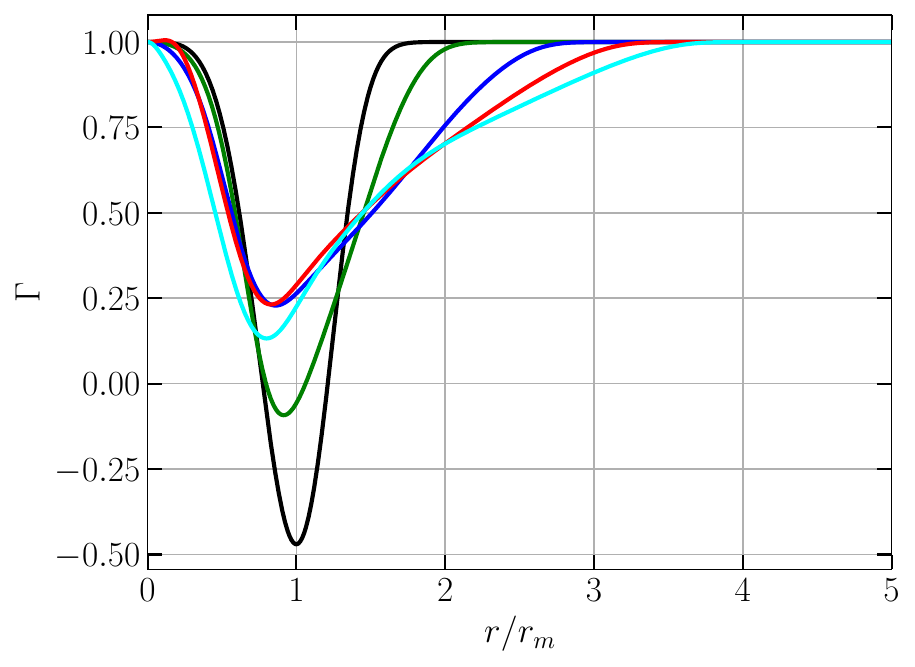}
\hspace*{-0.3cm}
\includegraphics[width=1.9 in]{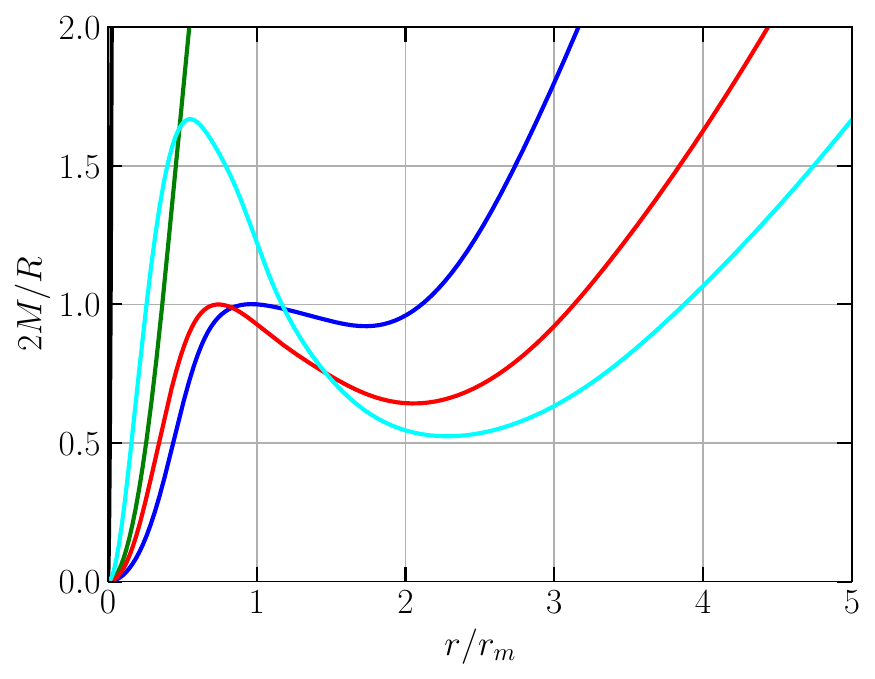}
\hspace*{-0.3cm}
\includegraphics[width=1.9 in]{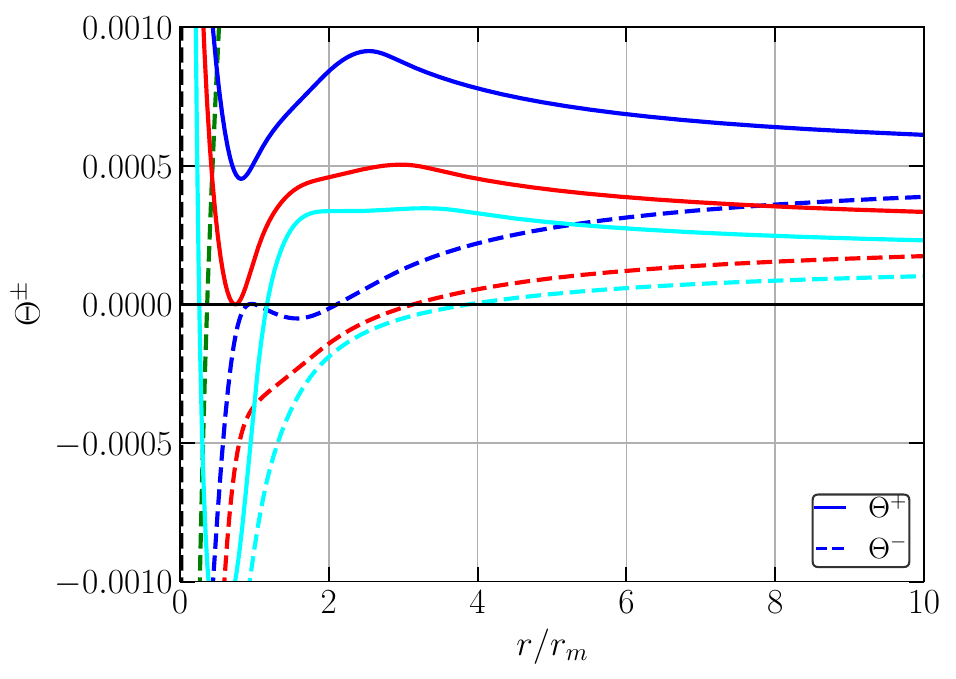}
\hspace*{-0.3cm}
\includegraphics[width=1.9 in]{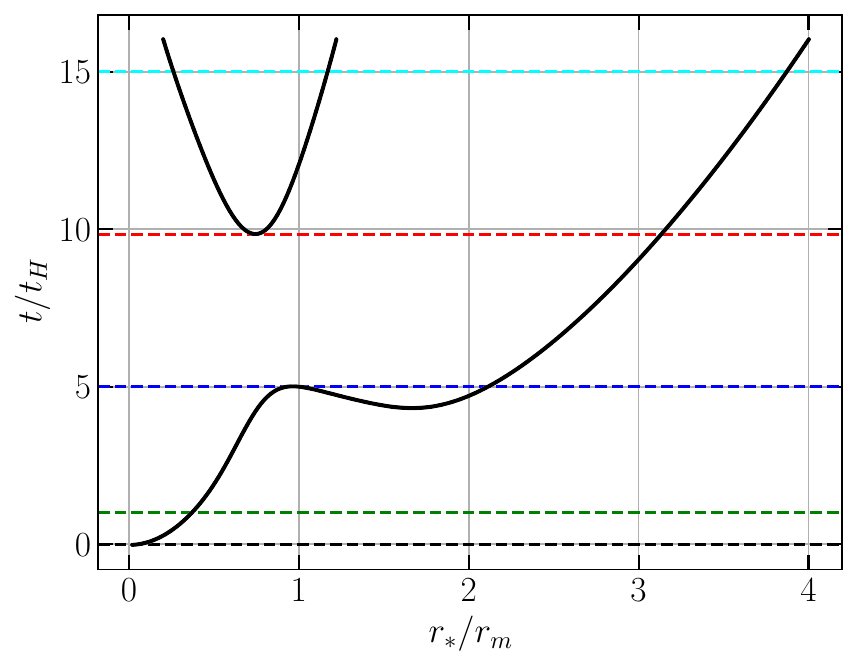}
\hspace*{-0.3cm}
\caption{Snapshots of the evolution of different quantities for the case $\beta=2$ with $\mu=1.0$ with $r_m = 20 R_{H}(t_0)$, corresponding to a case of PBH formation with type II-A PBH.}
\label{fig:dynamics1}
\end{figure}

First, we present the results in Fig.\ref{fig:dynamics1} for the parameters $\beta = 2$ and $\mu = 1.0$, corresponding to a scenario where the fluctuation amplitude is significantly above the threshold, resulting in the formation of type II-A PBHs. At the initial time $t_0$, we observe the throat-neck structure in the areal radius $R$, which causes the function $\Gamma$ to take negative values. The initial compaction function exhibits two distinct peaks, with a local minimum at $r_m$. As the fluctuation reenters the cosmological horizon, the non-linear regime of gravitational collapse begins, and we observe a behaviour similar to that of over-threshold type-I fluctuations: a collapsing process in which the peak of the energy density increases continuously over time, accompanied by a negative Eulerian velocity near the central region. We also note that the Misner-Sharp mass develops a local minimum during a period of the evolution. This behaviour is a consequence of the non-monotonicity of the areal radius, which causes the Misner-Sharp mass not to increase monotonically with $R$ (see Eq.\eqref{eq:misner_sharp}), specifically, when $\Gamma<0$, the integrand in Eq.~\eqref{eq:misner_sharp} ($4 \pi \tilde{R}^2 \rho \Gamma B$), taken over the comoving radial coordinate $r$ includes a locally "negative mass" component within that radius. At sufficiently late times, the compaction function develops a single peak and $\Gamma$ becomes positive throughout the entire radial domain (indicating that the non-monotonic behaviour of $R$ disappears). At later times, specifically around $t/t_H \approx 9.84$ , a black hole is formed (see the lower-middle and lower-right panels for the expansion of the congruences and the configuration of the horizons), corresponding to the typical trapping horizon configuration associated with type-A PBHs.

\begin{figure}[t]
\centering
\includegraphics[width=1.9 in]{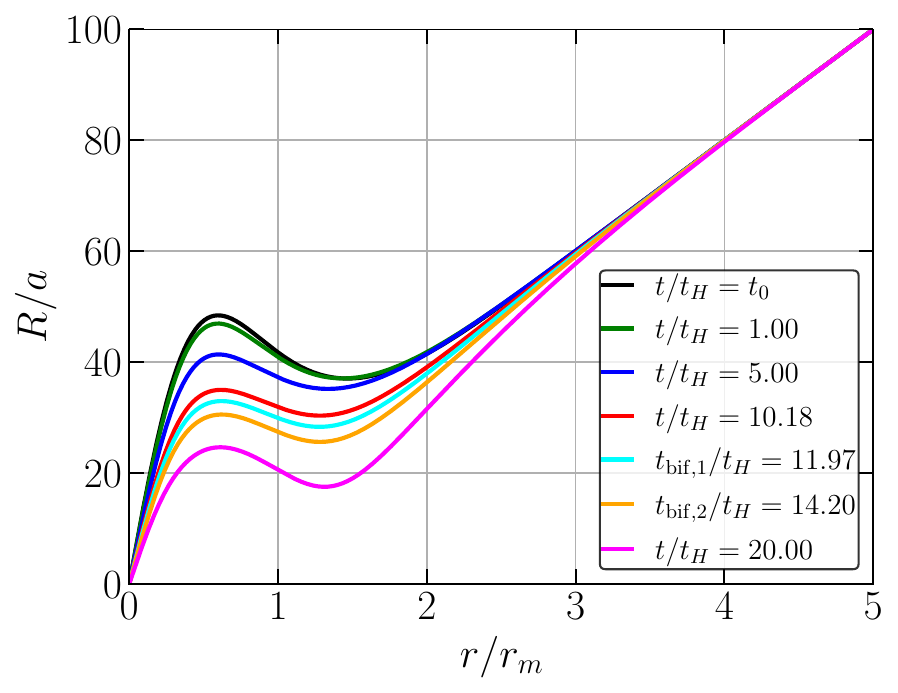}
\hspace*{-0.3cm}
\includegraphics[width=1.9 in]{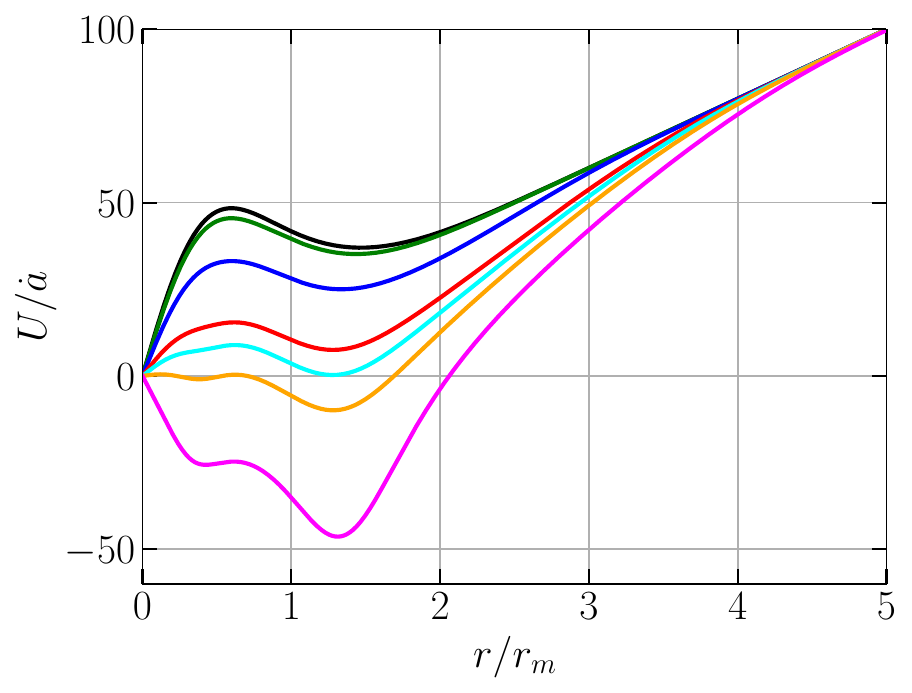}
\hspace*{-0.3cm}
\includegraphics[width=1.9 in]{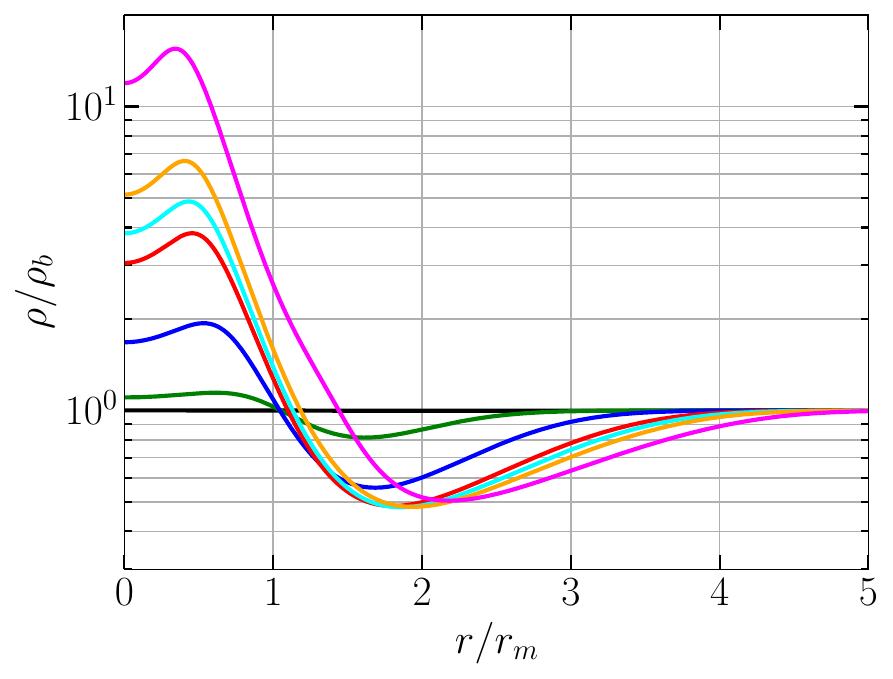}
\hspace*{-0.3cm}
\includegraphics[width=1.9 in]{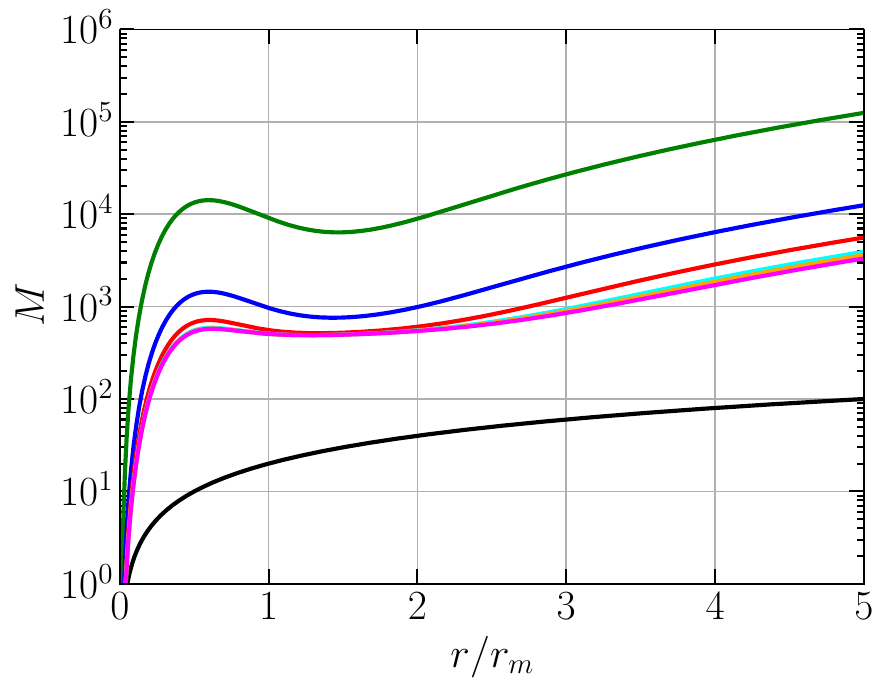}
\hspace*{-0.3cm}
\includegraphics[width=1.9 in]{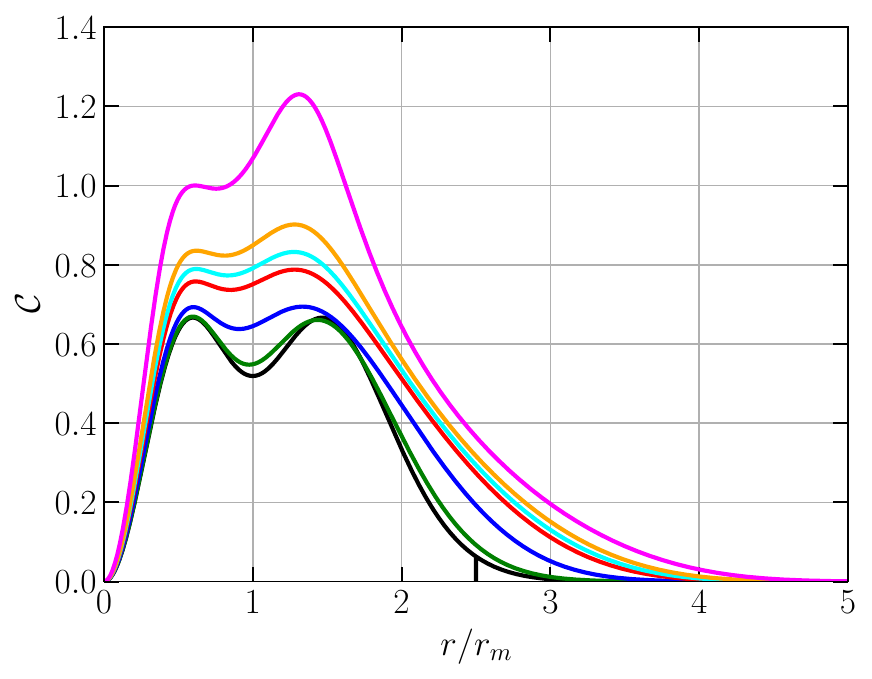}
\hspace*{-0.3cm}
\includegraphics[width=1.9 in]{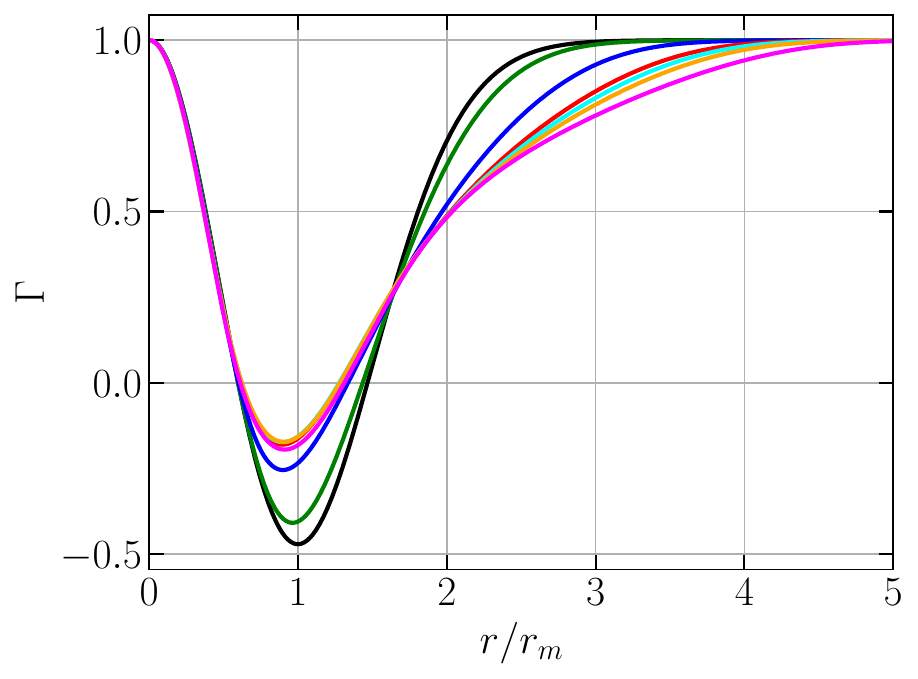}
\hspace*{-0.3cm}
\includegraphics[width=1.9 in]{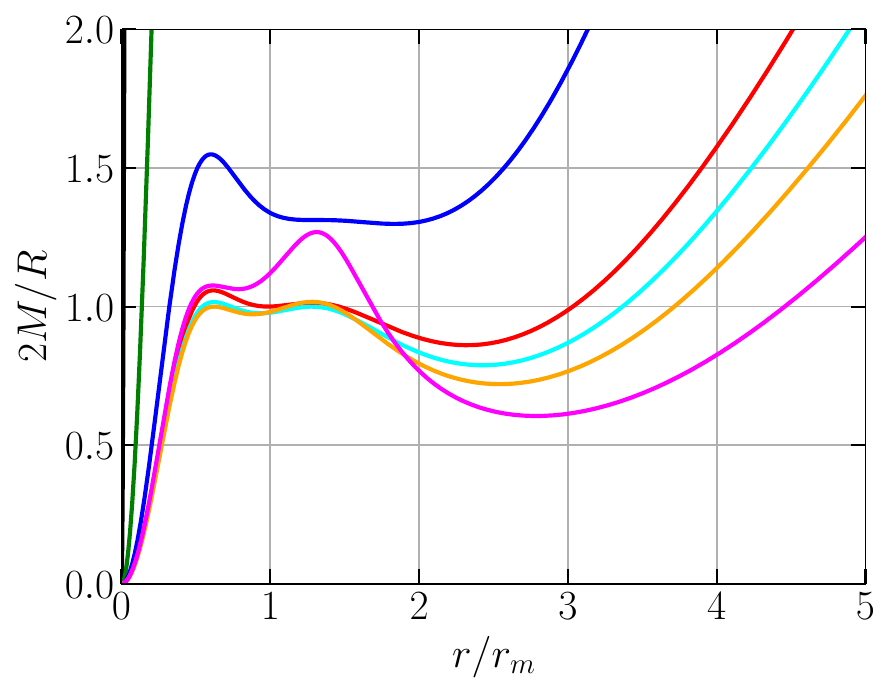}
\hspace*{-0.3cm}
\includegraphics[width=1.9 in]{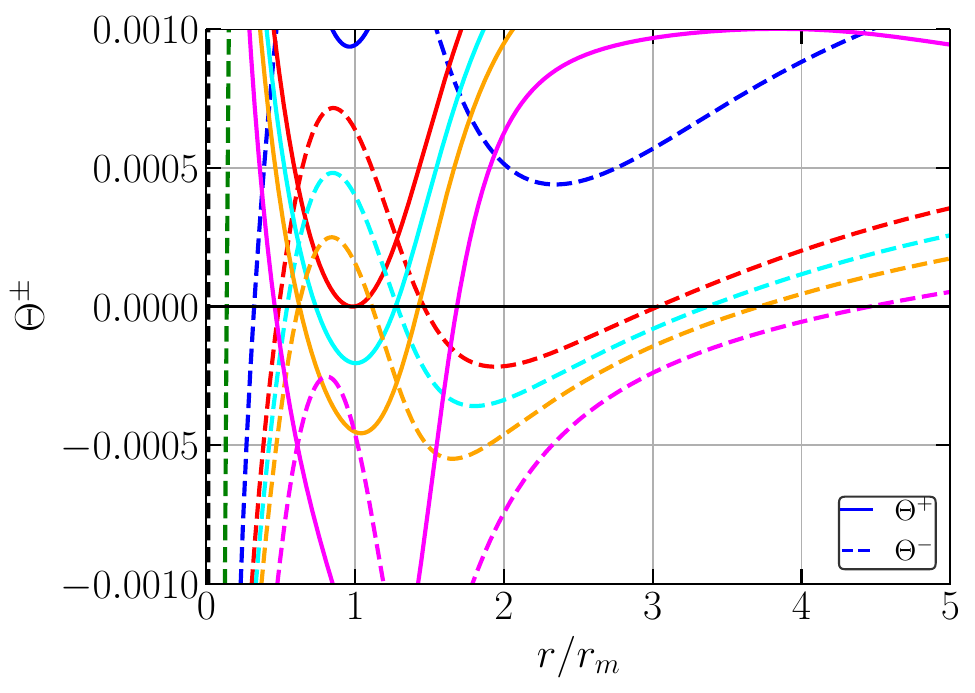}
\hspace*{-0.3cm}
\includegraphics[width=1.9 in]{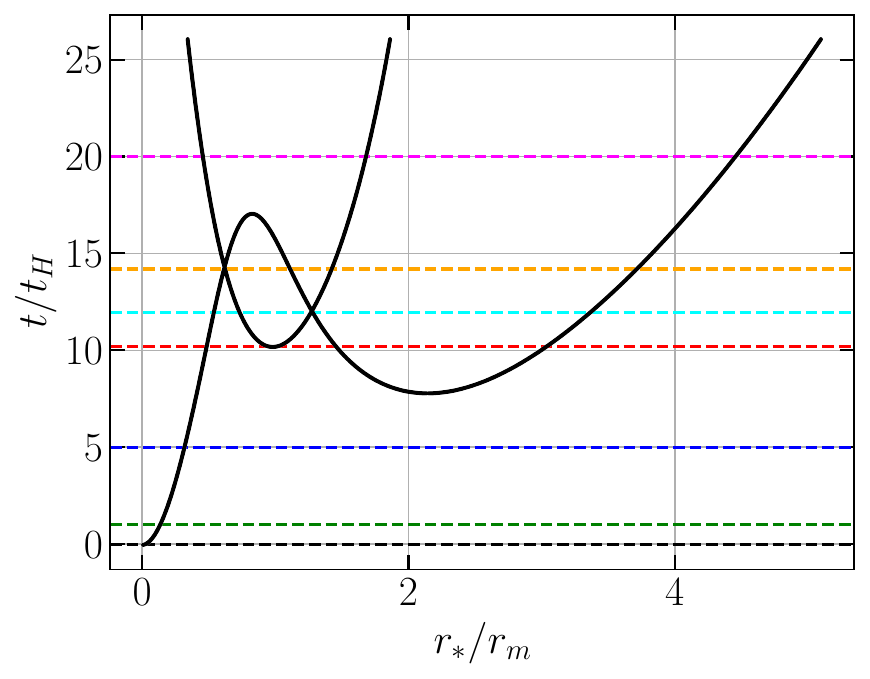}
\hspace*{-0.3cm}
\caption{Snapshots of the evolution of different quantities for the case $\beta=1$ with $\mu=2.0$ with $r_m = 10 R_{H}(t_0)$, corresponding to a case of PBH formation with type-II B PBH.}
\label{fig:dynamics2}
\end{figure}

In Fig.\ref{fig:dynamics2}, we present a case of type-II B PBH with $\beta = 1$, $\mu = 2.0$. The qualitative behavior is similar to that of the previous case involving collapsing fluctuations. The most distinctive feature is the formation of two bifurcated trapping horizons at times $t_{\rm bif,1}$ and $t_{\rm bif,2}$. By examining the expansion of the congruences $\Theta^{\pm}$, we observe that both vanish at the same radial coordinate, $r_*$, at these times. Thus, we consistently recover the result from \cite{Uehara:2024yyp}, where bifurcated trapping horizons were observed for the same profile but with a slightly lower amplitude ($\mu = 1.8$). The time-gap between the formation of the bifurcated horizons, $t_{\rm bif,2} - t_{\rm bif,1}$, increases as $\mu$ exceeds the threshold for the type-B PBH configuration. We also observe that the type-II feature is much more distinguishable compared to the previous case. For instance, at the time of the formation of the bifurcated trapping horizon, $\Gamma$ exhibits a clearly defined negative region. In contrast, in the previous example, the throat structure in $R$ has already disappeared. On the other hand, when $\mathcal{C} \approx 1$, it indicates black hole formation for type-II fluctuations, as it does for type-I \cite{Escriva:2019nsa} (which is used as an efficient numerical criterion to signal black hole formation), since at late times the Misner-Sharp mass at the location of the horizon becomes significantly larger than the background mass $M_b$ and therefore $\mathcal{C}=1 \Rightarrow 2M/R >1$.

\begin{figure}[t]
\centering
\includegraphics[width=1.9 in]{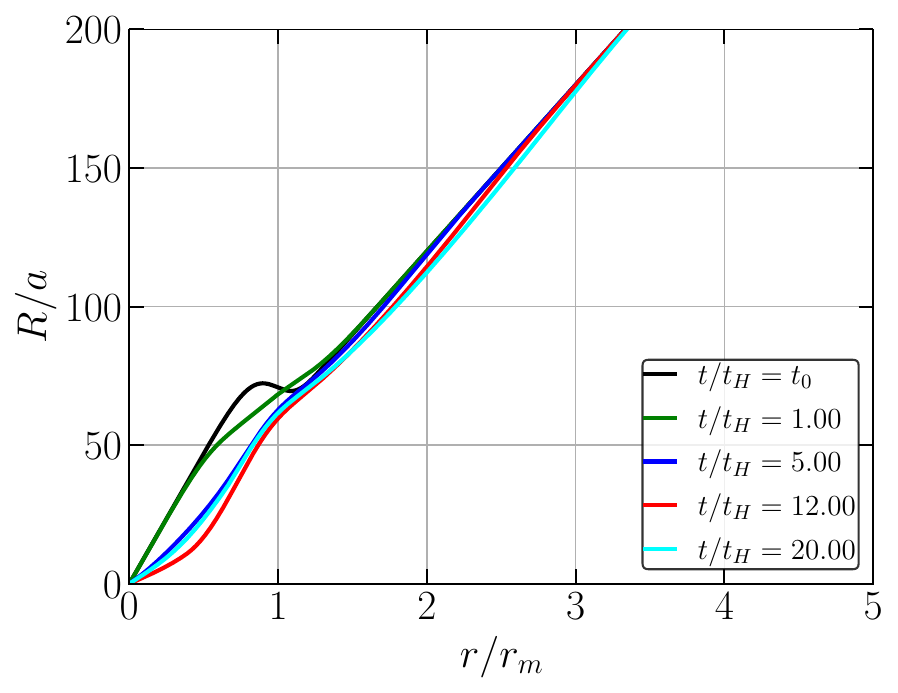}
\hspace*{-0.3cm}
\includegraphics[width=1.9 in]{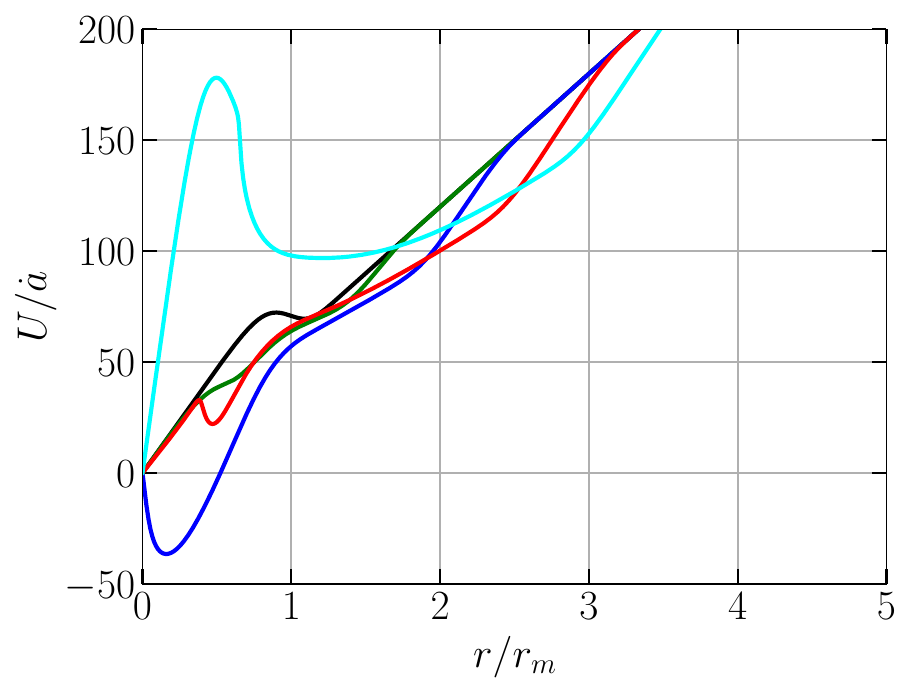}
\hspace*{-0.3cm}
\includegraphics[width=1.9 in]{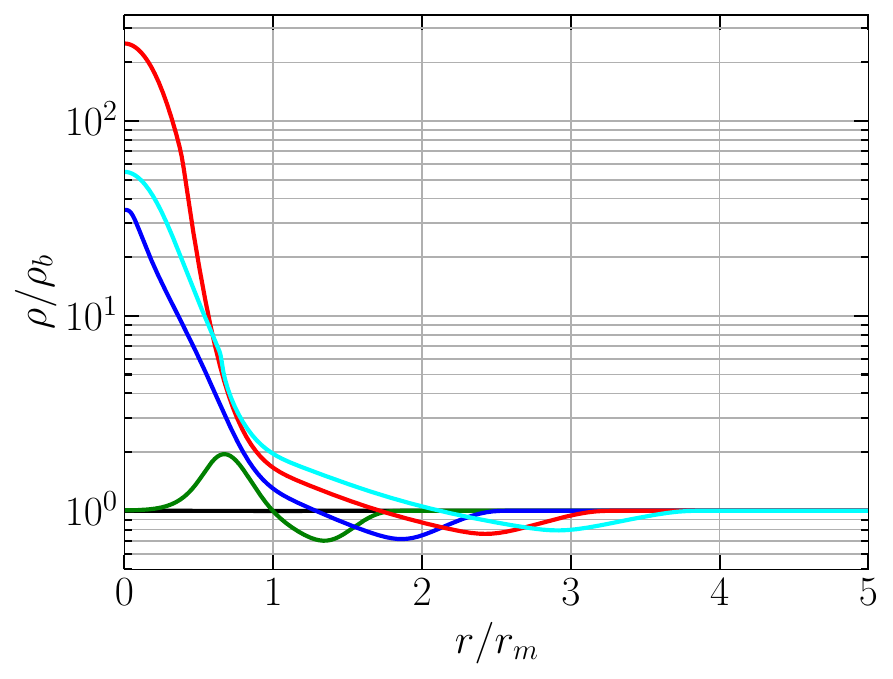}
\hspace*{-0.3cm}
\includegraphics[width=1.9 in]{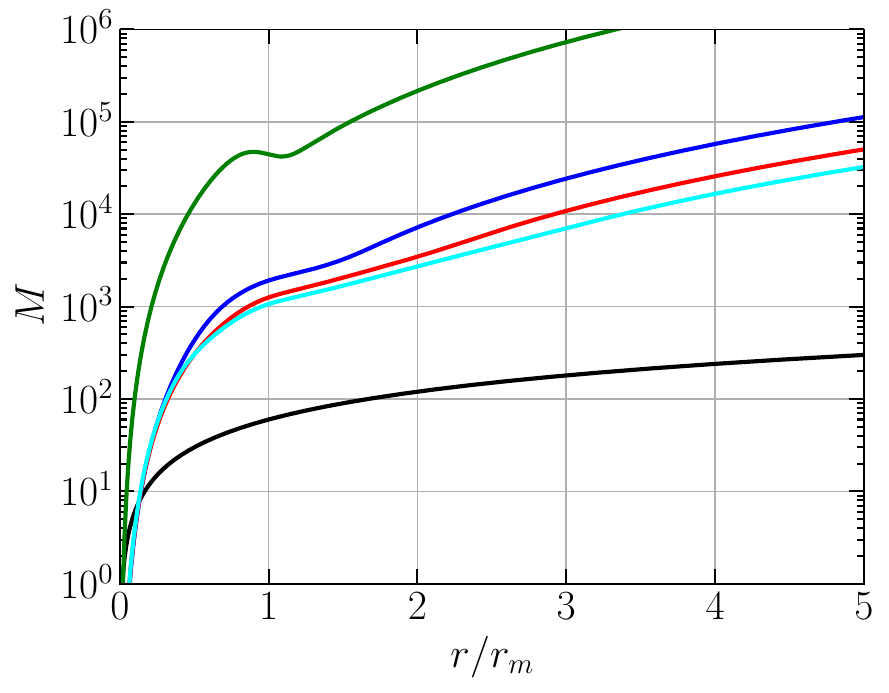}
\hspace*{-0.3cm}
\includegraphics[width=1.9 in]{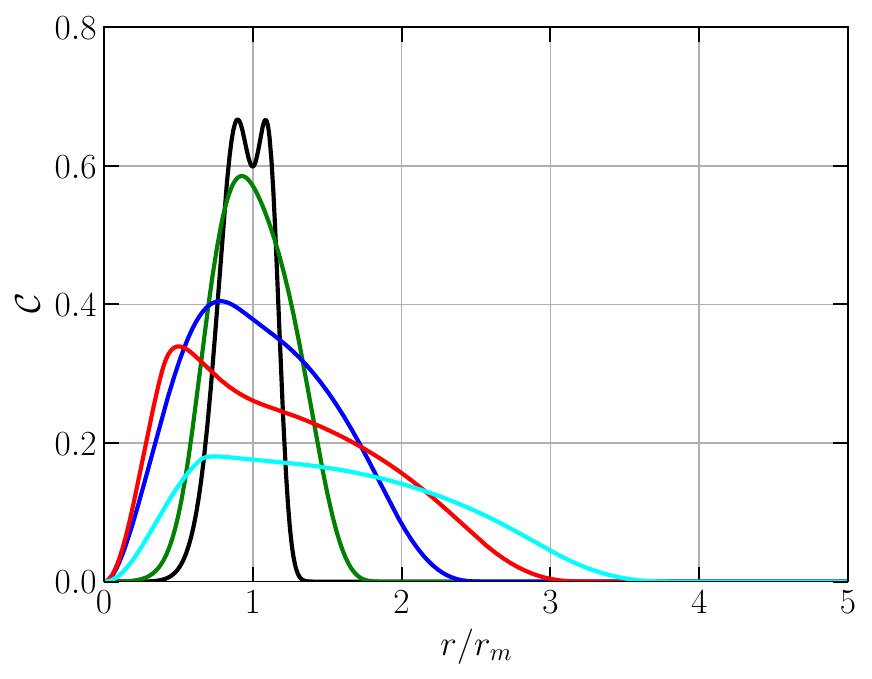}
\hspace*{-0.3cm}
\includegraphics[width=1.9 in]{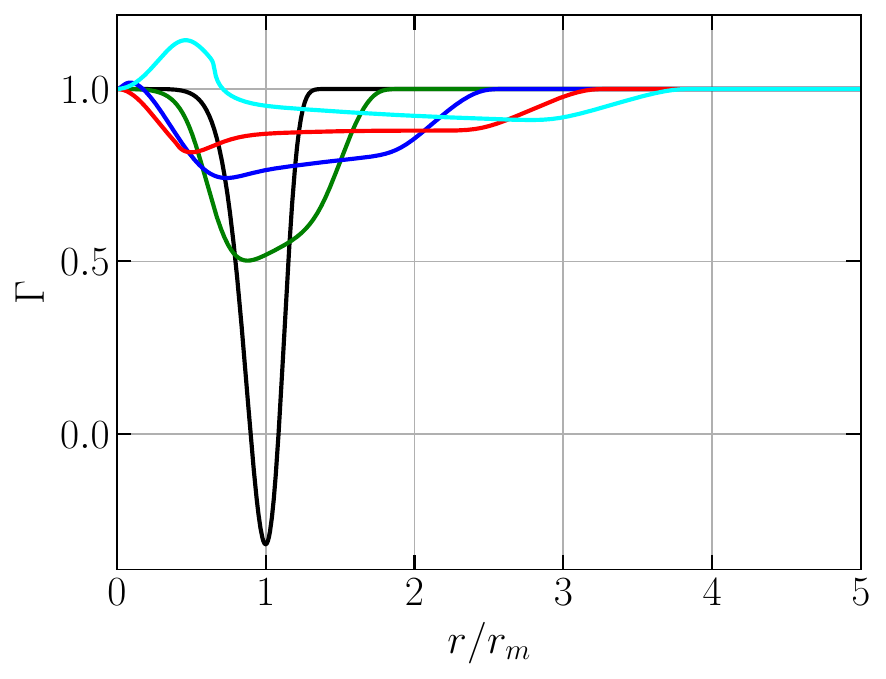}
\hspace*{-0.3cm}
\includegraphics[width=1.9 in]{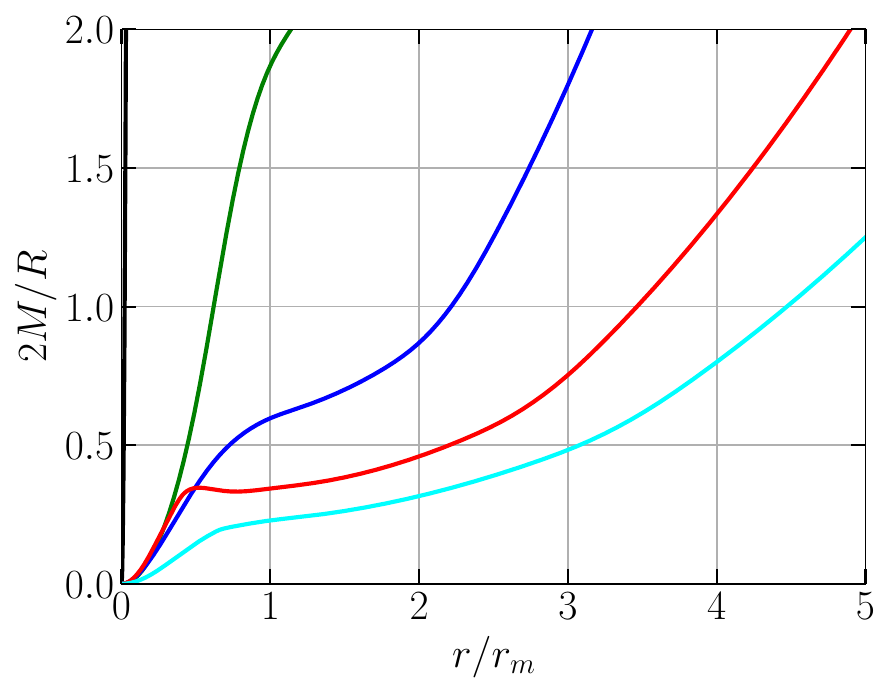}
\hspace*{-0.3cm}
\includegraphics[width=1.9 in]{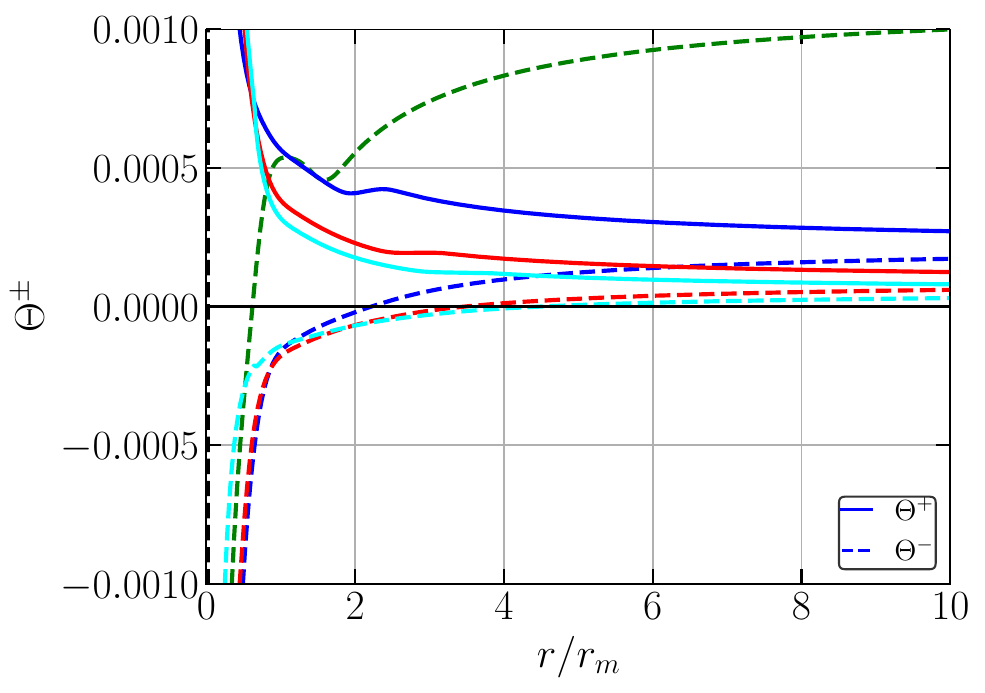}
\hspace*{-0.3cm}
\includegraphics[width=1.9 in]{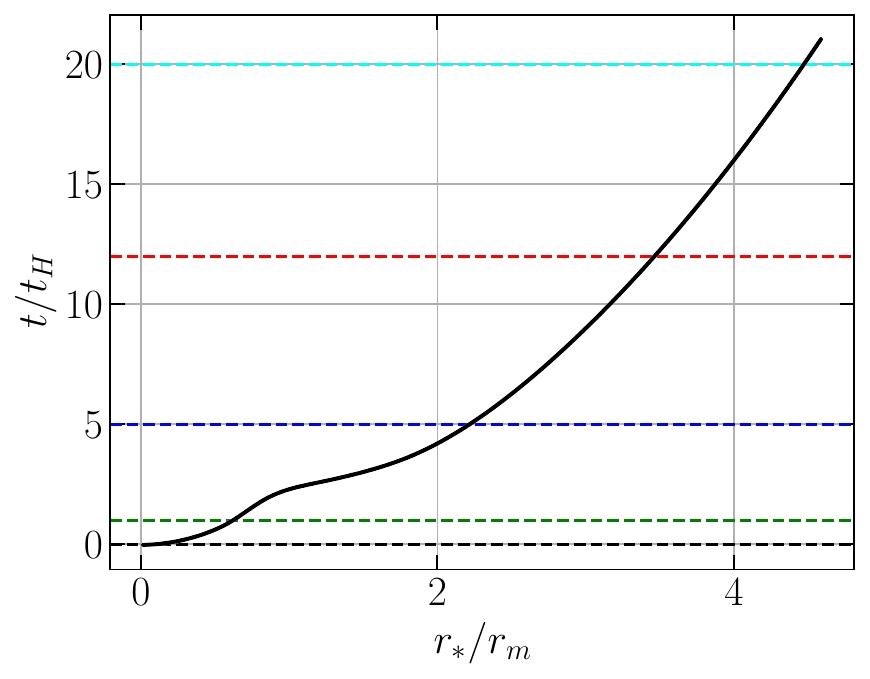}
\hspace*{-0.3cm}
\caption{Snapshots of the evolution of different quantities for the case $\beta=4$ with $\mu=0.45$ with $r_m = 30 R_{H}(t_0)$, corresponding to a case of type-II no PBH.}
\label{fig:dynamics3}
\end{figure}

Finally, in Fig.\ref{fig:dynamics3}, we present a case with parameters $\beta = 4$ and $\mu = 0.45$, which corresponds to a "type-II no PBH" scenario. Such scenario was previously identified in \cite{Shimada:2024eec,Inui:2024fgk} within the context of large negative non-Gaussianity models. This result contrasts with the expectations from the literature that type-II fluctuations collapse forming black holes. In our work, we identify new cases with the profile of Eq.\eqref{eq:profile_exp}, which suggest this scenario may not be as exceptional as previously thought. In our simulations, the initial configuration corresponds to a type-II fluctuation. As the cosmological evolution progresses and the fluctuation crosses the horizon, the fluctuation begins to disperse into the FLRW background without black hole formation. This is evident from the evolution of the compaction function, which decreases, and the mass excess disperses. Similarly, the peak value of the energy density increases to a maximum before decreasing at later times. This behaviour is accompanied by a positive Eulerian velocity $U$ at sufficiently late times, which prevents the collapse and disperses the fluctuation. Qualitatively, this behaviour is similar to that observed for fluctuations below the threshold for type-I fluctuations (see \cite{Escriva:2019nsa} for comparison).

\begin{figure}[t]
\centering
\includegraphics[width=1.8 in]{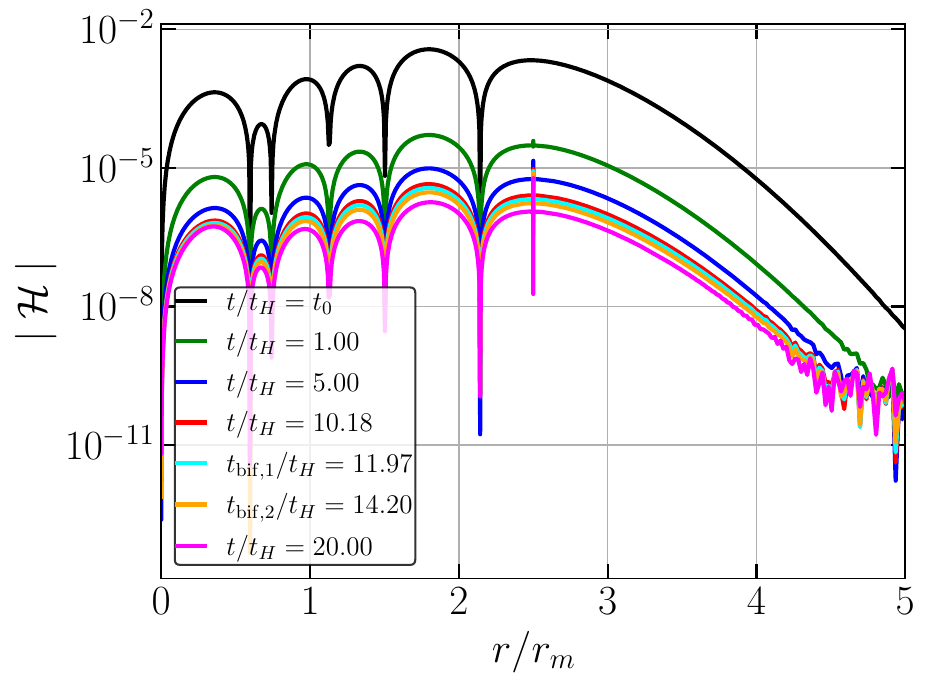}
\hspace*{-0.3cm}
\includegraphics[width=1.8 in]{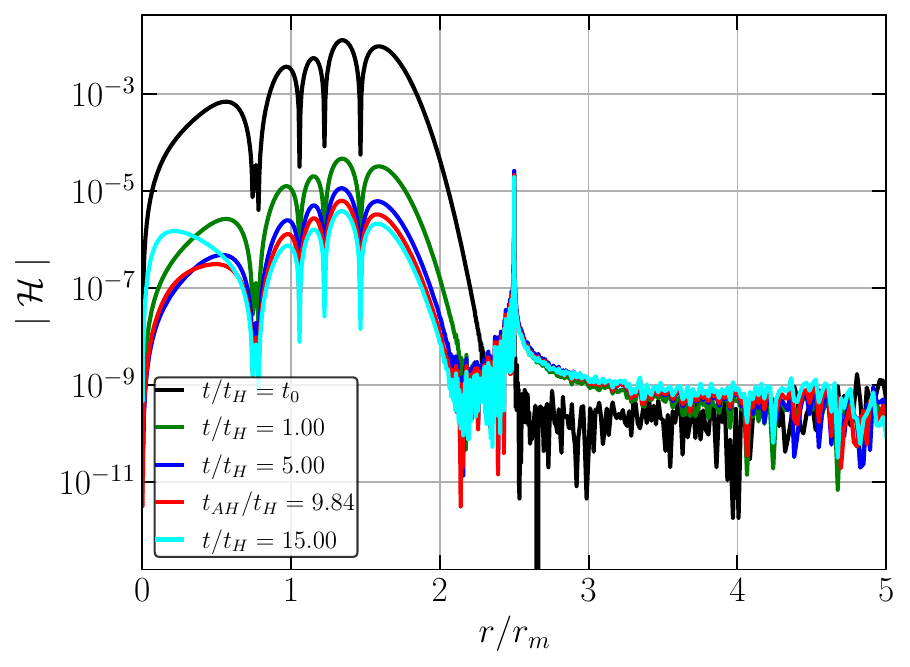}
\hspace*{-0.3cm}
\includegraphics[width=1.8 in]{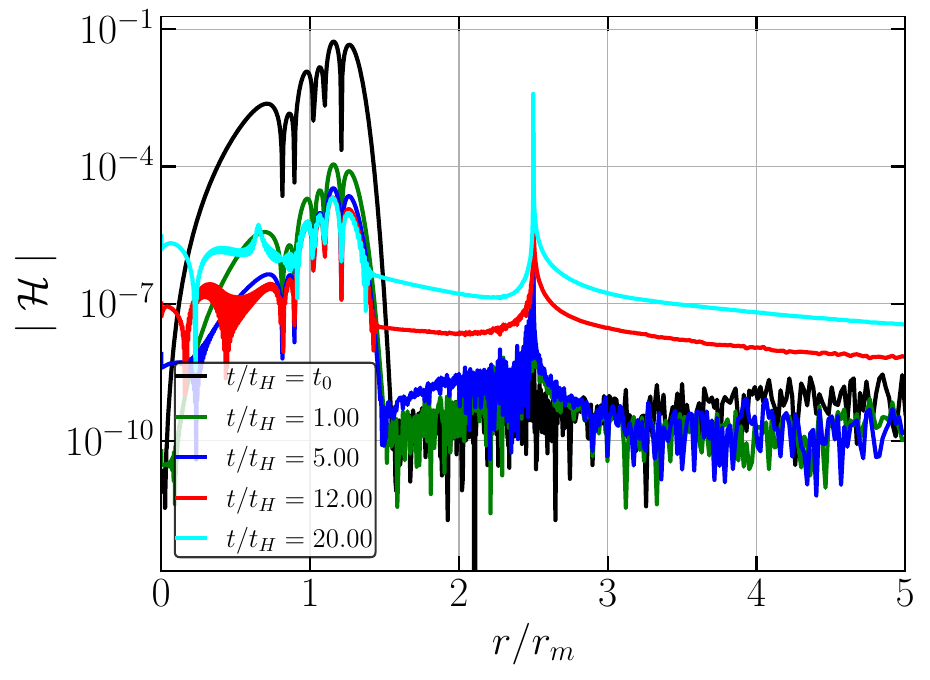}
\hspace*{-0.3cm}
\includegraphics[width=1.8 in]{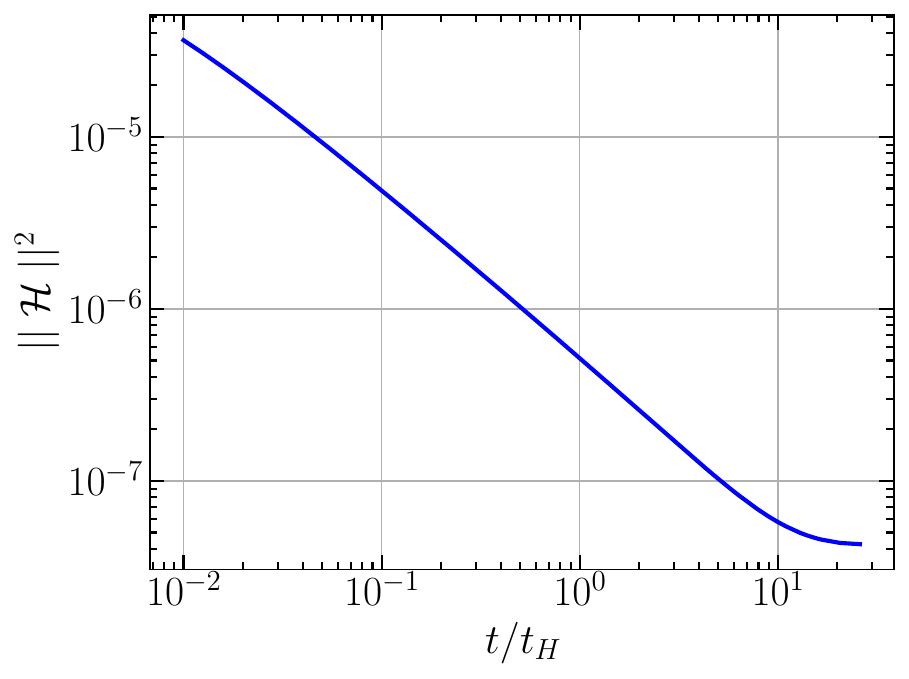}
\hspace*{-0.3cm}
\includegraphics[width=1.8 in]{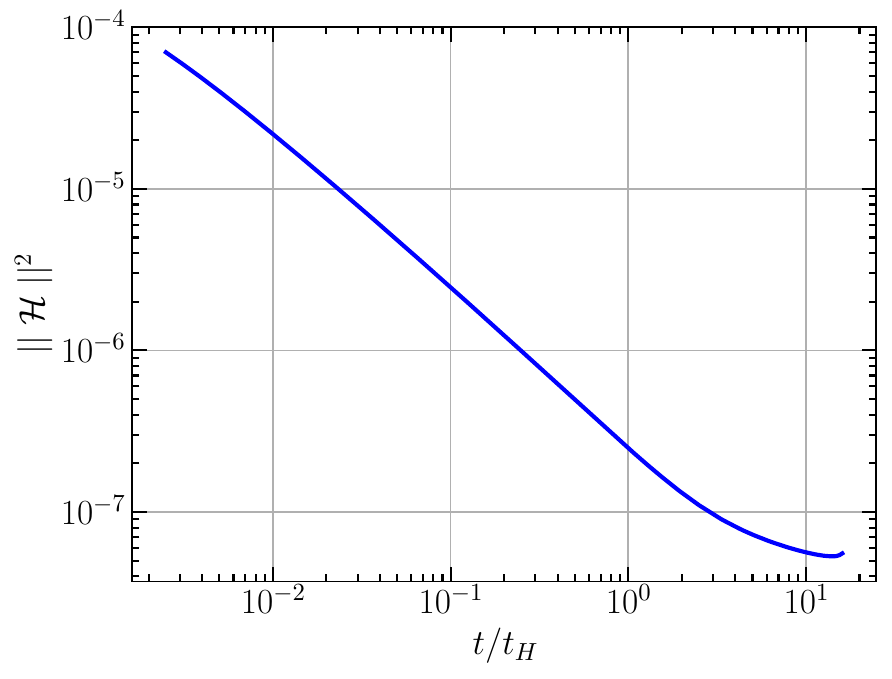}
\hspace*{-0.3cm}
\includegraphics[width=1.8 in]{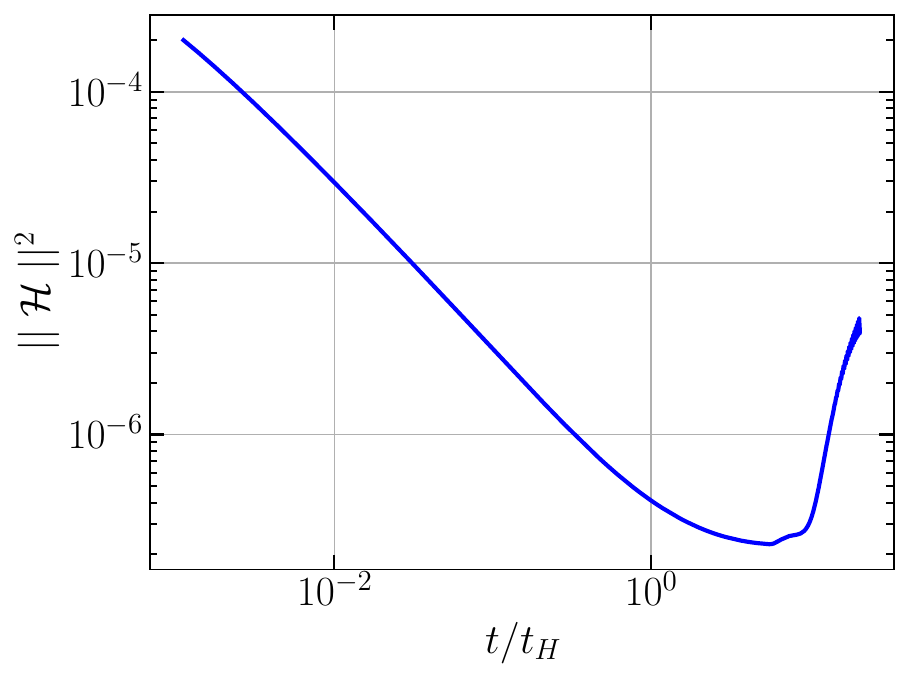}
\hspace*{-0.3cm}
\caption{Top panels: Snapshots of the Hamiltonian constraint $\mathcal{H}$ at specific times. Bottom panels: Average value of the Hamilton constraint evolution $\mid \mid \mathcal{H}\mid \mid^2$. }
\label{fig:Ham_constraint}
\end{figure}

In Fig.\ref{fig:Ham_constraint} we show some snapshots of the Hamiltonian constraint (top panels) and its averaged value (bottom panels) for the three cases previously considered. As can be observed, the Hamiltonian constraint is satisfied during the numerical evolution.

\subsection{Thresholds for type A/B PBH}

\begin{figure}[t]
\centering
\includegraphics[width=2.8 in]{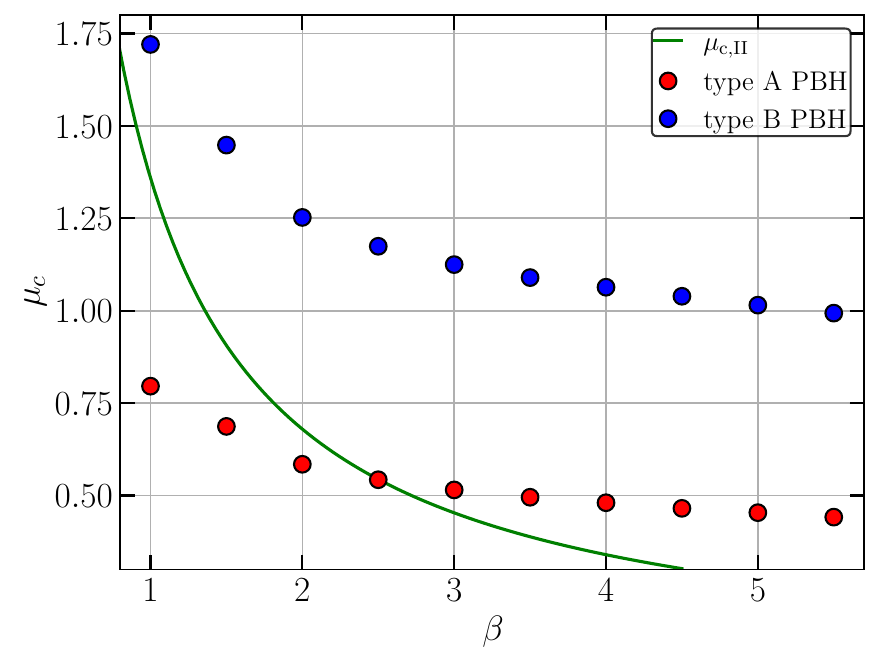}
\hspace*{-0.3cm}
\includegraphics[width=2.8 in]{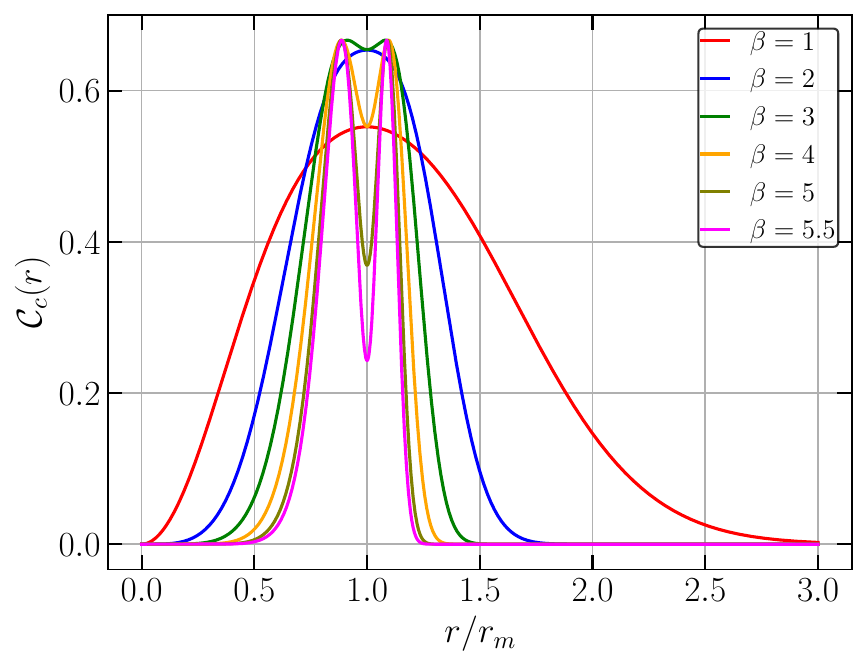}
\hspace*{-0.3cm}
\includegraphics[width=2.8 in]{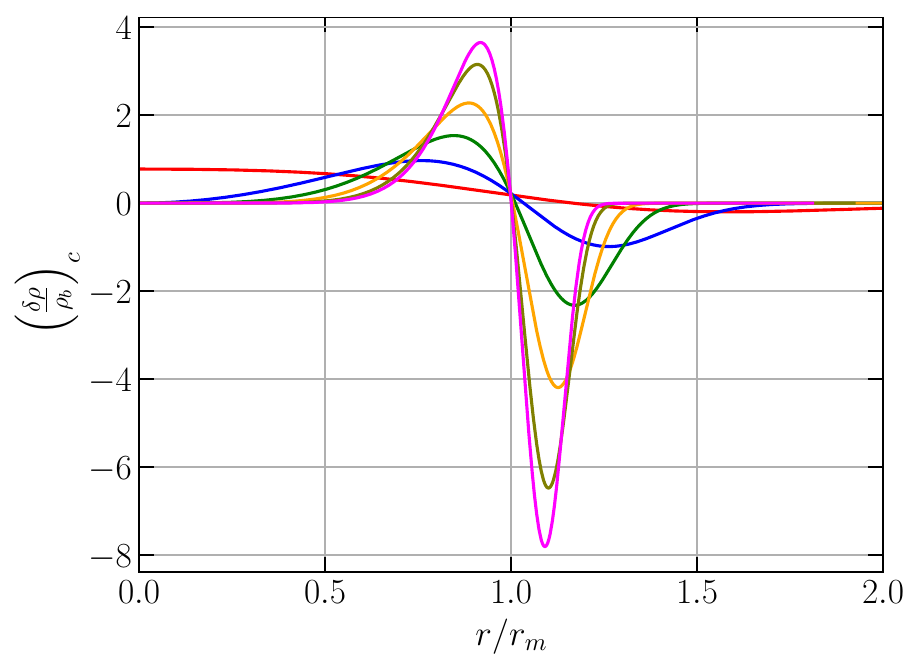}
\hspace*{-0.3cm}
\caption{Top-left panel: Thresholds in terms of the peak value of $\zeta$, i.e, $\mu_c = \zeta_c(r=0)$. The green line $\mu_{c,\rm II}$ separate the type-I region from the type-II region. Top-right panel: Critical compaction function profile for the different shapes $\beta$ with $\mu=\mu_{c,A}$. Bottom panel: Critical overdensity $(\rho-\rho_b)/\rho_b$ from Eq.\eqref{eq:rho_tilde} extrapolated at the time when $\epsilon=1$ with $\mu=\mu_{c,A}$.}
\label{fig:thresholds_cases}
\end{figure}

Using a bisection method, we can determine the threshold for black hole formation as well as the threshold that distinguishes type A and type B PBHs. First, to obtain the threshold for black hole formation for type A PBHs, we follow the procedure outlined in \cite{Escriva:2019nsa}, which involves checking the peak value of the compaction function. Additionally, we also compute the expansion of the congruences $ \Theta^{\pm} $ to identify and crosscheck the formation of trapped surfaces. If the peak value of the compaction function continuously decreases over time, it indicates that the fluctuation is dispersing into the FLRW background. Conversely, if the peak value continuously increases and reaches $ \mathcal{C} \approx 1 $, it signifies that a black hole has been formed\footnote{Equivalently, we can also check the lapse function at the origin: If it decreases continuously, a black hole is expected to be formed. Otherwise, if it decreases and then continuously increases after reaching a minimum (bouncing), it indicates dispersion of the fluctuation. All criteria we have tested in our simulations are consistent with each other.} (see Figs.~\ref{fig:dynamics1}, \ref{fig:dynamics2}, \ref{fig:dynamics3}, following the discussion in the previous subsection and Fig.6 in \cite{Escriva:2023nzn}). To distinguish between type A and type B PBHs, we compute the expansion of the congruences $ \Theta^{\pm} $ to identify trapping bifurcated horizons, where $ \Theta^{\pm}(r_*) = 0 $, which corresponds to an evolution leading to a type B PBH. If a trapping horizon is found with $ \Theta^{+}(r_*) = 0 $ and $ \Theta^{-}(r_*) < 0 $ without a bifurcated trapping horizon configuration, it corresponds to an evolution leading to a type A PBH. Our numerical results are shown in Fig.~\ref{fig:thresholds_cases}.

We have computed the formation thresholds of type A/B PBH within the parameter range $\beta \in [1,5.5]$ and with a relative resolution of $\delta \mu_c \sim \mathcal{O}(10^{-2}\,\%)$ (in absolute resolution $\delta \mu_c \sim \mathcal{O}(10^{-4})$). For $\beta < 1$, the curvature fluctuation does not satisfy the regularity conditions, as it introduces a sharp, divergent feature in the energy density at the center $r=0$. We discard these profiles, as we consider them physically unrealistic. For $\beta \gtrsim 5$, we found that large gradients develop in the dynamics when the fluctuation amplitude is very close to its threshold. Despite implementing a mesh refinement procedure, the refinement was insufficient to accurately resolve the dynamics and infer the formation thresholds for type A PBHs with the same resolution as in other cases. We leave the improvement of this procedure for future work.

We find that the thresholds $\mu_{c}$ for both type A ($\mu_{c,A}$) / B ($\mu_{c,B}$) PBHs decrease as $\beta$ increases. Our numerical result for $\beta =1$ is $\mu_{c,B}\approx 1.75$, which is consistent with the value inferred in \cite{Uehara:2024yyp}. A notable result is that, within our parameter range for $\beta$, we observe cases where type-II configurations do not collapse into black holes, contrary to expectations. The transition point that distinguishes the thresholds between the type-I and type-II regions occurs at $\beta \approx 2.5$. With the standard Misner-Sharp approach, we would only be able to run simulations for type-I fluctuations and obtain the thresholds for type A PBH for $\beta < 2.5$. For instance, in Ref.~\cite{Young:2019yug}, using simulations based on the standard Misner-Sharp approach, the threshold was assumed to saturate for $\beta \gtrsim 2.0$ in the type I region. Nevertheless, our simulations indicate that this previous assumption and behavior stems from numerical limitations in simulating type-II fluctuations. In our work, by using the trace of the extrinsic curvature $K$ as an auxiliary variable, we bypass the complications associated with the neck structure ($R'=0$), enabling us to simulate the type-II region and identify the PBH formation threshold for $\beta \gtrsim 2.5$.

The top-right panel of Fig.~\ref{fig:thresholds_cases} show the corresponding profiles of $\mathcal{C}{c}$ and the critical overdensity $(\delta \rho / \rho_b){c}$, extrapolated to the time when $\epsilon = 1$ (see Eq.~\eqref{eq:rho_tilde}). When the PBH formation threshold lies in the type-I region, $\mathcal{C}_{c}(r_m)$ exhibits a local maximum, whereas in the type-II region it becomes a local minimum. In these cases, we observe that the curvature of $\mathcal{C}_{c}(r_m)$ at the local minimum ($r_m$) increases as $\beta$ increases, and the shape of $\mathcal{C}$ at $r_m$ is characterized by a spiky valley.

By examining $(\delta \rho / \rho_b)_{c}$ (bottom panel), we observe that as $\beta$ increases, the central over-dense region is shifted, leading to a zero central over-density, making more difficult the collapse. Consequently, the critical profile becomes non-centrally peaked, and the cases we find correspond to large derivatives at $r_m$. This may indicate that the existence of the threshold in the type-II region is also related to the specific shape of $\mathcal{C}$ (or a related quantity) at the scale $r_m$. The work \cite{preparation} focuses on exploring the parameterization of the threshold for PBH formation in the type-II region with compaction functions, along with its analytical estimation.

\subsection{PBH mass}

Finally, we present the numerical results for the PBH mass for various values of the parameter $\beta$. We specifically examine fluctuation amplitudes in the critical regime \cite{Niemeyer:1999ak} and beyond.

First, we define $t_{\rm PBH}$ as the time when the PBH is formed. For type A PBHs, this corresponds to the moment when a marginally trapped surface at $r_*$ with $\Theta^{+}(r_*)=0, \, \Theta^{-}(r_*)<0$ is formed. In the case of type B PBHs, this time corresponds when the first bifurcated trapping horizon with $\Theta^{+}(r_*)= \Theta^{-}(r_*)=0$ at $t_{\rm bif,1}$ appears. After the formation of the PBH, an accretion process from the FLRW background follows, which can substantially increase the PBH mass by a few factors (see \cite{Escriva:2021pmf} for a study of profile dependence with type-I fluctuations). Following \cite{Escriva:2019nsa}, to obtain the final mass of the PBH\footnote{Another way to obtain the PBH mass is to use null coordinates with the Misner-Hernandez equations \cite{1966ApJ...143..452H} to avoid the formation of a singularity. In our case, however, we will use an excision technique.}, we use the Zeldovich-Novikov formula \cite{acreation1,acreation2,acreation3}\footnote{It is important to note that at early times, the PBH mass growth does not follow the Zeldovich formula, since the cosmological expansion is neglected \cite{Carr:2010wk}. Nevertheless, for sufficiently late times, when accretion becomes a stationary process, we can apply it to accurately infer the values of the PBH mass.}. In particular, the time evolution of the PBH mass can be obtained by solving:

\begin{equation}
\label{eq:ZN_formula}
\frac{dM}{dt} = 4 \pi F R^2_{\rm PBH} \rho_{b}(t)\  \Rightarrow M_{\rm PBH}(t) = \frac{1}{\frac{1}{M_{a}} + \frac{3}{2} F \left( \frac{1}{t} - \frac{1}{t_a} \right)}\ , \, M_{\rm PBH}(t \rightarrow \infty) = \left( \frac{1}{M_a} - \frac{3F}{2 t_a} \right)^{-1} \ ,
\end{equation}
where $M_a$ is the initial mass when the asymptotic approximation is used at the time $t_a$ and $F$ is an effective accretion constant, which is typically found to be of order $O(1)$. By the condition of the trapping horizon, $R_{\rm PBH} = 2M_{\rm PBH}$, the previous equation can be solved analytically to obtain $M_{\rm PBH}(t)$ and we can infer the final mass of the PBH as $M_{\rm PBH} \equiv M_{\rm PBH}(t\rightarrow \infty)$. Using an excision procedure \cite{excision,bookNR} based on the methodology outlined in \cite{Escriva:2019nsa} (for details, we refer the reader to that reference), we can remove the computational domain where a singularity is expected to form at late times, after the formation of the PBH, and continue evaluating the PBH mass, observing its subsequent increase. Compared with \cite{Escriva:2019nsa}, we apply the new refinement methodology described in section \ref{sec:method} for simulations in the critical regime.

We will find the parameters $M_a$, $t_a$, and $F$ by fitting the numerical evolution to the analytical Zeldovich formula for the PBH mass $M_{\rm PBH}(t)$ Eq.\eqref{eq:ZN_formula}. To find the time interval for fitting, we consider the criteria in \cite{Escriva:2019nsa} to account for the ratio of the black hole mass increment with respect to the Hubble scale, $\Psi = \dot{M} / (H M)$. In the regime described by Eq.\eqref{eq:ZN_formula}, we expect $\Psi < 1$. The range of numerical values used for the fit corresponds to those that satisfy $\Psi \lesssim 10^{-1}$, up until nearly the end of the numerical evolution, when the Hamiltonian constraint becomes significantly violated. Once $M_a$, $t_a$, and $F$ are found, the PBH mass is inferred as the asymptotic mass at $t \to \infty$.

In the bottom-left panel, we show the time evolution of the PBH mass for different amplitudes $\mu$, and in the bottom-right panel the time-evolution of the Hamiltonian constraint. The dotted lines correspond to the numerical fits of Eq.\eqref{eq:ZN_formula}, from which we infer the final PBH mass after the accretion process.

\begin{figure}[t]
\centering
\begin{minipage}{\textwidth}
    \centering
    \includegraphics[width=0.6\textwidth]{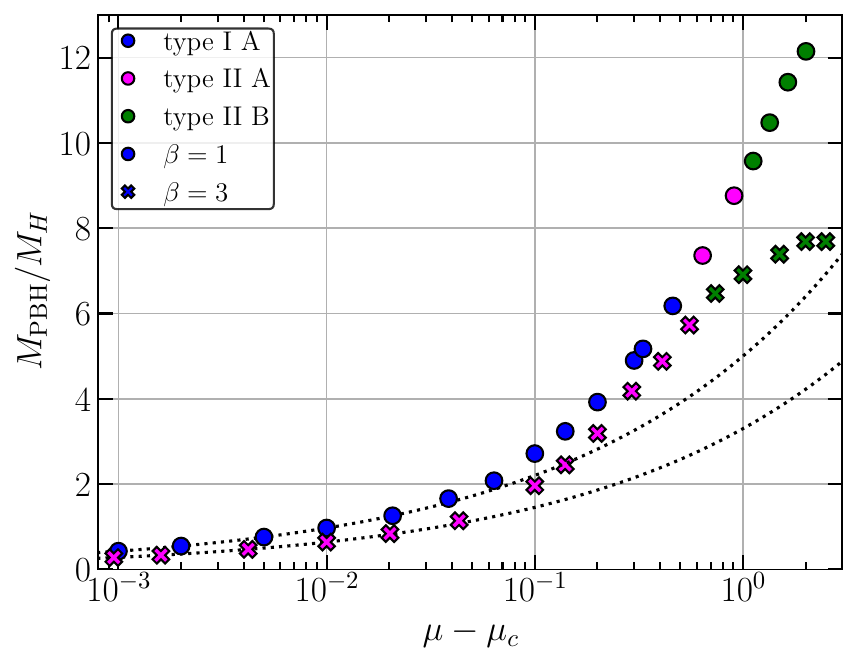} 
    \vspace{0.3cm} 
\end{minipage}

\begin{minipage}[t]{0.45\textwidth} 
    \centering
    \includegraphics[width=\textwidth]{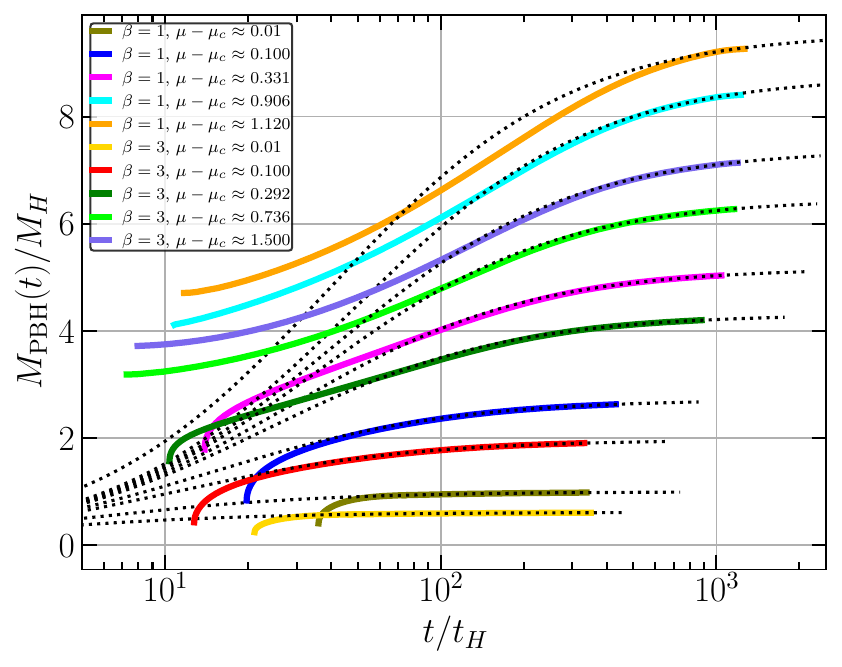}
\end{minipage}
\hfill 
\begin{minipage}[t]{0.48\textwidth} 
    \centering
    \includegraphics[width=\textwidth]{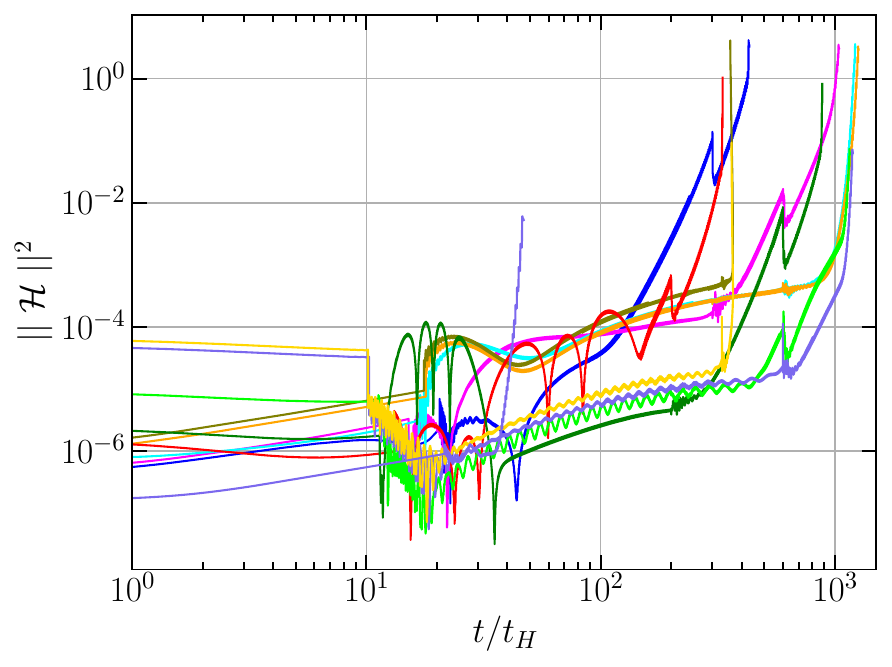}
\end{minipage}

\caption{
Top panel: Final values of the PBH mass $M_{\rm PBH}$ after the accretion process from the FLRW background for different cases. The different colors denote the type of PBH formed, and the symbols represent the profile: $\beta = 1$ (circle), $\beta = 3$ (crosses). The dotted-black line denotes the critical scaling law $M_{\rm PBH} \sim (\mu-\mu_c)^{\gamma}$. Bottom-left panel: Time evolution of the PBH mass for different cases of $\mu - \mu_c$. The dotted lines represent the numerical fit to Eq.~\eqref{eq:ZN_formula}. Bottom-right panel: Time evolution of the averaged Hamiltonian constraint.}
\label{fig:PBH_mass}
\end{figure}

Let's discuss the results regarding the black hole mass $M_{\rm PBH}$, which are shown in the top panel of Fig.\ref{fig:PBH_mass}. The results are classified using different colours according to the following categories: type-I A (blue), type-II A (magenta), and type-II B (green). This classification applies to the two profiles considered: $\beta = 1$ (solid dots) and $\beta = 3$ (crosses). We observe that for values of $\mu - \mu_c$ close to the threshold, within the range $\mu - \mu_c \in [10^{-3}, 10^{-2}]$, the numerical results exhibits a scaling law behavior. Specifically, we perform a linear fit in logarithmic scale for this range, using the relation $M_{\rm PBH} = \mathcal{K} M_H (\mu - \mu_c)^{\gamma}$. For this fitting, we find $\gamma(\beta = 1) \approx 0.353 \pm 5 \cdot 10^{-3}$, $\mathcal{K}(\beta = 1) \approx 5.0 \pm 10^{-1}$ and $\gamma(\beta = 3) \approx 0.351 \pm 4 \cdot 10^{-3}$, $\mathcal{K}(\beta = 3) \approx 3.3 \pm 10^{-1}$. This scaling behavior is characteristic of the PBH mass in the critical regime of gravitational collapse \cite{PhysRevLett.70.9,Evans:1994pj,Koike:1995jm,Niemeyer:1999ak,Musco:2012au}. The critical exponent, $\gamma\approx 0.356$, is universal, depending only on the equation of state parameter. Our results recover this expected value and confirms that the critical scaling law applies also to type-II fluctuations. We tote that our results for very large amplitudes $\mu \gg \mu_c$ correspond to cases beyond the critical collapse regime, where the scaling law for $M_{\rm PBH}$ is known to deviate \cite{Escriva:2019nsa}. We find that this deviation can be much more substantial than previously reported, $\sim \mathcal{O}(15 \%)$, whereas for the largest computed amplitude $\mu$ we now obtain $\sim \mathcal{O}(82 \%)$ for $\beta = 1$ and $\sim \mathcal{O}(62 \%)$ for $\beta = 3$, since only type-I fluctuations were considered in \cite{Escriva:2019nsa}.

On the other hand, we find that the PBH mass increases monotonically with the amplitude $\mu$ for the case $\beta=1$. The results are consistent quantitatively\footnote{There is a factor of $2$ difference between the numerical results for the final PBH mass. This is related to the different definitions of the mass of the cosmological horizon $M_H$. Here, we use $r_m$ to define the time of horizon croosing $t_H$, whereas in \cite{Uehara:2024yyp}, the comoving wavelength mode $k$ is used. Both scales are related to each other by $k = \sqrt{2}/r_m$. The difference between the two when computing $M_H$ in Eq.\eqref{eq:background_tH_MH} results in the factor of $2$.} with those reported in \cite{Uehara:2024yyp} for the Gaussian-shape profile $\beta=1$ for the range of $\mu$ compared. We observe that the PBH mass values are larger for $\beta = 1$ than for $\beta = 3$ at the same $\mu - \mu_c$. This is attributed to the effect of the pressure gradient, which is stronger for $\beta = 3$ and suppresses accretion from the FLRW background \cite{Escriva:2021pmf}.

For the profile with $\beta = 3$, the PBH mass reaches a maximum value and then decreases for the highest value of $\mu$ computed corresponding to type-II B PBH, resulting in a non-monotonic behavior for the mass in terms of $\mu$ (see also \cite{Uehara:2024yyp} for the three-zone model profile and \cite{Shimada:2024eec} for a model with non-Gaussianity). The existence of the throat structure in the initial data with $\mu-\mu_{\rm c,II}>0$, combined with the specific shape dependence of the profile (pressure gradient effects), significantly influences the PBH mass behavior. Our results suggest that increasing the pressure gradients associated with the curvature profile can lead to a non-monotonic mass spectrum as a function of the fluctuation amplitude. In this regard, further studies on the profile dependence of the PBH mass for type-II fluctuations are necessary, as they are relevant for improving statistical estimations of PBH mass functions, for instance, a non-monotonic behavior of the PBH mass can imprint a divergent feature in the mass function due to the Jacobian of the transformation between the PBH mass and the fluctuation amplitude \cite{Kitajima:2021fpq,Yoo:2022mzl}.

\section{Conclusions}

In this work, we present a new approach based on the Misner-Sharp formalism for conducting numerical simulations of spherical PBH formation resulting from the collapse of generic curvature fluctuations. In contrast with the standard methodology, we introduce the trace of the extrinsic curvature $K$ as an auxiliary variable to address issues associated with type-II fluctuations, particularly the appearance of divergent terms of the form $(0/0)$ due to the presence of a throat structure characterized by the non-monotonic behavior of the areal radius $R$. By doing that, we have obtained a set of time-evolution equations with the corresponding initial conditions, free from divergences associated with type-II fluctuations and consistent with the quasi-homogeneus solution. Our results constitute the first successful numerical simulation of type-II curvature fluctuations within the Misner–Sharp formalism.

Following our new approach, we have developed a simple and efficient numerical code that builds upon and updates the numerical method used in \cite{Escriva:2019nsa}, employing a pseudospectral collocation scheme. Using this code, we successfully simulated the evolution and collapse of super-horizon type-II curvature fluctuations, for what we have done a detailed study of the dynamics of the gravitational collapse for three representative cases using standard exponential-shaped curvature profiles Eq.\eqref{eq:profile_exp}.

The numerical results presented in Section \ref{sec:num_results} (which agree with those reported in \cite{Uehara:2024yyp} for the case of $\beta=1$ when compared) demonstrate the robustness and effectiveness of our approach, demonstrating its reliability and applicability in simulating type-II fluctuations. Therefore, our work demonstrates that the Misner–Sharp formalism in the comoving gauge does not pose any issues for simulating type-II fluctuations, contrary to previous expectations. Rather, type-II simulations can be effectively conducted by avoiding divergent terms, which we achieved in this work by using $K$.

By implementing a bisection method, we determined the threshold for black hole formation for both type A and type B PBHs. Notably, we identify cases in which type-II fluctuations do not necessarily lead to black hole formation. Specifically, we find that the threshold transitions into the type-II region at approximately $\beta \approx 2.5$, which differs from previous results in the literature. This finding motivates further exploration of the profile dependence of type-II fluctuations.

We have also studied the PBH mass behaviour for the profiles $\beta = 1$ and $\beta = 3$. We successfully recovered the critical behaviour of the collapse for $\mu \ll \mu_c$ for both type-I and type-II fluctuations. The presence of a throat structure in type-II fluctuations and the profile dependence can significantly affect the PBH mass, particularly leading to a non-monotonic increasing behaviour for the black hole mass as a function of $\mu$ for sufficiently large values of $\mu - \mu_c$ for the case $\beta = 3$, which may be related to the stronger pressure gradient.

Potential directions for further research may include extending this study to equations of state with parameters different from the radiation case $w=1/3$ and explore different scenarios, such as those involving thermal transitions or the inclusion of multi-component fluids and scalar fields. Additionally, improving the mesh refinement procedure could enable us to investigate cases with sharp gradients characterized by large $\beta$ values.

\begin{acknowledgments}
I thank Tomohiro Harada, Daiki Saito, Masaaki Shimada, Koichiro Uehara, and Chul-Moon Yoo for discussions on type-II fluctuations and PBH formation. I thank the support from the YLC program at the Institute for Advanced Research, Nagoya University.
\end{acknowledgments}

\appendix

\section{Sketch of the derivation of Eq.\eqref{eq:K_simply}}
\label{apendix_K}

In general, the line element of a $3+1$ spacetime can be written as:

\begin{equation}
    ds^2 = -A^{2} dt^{2} + \gamma_{ij} \left( dx^{i} + \beta^{i} dt \right) \left( dx^{j} + \beta^{j} dt \right)
\end{equation}
where $A$ is the lapse function, $\gamma_{ij}$ is the spatial metric, and $\beta^{i}$ is the shift vector. Under the assumption of spherical symmetry and the comoving threading $\beta^{i} = 0$, we have Eq.\eqref{2_metricsharp}. The extrinsic curvature of the spacetime is defined as

\begin{equation}
    K_{ij} = \frac{\mathcal{L}_{\beta}\gamma_{ij} - \partial_{t}\gamma_{ij}}{2 A},
\end{equation}
where its trace is given by $K = \gamma^{ij} K_{ij}$. The time derivative of $K$ reads as \cite{2016nure.book.....S,Baumgarte_Shapiro_2021}, 

\begin{equation}
\label{dot_K}
\dot{K} = -\nabla^i \nabla_{i} A + A \left( K_{ij} K^{ij} + 4 \pi (\rho + S) + \beta^{i} \nabla_{i} K \right),
\end{equation}
where $S = \gamma^{ij} T_{ij} = 3p$ taking into account that we are considering a perfect fluid (see Eq.\eqref{eq:tensor_energy}), and $\nabla_{i}$ is the covariant derivative with respect to the spatial index $i$. From now, consider the comoving threading $\beta^{i} =0$. The components $K_{ij}$ and $K^{ij}$ are given by

\begin{equation}
    K_{ij} = -\frac{\partial_{t} \gamma_{ij}}{2 A}, \, \,\,K^{\tilde{i}\tilde{j}} = \gamma^{\tilde{i}i}\gamma^{\tilde{j}j}K_{ij}.
\end{equation}

The different components in spherical coordinates read as,

\begin{align}
    \gamma_{rr} &= B^{2}, \quad \gamma_{\theta \theta} = R^{2}, \quad \gamma_{\phi \phi} = R^{2} \sin^{2} \theta \\
    K_{rr} &= -\frac{B \dot{B}}{A}, \quad K_{\theta \theta} = -\frac{R \dot{R}}{A}, \quad K_{\phi \phi} = -\frac{R \dot{R} \sin^{2} \theta}{A} \\
    \gamma^{rr} &= \frac{1}{B^{2}}, \quad \gamma^{\theta \theta} = \frac{1}{R^{2}}, \quad \gamma^{\phi \phi} = \frac{1}{R^{2} \sin^{2} \theta} \\
    K^{rr} &= -\frac{\dot{B}}{A B^{3}}, \quad K^{\theta \theta} = -\frac{\dot{R}}{A R^{3}}, \quad K^{\phi \phi} = -\frac{\dot{R}}{A R^{3} \sin^{2} \theta}
\end{align}

Then, for $K_{ij} K^{ij}$, we obtain:

\begin{equation}
\label{eq:contraction_K}
    K_{ij} K^{ij} = \frac{1}{A^{2}} \left[ \left( \frac{\dot{B}}{B} \right)^{2} + 2 \left( \frac{\dot{R}}{R} \right)^{2} \right] = \left( K + \frac{2U}{R} \right)^{2} + 2 \left( \frac{U}{R} \right)^{2},
\end{equation}
where, in the last step, we have used $\dot{B} = -BA (K + 2U/R)$ obtained from Eqs.\eqref{eq:K} and \eqref{BB}.

The D'Alembertian term $\nabla^{i} \nabla_{i} A$ reads as:

\begin{equation}
\label{eq:dalambertian}
    \nabla^{i} \nabla_{i} A = \frac{1}{\sqrt{\textrm{det}(\gamma_{ij})}} \partial_{i} \left( \sqrt{\textrm{det}(\gamma_{ij})} \gamma^{ij} \partial_{j} A \right) = \frac{1}{B R^{2}} \partial_{r} \left( \frac{A' R^{2}}{B} \right) = \frac{1}{B^{2}} \left[ A'' + A' \left( \frac{2 R'}{R} - \frac{B'}{B} \right) \right],
\end{equation}
where we have used the fact that the functions depend only on the $r,t$ coordinates and $ \textrm{det}(\gamma_{ij}) = B^{2} R^{4} \sin^{2} \theta$. Introducing Eqs.\eqref{eq:dalambertian} and \eqref{eq:contraction_K} into Eq.\eqref{dot_K} and taking into account that the last term vanishes because of $\beta^{i} = 0$, we obtain Eq.\eqref{eq:K_simply} in the main text.

\section{Sketch of the derivation of Eq.\eqref{eq:lie_congruences}}
\label{apendix_L}

In this section, we provide some details to derive Eq.\eqref{eq:lie_congruences}. Let's define the Lie derivative of $\Theta^{q} =(\Theta^{+},\Theta^{-})$ along the congruence $k_{p}^{\mu}= (k^{\mu(+)},k^{\mu(-)})$ as $\mathcal{L}_{p} \Theta^{q} = k_{p}^{\mu} \partial_{\mu} \Theta^{q}= (D_{t}+p D_{r})\Theta^{q}$, where $q,p$ take the values $(+,-)$.
\begin{align}
\label{eq:cosa}
    \mathcal{L}_{p} \Theta^{q} &= k_{p}^{\mu} \partial_{\mu} \Theta^{q}= (D_{t}+p D_{r})\Theta^{q}=(D_{t}+p D_{r}) \left[ \frac{2}{R}(U+q \,\Gamma)\right] \\ \nonumber
    &= \frac{2}{R}\left[ (D_{t}+p D_{r})(U+q \Gamma)-\frac{(U+q \Gamma)}{R}(D_t+p D_r)R  \right]
\end{align}
Using Eqs.\eqref{eq:u_simply},\eqref{eq:G_simply} and the analytical solution for the lapse function Eq.\eqref{eq:lapse} we find,
\begin{equation}
    D_t U + q \, D_t \Gamma = -\frac{M}{R^{2}}-4 \pi R w \rho-\frac{w}{w+1}\frac{\rho'}{\rho\,B} (\Gamma+q \,U)
\end{equation}
Differentiating Eq.\eqref{eq:gamma_constraint}, we can obtain an expression for $D_r \Gamma$, and then we have:
\begin{align}
    D_{r}U +q D_{r} \Gamma = \frac{U'}{R'}(q U+\Gamma)+q\left( \frac{M}{R^{2}}-4 \pi \rho R\right) \\
    =-\left(K+\frac{2 U}{R}\right)(q U+\Gamma)+q\left( \frac{M}{R^{2}}-4 \pi \rho R\right)
\end{align}
where we have used Eq.\eqref{eq:K} to avoid the divergent term $U'/R'$ for type-II fluctuations. Putting together the previous equations into Eq.\eqref{eq:cosa}, we obtain Eq.\eqref{eq:lie_congruences}.

\section{Sketch of the derivation of the initial conditions for PBH formation}
\label{sec:derivation_QHS}

In this appendix, we provide further details on the derivation of Eqs.\eqref{eq:U_tilde}–\eqref{eq:tilde_K}, which are used to set the initial conditions for the numerical simulations. We consider the spherically symmetric spacetime given by Eq.\eqref{2_metricsharp}, as well as the spacetime metric at zeroth order in the long-wavelength approximation in the limit $\epsilon \rightarrow 0$, given in Eq.\eqref{eq:spacetime_metric} with the conformally flat coordinate $\hat{r}$ and is characterised by the curvature perturbation $\zeta$. To construct the initial conditions, we substitute Eqs.\eqref{eq_ms} into the Misner–Sharp equations and expand in powers of $\epsilon$, keeping terms up to leading order in $\epsilon^2$ \cite{Lyth:2004gb,Polnarev:2006aa,Polnarev:2012bi,Harada:2015yda}, to find the tilde variables.

\begin{equation}
    \frac{\dot{\epsilon}}{\epsilon}= \frac{(1+3w)}{2}H\, ; \, \xi = \ln a ; \, \, \epsilon= \frac{R}{R_m}=\frac{1}{a H r_m e^{\zeta(r_m)}}.
\end{equation}

For the first equation, we have used the fact that $\dot{H} = -\frac{3}{2}(1 + w)H^2$. Additionally, we take into account that

\begin{equation}
    \frac{\partial (\epsilon^2 X)}{\partial t}\equiv\dot{(\epsilon^2 X )} = 2 \epsilon \dot{\epsilon}\tilde{X}+\epsilon^2 \dot{\tilde{X}} = (1+3w)H\epsilon^2 \tilde{X}+\epsilon^2 \frac{\partial \tilde{X}}{\partial t}=\epsilon^2 H \left[ (1+3w) \tilde{X}+\frac{\partial  \tilde{X}}{\partial \xi}\right].
\end{equation}

First of all, $\tilde{A}$ can be obtained by using the analytical solution of Eq.~\eqref{eq:lapse}:

\begin{align}
&A=1+\epsilon^{2} \tilde{A}+\mathcal{O}(\epsilon^4), \\
&\left(\frac{\rho_b}{\rho}\right)^{\frac{w}{w+1}}= 1-\frac{w}{1+w}\epsilon^{2}\tilde{\rho}+\mathcal{O}(\epsilon^4),\\
&\Rightarrow\tilde{A} = -\frac{w}{1+w}\tilde{\rho}.
\end{align}

On the other hand, from the constraint equation for $\Gamma$, Eq.~\eqref{eq:gamma_constraint}, and using $R'/B = \Gamma$, we can expand both terms in $\epsilon^2$ to obtain $\tilde{U}$

\begin{align}
&\left( \frac{R'}{B}\right)^{2} = (1+r\zeta')^{2}+2(1+r \zeta') \left[ \tilde{R}-\tilde{B}+r\tilde{R}'+r\zeta'(\tilde{R}-\tilde{B})\right]\epsilon^2+\mathcal{O}(\epsilon^4),\\
&\Gamma^2 = 1+U^2-\frac{2M}{R}= 1-e^{2(\zeta-\zeta_m)}\frac{r^2}{r^2_m}(\tilde{M}-2\tilde{U})-\epsilon^2\frac{r^2}{r^2_m}e^{2(\zeta-\zeta_m)}\left( 2\tilde{M}\tilde{R}-\tilde{U}(4\tilde{R}+\tilde{U}) \right)+\mathcal{O}(\epsilon^4).
\end{align}
From comparing the first terms at zeroth order in $\epsilon$, we obtain the relation:
\begin{align}
&(1+r\zeta')^{2}= 1-\left(\frac{r}{r_m}\right)^{2}e^{2 (\zeta-\zeta_m)}(\tilde{M}-2\tilde{U}),\\
\label{eq:M_tilde_cosa}
&\Rightarrow \tilde{U}= \frac{1}{2}\left[ \tilde{M}+e^{-2(\zeta-\zeta_m)}r^2_m \frac{\zeta'(2+r \zeta')}{r}\right].
\end{align}
where $\zeta_m \equiv \zeta(r_m)$.

Let's take Eq.~\eqref{eq:r_simply}. Expanding both sides of the equation, we get
\begin{align}
&\dot{R} = aHre^{\zeta}(1+\epsilon^2 \tilde{R})+are^{\zeta}\dot{(\epsilon^2 \tilde{R})}+\mathcal{O}(\epsilon^4)=aHr e^{\zeta}\left[ 1+\epsilon^2\left((2+3w)\tilde{R}+\frac{\partial \tilde{R}}{\partial \xi} \right)\right]+\mathcal{O}(\epsilon^4),\\
&AUR = aHre^{\zeta}(1+\epsilon^2 \tilde{A})(1+\epsilon^2 \tilde{U})(1+\epsilon^2 \tilde{R})+\mathcal{O}(\epsilon^4),\\
\Rightarrow&(1+3 w)\tilde{R}+\frac{\partial \tilde{R}}{\partial \xi} = \tilde{A}+\tilde{U},\\
&\frac{\partial \tilde{R}}{\partial \xi}=0 \Rightarrow \tilde{R} = \frac{1}{1+3w} (\tilde{A}+\tilde{U}).
\end{align}
In the last term, we set $\partial \tilde{R}/\partial \xi = 0$ since we are considering a constant equation of state $w$.

We can use now Eq.\eqref{eq:M_simply} to obtain $\tilde{M}$

\begin{align}
&\dot{M} = \frac{4\pi}{3}\left(\dot{\rho}_b R^3 +3 \rho_b R^2 \dot{R}\right)(1+\epsilon^2\tilde{M})+\frac{4\pi}{3}(\rho_b R^3)\dot{(\epsilon^2\tilde{M})}+\mathcal{O}(\epsilon^4),\\
&= \frac{4 \pi}{3}\rho_b R^3 \left[(1+\epsilon^2 \tilde{M})\left( \frac{\dot{\rho}_b}{\rho_b}+3\frac{\dot{R}}{R} \right)+\epsilon^2 H(1+3w) \tilde{M}+\epsilon^2 H \frac{\partial \tilde{M}}{\partial \xi}\right]+\mathcal{O}(\epsilon^4),\\
&-4\pi w \rho R^2 A U=-4\pi w \rho R^2 \dot{R}=-\frac{4\pi}{3} \rho_b R^3 \left(3 w(1+\epsilon^2 \tilde{\rho})\frac{\dot{R}}{R}\right)+\mathcal{O}(\epsilon^4).
\end{align}

Then, taking into account the FLRW equations $\dot{\rho}_b=-3H(1+w)\rho_b$ and $\dot{R}/R = H\left[ 1+\epsilon^2(\tilde{U}+\tilde{A})\right]+\mathcal{O}(\epsilon^4)$
\begin{align}
&\dot{M} = \frac{4 \pi}{3}\rho_b R^3 \left( -3w H+H\epsilon^2\left[\frac{\partial \tilde{M}}{\partial \xi}+3(\tilde{A}+\tilde{U})+\tilde{M} \right] \right)+\mathcal{O}(\epsilon^4),\\
&-4\pi w \rho R^2 A U=-\frac{4\pi}{3}\rho_b R^3\left[ 3wH(1+\epsilon^2 \tilde{\rho})(1+\epsilon^2(\tilde{A}+\tilde{U})) \right]+\mathcal{O}(\epsilon^4),\\
\label{eq:1234}
&\Rightarrow\tilde{M}+\frac{\tilde{M}}{\partial \xi}=-3(1+w)\tilde{U}
\Rightarrow\tilde{M}=-3(1+w)\tilde{U},
\end{align}
where in the last equation we have used $\tilde{A}=-w \tilde{\rho}/(1+w)$.

We can now use Eq.\eqref{eq:M_tilde_cosa} and Eq.\eqref{eq:1234} to obtain $\tilde{U}$ independently of the other tilde variables:
\begin{equation}
\label{eq:4567}
\tilde{U} = \frac{1}{5+3w} e^{-2 (\zeta-\zeta_m)} \zeta' \left( \frac{2}{r}+\zeta' \right) r^2_m.
\end{equation}

Let's now use the Hamiltonian constraint equation Eq.\eqref{eq:H_constraint}. making the expansion we first have,
\begin{align}
&M'=\frac{4 \pi}{3} \rho_bR^3 \left( 3 \frac{R'}{R} (1+\epsilon^2 \tilde{M})+\epsilon^2  \tilde{M}'\right) +\mathcal{O}(\epsilon^4),\\
&4\pi \rho R^2 R'=\frac{4 \pi}{3} \rho_bR^3\left[3(1+\epsilon^2\tilde{\rho})\frac{R'}{R}\right]+\mathcal{O}(\epsilon^4),\\
&\Rightarrow \frac{1}{3}\tilde{M}=\frac{R'}{R}\left(\tilde{\rho}-\tilde{M}\right),\\
&\Rightarrow  \tilde{\rho} = \tilde{M}+\frac{r \tilde{M}'}{3(1+r\zeta')},\\
&\tilde{\rho} = \frac{-2(1+w)}{5+3w}e^{-2 (\zeta-\zeta_m)} \left[ \zeta''+\zeta'\left(\frac{2}{r}+\frac{\zeta'}{2}\right) \right]r^2_m ,
\end{align}
where in the last sep we have used Eq.\eqref{eq:1234} and Eq.\eqref{eq:4567}. The variables $\tilde{A}, \tilde{U}, \tilde{R}, \tilde{\rho}, \tilde{M}$ can then be expressed in terms of $\zeta$ and its derivatives. The result matches that of \cite{Polnarev:2006aa, Musco:2018rwt}.

We now move on to obtain the equation for $\tilde{B}$. By using Eq.\eqref{BB} with $\dot{B}=AU'B/R'$ we have
\begin{align}
&\dot{B} = a e^{\zeta}H \left[1+\epsilon^2 \left( \tilde{B}(2+3w)+\frac{\partial \tilde{B}}{\partial \xi}\right) \right]+\mathcal{O}(\epsilon^4),\\
&\frac{AU'B}{R'} = aHe^{\zeta}\left[1+\epsilon^2\left( \tilde{A}+\tilde{B}+\tilde{U}+\frac{r\tilde{U}'}{1+r\zeta'} \right)\right]+\mathcal{O}(\epsilon^4),
\end{align}
putting together the previous equations up to the order $\epsilon^2$ we get,
\begin{equation}
  \tilde{B}(1+3w)+\frac{\partial \tilde{B}}{\partial \xi}=\tilde{A}+\tilde{U}+\frac{r \tilde{U}'}{1+r\zeta'}.
\end{equation}

Taking into account the perturbation variables $\tilde{U}$ and $\tilde{\rho}$ obtained in the previous steps (which correspond to Eq.\eqref{eq:U_tilde} and Eq.\eqref{eq:rho_tilde}), we obtain the following relation:

\begin{equation}
\label{eq:cosaaa}
\tilde{U}'= \frac{-(1+r \zeta')}{r}\left[ \frac{\tilde{\rho}}{1+w}+3 \tilde{U} \right].
\end{equation}
Substituting this into the previous equation and taking into account the expression for $\tilde{A}$ obtained earlier (which corresponds to Eq.~\eqref{eq:tilde_A}), we get:

\begin{equation}
\label{eq:cosita}
\tilde{B} = \frac{-1}{1+3w}(\tilde{\rho}+2\tilde{U}),
\end{equation}

which corresponds to Eq.~\eqref{eq:Btilde} in the main text. Finally, let us obtain the term $\tilde{K}$. From $K = -(\dot{B}/B + 2\dot{R}/R)/A$ and expanding to order $\epsilon^2$, we obtain:

\begin{align}
\label{eq:K_expansion}
&K = -3H(1+\epsilon^2 \tilde{K})+\mathcal{O}(\epsilon^4),\\
&-\frac{1}{A}(\dot{B}/B+2\dot{R}/R) = -3H \left[1+\epsilon^2 \left( \frac{\tilde{B}(1+3w)+2\tilde{U}-\tilde{A}}{3} \right)\right]+\mathcal{O}(\epsilon^4),
\end{align}
where we took $\partial \tilde{X}/\partial \xi=0$. Matching the $\epsilon^2$ terms, we get:
\begin{equation}
    \tilde{K} = \frac{1}{3}\left(2 \tilde{U}-\tilde{A}+\tilde{B}(1+3w)\right)=\frac{-\tilde{\rho}}{3(1+w)},
\end{equation}
therefore obtaining Eq.\eqref{eq:tilde_K}, where we have used the variables $\tilde{B}$ and $\tilde{A}$ obtained previously. Notice that the same result can be obtained by using Eq.\eqref{eq:K} instead. Taking that into account,

\begin{align}
&\frac{U}{R} = H\left( 1+\epsilon^2 \tilde{U}\right)+\mathcal{O}(\epsilon^4), \\
&\frac{U'}{R'} =  \frac{H\left[R'(1+\epsilon^2 \tilde{U})+R \epsilon^2 \tilde{U}'\right]}{R'}= H \left[ 1-\epsilon^2 \left( \frac{\tilde{\rho}}{1+w} + 2\tilde{U}  \right)\right]+\mathcal{O}(\epsilon^4),\\
&K = -\left( \frac{U'}{R'}+2\frac{U}{R}\right) = -3 H +\epsilon^2 H  \frac{\tilde{\rho}}{1+w}+\mathcal{O}(\epsilon^4),
\end{align}
where we have used Eq.\eqref{eq:cosaaa} and the fact that
\begin{equation}
\frac{R}{R'} = \frac{r}{1+r \zeta'}+\mathcal{O}(\epsilon^2).
\end{equation}
Then, by comparing the $\epsilon^2$ terms, we obtain again: 
\begin{equation}
    \tilde{K} = -\frac{\tilde{\rho}}{3(1+w)},
\end{equation}
which also gives the same result if we take Eq.~\eqref{eq:K_simply} and expand the right-hand side of the equation to order $\epsilon^2$. The equation for $\tilde{B}$ can also be obtained by using the definition of $\dot{B}=-A B (K+2U/R)$ to isolate $\tilde{B}$, using the result for $\tilde{K}$, which yields the same result obtained below Eq.\eqref{eq:cosita}.

Equation \eqref{eq:tilde_K} matches the result obtained in \cite{Harada:2015yda} using the gradient expansion method for the long-wavelength solution in the comoving gauge (see Eqs.(4.3) and (4.67) in \cite{Harada:2015yda}), thereby proving the consistency of our quasi-homogeneous expansion. However, the result for $\tilde{B}=-(\tilde{\rho}+2\tilde{U})/(1+3w)$ differs from that of Ref.~\cite{Musco:2018rwt}. To clarify, let's try compare our result for $\tilde{B}$ in Eq.~\eqref{eq:Btilde} with the expression given in \cite{Musco:2018rwt} (Eq.(34) in that reference) that we denote as $B_{C}(r,t)$,
\begin{align}
\label{eq:numero_cosas}
&B_C(r,t) =a e^{\zeta}(1+\epsilon^2 \tilde{B}_C(r)),\\ \nonumber
&\tilde{B}_{C}(r)=\frac{w}{(1+3w)(1+w)}\frac{r}{1+r\zeta'}\tilde{\rho}'.  
\end{align}
We can make a coordinate transformation from the areal radial coordinate $\hat{r}$ to the conformally flat coordinate $r$. Accordingly, we define $B_{\rm A}$ following \cite{Polnarev:2006aa} using the areal radial coordinate $\hat{r}$ as follows\footnote{We use the prime symbol (') to denote derivatives with respect to $r$ only, and not with respect to $\hat{r}$.}:

\begin{align}
&\hat{B}_{\rm A}(\hat{r}, t) = 
\frac{\partial_{\hat{r}} R(\hat{r}, t)}{\sqrt{1 - \hat{r}^2 \chi(\hat{r})}} 
\left(1 + \epsilon^2 \tilde{B}_{\rm A}(\hat{r})\right),\\ \nonumber
&\tilde{B}_{\rm A}(\hat{r})=\frac{w}{(1+3w)(1+w)}\hat{r} \partial_{\hat{r}}\tilde{\rho}(\hat{r}).
\end{align}
The coordinate $\hat{r}$ is defined through the spatial metric in the long-wavelength limit $\epsilon \rightarrow 0$, which resembles the flat FLRW metric with a non-homogeneous curvature function $\chi(\hat{r})$.
\begin{align}
    &ds^2 =-\hat{A}^2(\hat{r},t)dt^2+\hat{B}^2(\hat{r},t)d\hat{r}^2+\hat{R}^2(\hat{r},t)d\Omega^2,\\ \nonumber
    &d\hat{\Sigma}^{2}(\epsilon\rightarrow 0) = a^2 \left(\frac{d\hat{r}^2}{1 - \chi(\hat{r})\hat{r}^2} + \hat{r}^2 d{\Omega}^2\right).
\end{align}
Let us now change the coordinate $\hat{r}$ to the conformally flat coordinate $r$, in which the spatial metric is characterized by the curvature function $\zeta(r)$.
\begin{align}
&ds^2 =-A^2(r,t)dt^2+B^2(r,t)dr^2+R^2(r,t)d\Omega^2,\\ \nonumber
&d\Sigma^{2}(\epsilon\rightarrow 0)=a^2 e^{2\zeta(r)}\left( dr^2+r^2 d\Omega^2\right).
\end{align}
The coordinate transformation is described in detail in \cite{Harada:2015yda}, and we refer the reader to that work for further details. However, we present here some definitions that we use:
\begin{align}
&\hat{r}= re^{\zeta},\\
&\frac{d\hat{r}}{dr} = e^{\zeta}(1+r\zeta'),\\
&\sqrt{1-\hat{r}^2 \chi(\hat{r})}=1+r\zeta'.
\end{align}
To obtain $B_{\rm A}(r, t)$ from a coordinate transformation $\hat{r} \rightarrow r$ applied to $\hat{B}_{\rm A}(\hat{r}, t)$, we need to account for the change of coordinates in the spacetime metric. The components of the metric tensor transform under a coordinate change as
\begin{align}
   &g_{\mu\nu} = g_{\hat{ \mu}\hat{\nu}}\frac{\partial x^{\hat{\mu}}}{\partial x^{\mu}}\frac{\partial x^{\hat{\nu}}}{\partial x^{\nu}}, \\ \nonumber
   &g_{rr} = g_{\hat{r}\hat{r}}\left(\frac{d\hat{r}}{dr}\right)^2 \Rightarrow B_{\rm A}(r,t)=\hat{B}_{\rm A}(\hat{r},t)\frac{d\hat{r}}{dr}.
\end{align}
Only the radial component of the metric is affected due to the one-to-one correspondence $\hat{r} \rightarrow r$. Applying this transformation to find $B_{\rm A}(r, t)$, we obtain:
\begin{align}
&B_{\rm A}(r,t)=\hat{B}_{\rm A}(\hat{r}(r),t)\frac{d\hat{r}}{dr}=\frac{d\hat{r}}{dr}\frac{dr}{d\hat{r}}\frac{ R'(\hat{r}(r),t)}{1+r\zeta'}(1+\epsilon^2\tilde{B}_{\rm A}(\hat{r}(r))),\\
&R'(r,t) = \partial_r(ar e^{\zeta}(1+\epsilon^2 \tilde{R}))=ae^{\zeta}\left[(1+r\zeta')+\epsilon^2 \left( (1+r\zeta')\tilde{R}+r \tilde{R}' \right) \right],\\
\label{eq:unadostres}
&\Rightarrow B_{\rm A}(r,t)=ae^{\zeta}\left[1+\epsilon^2 \left( \tilde{R}+\frac{r \tilde{R}'}{1+r\zeta'}\right) \right]\left[ 1+\epsilon^2\tilde{B}_{\rm A}(\hat{r}(r))\right].
\end{align}

By performing some additional calculations, we obtain
\begin{align}
&\tilde{B}_{\rm A}(\hat{r}(r))=\frac{w}{(1+3w)(1+w)}\hat{r} \partial_{\hat{r}}\tilde{\rho}=\frac{w}{(1+3w)(1+w)}\frac{r}{1+r \zeta'} \tilde{\rho}',\\
\label{eq_intermedio}
&\tilde{R}'  = \frac{-1}{r(1+w)(1+3w)}\left[ r w\tilde{\rho}'+(1+r\zeta')(3(1+w)\tilde{U}+\tilde{\rho}) \right], \\
\label{eq:B_final_final}
&\Rightarrow B_{\rm A}(r,t) = a e^{\zeta}\Biggl[1-\epsilon^2 \frac{(\tilde{\rho}+2\tilde{U})}{1+3w}\\
\nonumber
&-\epsilon^4 \frac{
 r\, w\, \tilde{\rho}' \left[
  r\, w\, \tilde{\rho}' +
  2(1 + w)\, \tilde{U}\, \left(1 + r\, \zeta'\right) +
  (1 + w)\, \tilde{\rho}\, \left(1 + r\,  \zeta'\right)
\right]\, 
}{
(1 + w)^2 (1 + 3w)^2 \left(1 + r\, \zeta'\right)^2
}
 \biggr].
\end{align}

where we have used Eq.\eqref{eq:cosaaa} in Eq.\eqref{eq_intermedio}. When comparing Eq.\eqref{eq:B_final_final} with our Eqs.\eqref{eq_ms} and \eqref{eq:Btilde} for $B, \tilde{B}$ in the main text, we recover the result from our work, $\tilde{B} = - (\tilde{\rho} + 2\tilde{U}) / (1 + 3w)$ at order $\epsilon^2$. The term $\epsilon^4$ in Eq.\eqref{eq:B_final_final} comes from defining the asymptotic solution at $\epsilon \rightarrow 0$ in terms of $\partial_{\hat{r}} R(\hat{r},t) / \sqrt{(1 - \hat{r}^2 \chi(\hat{r}))}$, which has a divergence for type-II fluctuations (when $1 + r_{II} \zeta'(r_{II}) = 0$) and does not cancel out. Our calculations confirm that our initial conditions are consistent with the quasi-homogeneous solution. In particular, we clarified how to set the initial condition for the function $\tilde{B}(r)$ in the conformally flat coordinate $r$.

\bibliographystyle{JHEP}
\bibliography{references.bib}

@article{Zeldovich:1967lct,
    author = "Zel'dovich, Ya. B. and Novikov, I. D.",
    title = "{The Hypothesis of Cores Retarded during Expansion and the Hot Cosmological Model}",
    journal = "Sov. Astron.",
    volume = "10",
    pages = "602",
    year = "1967"
}

@article{Hawking:1971ei,
    author = "Hawking, Stephen",
    title = "{Gravitationally collapsed objects of very low mass}",
    doi = "10.1093/mnras/152.1.75",
    journal = "Mon. Not. Roy. Astron. Soc.",
    volume = "152",
    pages = "75",
    year = "1971"
}

@article{Carr:1974nx,
    author = "Carr, Bernard J. and Hawking, S. W.",
    title = "{Black holes in the early Universe}",
    doi = "10.1093/mnras/168.2.399",
    journal = "Mon. Not. Roy. Astron. Soc.",
    volume = "168",
    pages = "399--415",
    year = "1974"
}

@article{Carr:1975qj,
    author = "Carr, Bernard J.",
    title = "{The Primordial black hole mass spectrum}",
    doi = "10.1086/153853",
    journal = "Astrophys. J.",
    volume = "201",
    pages = "1--19",
    year = "1975"
}

@article{Aurrekoetxea:2024mdy,
    author = "Aurrekoetxea, Josu C. and Clough, Katy and Lim, Eugene A.",
    title = "{Cosmology using numerical relativity}",
    eprint = "2409.01939",
    archivePrefix = "arXiv",
    primaryClass = "gr-qc",
    month = "9",
    year = "2024"
}

@article{Chapline:1975ojl,
	author = {Chapline, George F.},
	doi = {10.1038/253251a0},
	journal = {Nature},
	number = {5489},
	pages = {251--252},
	title = {{Cosmological effects of primordial black holes}},
	volume = {253},
	year = {1975}}

@ARTICLE{2016PhRvX...6d1015A,
       author = {{LIGO Scientific Collaboration} and {Virgo Collaboration} and B. Abbott and others},
        title = "{Binary Black Hole Mergers in the First Advanced LIGO Observing Run}",
      journal = {Physical Review X},
         year = 2016,
        month = oct,
       volume = {6},
       number = {4},
          eid = {041015},
        pages = {041015},
          doi = {10.1103/PhysRevX.6.041015},
archivePrefix = {arXiv},
       eprint = {1606.04856},
 primaryClass = {gr-qc},
       adsurl = {https://ui.adsabs.harvard.edu/abs/2016PhRvX...6d1015A}
}

@ARTICLE{2023PhRvX..13d1039A,
       author = {{LIGO Scientific Collaboration} and {Virgo Collaboration} and {KAGRA Collaboration} and {Abbott}, R. and others},
        title = "{GWTC-3: Compact Binary Coalescences Observed by LIGO and Virgo during the Second Part of the Third Observing Run}",
      journal = {Physical Review X},
         year = 2023,
        month = oct,
       volume = {13},
       number = {4},
          eid = {041039},
        pages = {041039},
          doi = {10.1103/PhysRevX.13.041039},
archivePrefix = {arXiv},
       eprint = {2111.03606},
 primaryClass = {gr-qc},
       adsurl = {https://ui.adsabs.harvard.edu/abs/2023PhRvX..13d1039A}
}

@article{Carr:2019kxo,
    author = {Carr, Bernard and Clesse, Sebastien and Garc\'\i{}a-Bellido, Juan and K\"uhnel, Florian},
    title = "{Cosmic conundra explained by thermal history and primordial black holes}",
    eprint = "1906.08217",
    archivePrefix = "arXiv",
    primaryClass = "astro-ph.CO",
    doi = "10.1016/j.dark.2020.100755",
    journal = "Phys. Dark Univ.",
    volume = "31",
    pages = "100755",
    year = "2021"
}

@article{Carr:2023tpt,
    author = "Carr, Bernard and Clesse, Sebastien and Garcia-Bellido, Juan and Hawkins, Michael and Kuhnel, Florian",
    title = "{Observational evidence for primordial black holes: A positivist perspective}",
    eprint = "2306.03903",
    archivePrefix = "arXiv",
    primaryClass = "astro-ph.CO",
    doi = "10.1016/j.physrep.2023.11.005",
    journal = "Phys. Rept.",
    volume = "1054",
    pages = "1--68",
    year = "2024"
}

@article{Escriva:2022duf,
    author = "Escriv\`a, Albert and Kuhnel, Florian and Tada, Yuichiro",
    title = "{Primordial Black Holes}",
    eprint = "2211.05767",
    archivePrefix = "arXiv",
    primaryClass = "astro-ph.CO",
    doi = "10.1016/B978-0-32-395636-9.00012-8",
    month = "11",
    year = "2022"
}

@article{Escriva:2021aeh,
    author = "Escriv\`a, Albert",
    title = "{PBH Formation from Spherically Symmetric Hydrodynamical Perturbations: A Review}",
    eprint = "2111.12693",
    archivePrefix = "arXiv",
    primaryClass = "gr-qc",
    doi = "10.3390/universe8020066",
    journal = "Universe",
    volume = "8",
    number = "2",
    pages = "66",
    year = "2022"
}

@article{Escriva:2019nsa,
    author = "Escriv\`a, Albert",
    title = "{Simulation of primordial black hole formation using pseudo-spectral methods}",
    eprint = "1907.13065",
    archivePrefix = "arXiv",
    primaryClass = "gr-qc",
    reportNumber = "ICCUB-19-012",
    doi = "10.1016/j.dark.2020.100466",
    journal = "Phys. Dark Univ.",
    volume = "27",
    pages = "100466",
    year = "2020"
}

@article{Harada:2024jxl,
    author = "Harada, Tomohiro",
    title = "{Primordial Black Holes: Formation, Spin and Type II}",
    eprint = "2409.01934",
    archivePrefix = "arXiv",
    primaryClass = "gr-qc",
    reportNumber = "RUP-24-16",
    doi = "10.3390/universe10120444",
    journal = "Universe",
    volume = "10",
    number = "12",
    pages = "444",
    year = "2024"
}

@article{Milligan:2025zbu,
    author = "Milligan, Ethan and Padilla, Luis E. and Mulryne, David J. and Hidalgo, Juan Carlos",
    title = "{Primordial Black Hole Formation in a Scalar Field Dominated Universe}",
    eprint = "2504.02600",
    archivePrefix = "arXiv",
    primaryClass = "astro-ph.CO",
    month = "4",
    year = "2025"
}

@article{Escriva:2023qnq,
    author = "Escriv\`a, Albert and Yoo, Chul-Moon",
    title = "{Primordial Black hole formation from overlapping cosmological fluctuations}",
    eprint = "2310.16482",
    archivePrefix = "arXiv",
    primaryClass = "gr-qc",
    doi = "10.1088/1475-7516/2024/04/048",
    journal = "JCAP",
    volume = "04",
    pages = "048",
    year = "2024"
}

@article{Kitajima:2021fpq,
    author = "Kitajima, Naoya and Tada, Yuichiro and Yokoyama, Shuichiro and Yoo, Chul-Moon",
    title = "{Primordial black holes in peak theory with a non-Gaussian tail}",
    eprint = "2109.00791",
    archivePrefix = "arXiv",
    primaryClass = "astro-ph.CO",
    reportNumber = "TU-1130",
    doi = "10.1088/1475-7516/2021/10/053",
    journal = "JCAP",
    volume = "10",
    pages = "053",
    year = "2021"
}

@article{Yoo:2022mzl,
    author = "Yoo, Chul-Moon",
    title = "{The Basics of Primordial Black Hole Formation and Abundance Estimation}",
    eprint = "2211.13512",
    archivePrefix = "arXiv",
    primaryClass = "astro-ph.CO",
    doi = "10.3390/galaxies10060112",
    journal = "Galaxies",
    volume = "10",
    number = "6",
    pages = "112",
    year = "2022"
}

@article{Escriva:2022yaf,
    author = "Escriv\`a, Albert and Subils, Javier G.",
    title = "{Primordial black hole formation during a strongly coupled crossover}",
    eprint = "2211.15674",
    archivePrefix = "arXiv",
    primaryClass = "astro-ph.CO",
    reportNumber = "NORDITA 2022-082",
    doi = "10.1103/PhysRevD.107.L041301",
    journal = "Phys. Rev. D",
    volume = "107",
    number = "4",
    pages = "L041301",
    year = "2023"
}

@ARTICLE{1980AZh....57..250N,
       author = {{Novikov}, I.~D. and {Polnarev}, A.~G.},
        title = "{The hydrodynamics of primordial black hole formation - Dependence on the equation of state}",
      journal = {\azh},
         year = 1980,
        month = apr,
       volume = {57},
        pages = {250-258},
       adsurl = {https://ui.adsabs.harvard.edu/abs/1980AZh....57..250N}
}

@ARTICLE{1978SvA....22..129N,
       author = {{Nadezhin}, D.~K. and {Novikov}, I.~D. and {Polnarev}, A.~G.},
        title = "{The hydrodynamics of primordial black hole formation}",
      journal = {\sovast},
         year = 1978,
        month = apr,
       volume = {22},
        pages = {129-138},
       adsurl = {https://ui.adsabs.harvard.edu/abs/1978SvA....22..129N}
}

@article{Musco:2008hv,
    author = "Musco, Ilia and Miller, John C. and Polnarev, Alexander G.",
    title = "{Primordial black hole formation in the radiative era: Investigation of the critical nature of the collapse}",
    eprint = "0811.1452",
    archivePrefix = "arXiv",
    primaryClass = "gr-qc",
    doi = "10.1088/0264-9381/26/23/235001",
    journal = "Class. Quant. Grav.",
    volume = "26",
    pages = "235001",
    year = "2009"
}

@article{Musco:2012au,
    author = "Musco, Ilia and Miller, John C.",
    title = "{Primordial black hole formation in the early universe: critical behaviour and self-similarity}",
    eprint = "1201.2379",
    archivePrefix = "arXiv",
    primaryClass = "gr-qc",
    doi = "10.1088/0264-9381/30/14/145009",
    journal = "Class. Quant. Grav.",
    volume = "30",
    pages = "145009",
    year = "2013"
}

@article{Yoo:2021fxs,
    author = "Yoo, Chul-Moon and Harada, Tomohiro and Hirano, Shin'ichi and Okawa, Hirotada and Sasaki, Misao",
    title = "{Primordial black hole formation from massless scalar isocurvature}",
    eprint = "2112.12335",
    archivePrefix = "arXiv",
    primaryClass = "gr-qc",
    reportNumber = "YITP-21-161, RUP-21-23",
    doi = "10.1103/PhysRevD.105.103538",
    journal = "Phys. Rev. D",
    volume = "105",
    number = "10",
    pages = "103538",
    year = "2022"
}

@article{Escriva:2024aeo,
    author = "Escriv\`a, Albert and Yoo, Chul-Moon",
    title = "{Non-spherical effects on the mass function of Primordial Black Holes}",
    eprint = "2410.03451",
    archivePrefix = "arXiv",
    primaryClass = "gr-qc",
    month = "10",
    year = "2024"
}

@article{Escriva:2024lmm,
    author = "Escriv\`a, Albert and Yoo, Chul-Moon",
    title = "{Simulations of Ellipsoidal Primordial Black Hole Formation}",
    eprint = "2410.03452",
    archivePrefix = "arXiv",
    primaryClass = "gr-qc",
    month = "10",
    year = "2024"
}

@article{Yoo:2024lhp,
    author = "Yoo, Chul-Moon",
    title = "{Primordial black hole formation from a nonspherical density profile with a misaligned deformation tensor}",
    eprint = "2403.11147",
    archivePrefix = "arXiv",
    primaryClass = "gr-qc",
    doi = "10.1103/PhysRevD.110.043526",
    journal = "Phys. Rev. D",
    volume = "110",
    number = "4",
    pages = "043526",
    year = "2024"
}

@article{deJong:2023gsx,
    author = "de Jong, Eloy and Aurrekoetxea, Josu C. and Lim, Eugene A. and Fran\c{c}a, Tiago",
    title = "{Spinning primordial black holes formed during a matter-dominated era}",
    eprint = "2306.11810",
    archivePrefix = "arXiv",
    primaryClass = "astro-ph.CO",
    reportNumber = "KCL-PH-TH/2023-35",
    doi = "10.1088/1475-7516/2023/10/067",
    journal = "JCAP",
    volume = "10",
    pages = "067",
    year = "2023"
}

@article{deJong:2021bbo,
    author = "de Jong, Eloy and Aurrekoetxea, Josu C. and Lim, Eugene A.",
    title = "{Primordial black hole formation with full numerical relativity}",
    eprint = "2109.04896",
    archivePrefix = "arXiv",
    primaryClass = "astro-ph.CO",
    reportNumber = "KCL-PH-TH/2021-65",
    doi = "10.1088/1475-7516/2022/03/029",
    journal = "JCAP",
    volume = "03",
    number = "03",
    pages = "029",
    year = "2022"
}

@article{PhysRevD.83.124025,
  title = {Separate universes do not constrain primordial black hole formation},
  author = {Kopp, Michael and Hofmann, Stefan and Weller, Jochen},
  journal = {Phys. Rev. D},
  volume = {83},
  issue = {12},
  pages = {124025},
  numpages = {20},
  year = {2011},
  month = {Jun},
  publisher = {American Physical Society},
  doi = {10.1103/PhysRevD.83.124025},
  url = {https://link.aps.org/doi/10.1103/PhysRevD.83.124025}
}

@article{Okawa:2014nda,
    author = "Okawa, Hirotada and Witek, Helvi and Cardoso, Vitor",
    title = "{Black holes and fundamental fields in Numerical Relativity: initial data construction and evolution of bound states}",
    eprint = "1401.1548",
    archivePrefix = "arXiv",
    primaryClass = "gr-qc",
    doi = "10.1103/PhysRevD.89.104032",
    journal = "Phys. Rev. D",
    volume = "89",
    number = "10",
    pages = "104032",
    year = "2014"
}

@article{Yoo:2014boa,
    author = "Yoo, Chul-Moon and Okawa, Hirotada",
    title = "{Black hole universe with a cosmological constant}",
    eprint = "1404.1435",
    archivePrefix = "arXiv",
    primaryClass = "gr-qc",
    doi = "10.1103/PhysRevD.89.123502",
    journal = "Phys. Rev. D",
    volume = "89",
    number = "12",
    pages = "123502",
    year = "2014"
}

@article{PhysRevD.59.024007,
  title = {Numerical integration of Einstein's field equations},
  author = {Baumgarte, Thomas W. and Shapiro, Stuart L.},
  journal = {Phys. Rev. D},
  volume = {59},
  issue = {2},
  pages = {024007},
  numpages = {7},
  year = {1998},
  month = {Dec},
  publisher = {American Physical Society},
  doi = {10.1103/PhysRevD.59.024007},
  url = {https://link.aps.org/doi/10.1103/PhysRevD.59.024007}
}

@article{PhysRevD.52.5428,
  title = {Evolution of three-dimensional gravitational waves: Harmonic slicing case},
  author = {Shibata, Masaru and Nakamura, Takashi},
  journal = {Phys. Rev. D},
  volume = {52},
  issue = {10},
  pages = {5428--5444},
  numpages = {0},
  year = {1995},
  month = {Nov},
  publisher = {American Physical Society},
  doi = {10.1103/PhysRevD.52.5428},
  url = {https://link.aps.org/doi/10.1103/PhysRevD.52.5428}
}

@ARTICLE{1986ApJ...304...15B,
       author = {{Bardeen}, J.~M. and {Bond}, J.~R. and {Kaiser}, N. and {Szalay}, A.~S.},
        title = "{The Statistics of Peaks of Gaussian Random Fields}",
      journal = {\apj},
         year = 1986,
        month = may,
       volume = {304},
        pages = {15},
          doi = {10.1086/164143},
       adsurl = {https://ui.adsabs.harvard.edu/abs/1986ApJ...304...15B}
}

@article{escriva2023formation,
    author = "Escriv\`a, Albert and Atal, Vicente and Garriga, Jaume",
    title = "{Formation of trapped vacuum bubbles during inflation, and consequences for PBH scenarios}",
    eprint = "2306.09990",
    archivePrefix = "arXiv",
    primaryClass = "astro-ph.CO",
    doi = "10.1088/1475-7516/2023/10/035",
    journal = "JCAP",
    volume = "10",
    pages = "035",
    year = "2023"
}

@article{Gow:2022jfb,
    author = "Gow, Andrew D. and Assadullahi, Hooshyar and Jackson, Joseph H. P. and Koyama, Kazuya and Vennin, Vincent and Wands, David",
    title = "{Non-perturbative non-Gaussianity and primordial black holes}",
    eprint = "2211.08348",
    archivePrefix = "arXiv",
    primaryClass = "astro-ph.CO",
    doi = "10.1209/0295-5075/acd417",
    journal = "EPL",
    volume = "142",
    number = "4",
    pages = "49001",
    year = "2023"
}

@article{Fumagalli:2024kxe,
    author = "Fumagalli, Jacopo and Garriga, Jaume and Germani, Cristiano and Sheth, Ravi K.",
    title = "{The unexpected shape of the primordial black hole mass function}",
    eprint = "2412.07709",
    archivePrefix = "arXiv",
    primaryClass = "astro-ph.CO",
    month = "12",
    year = "2024"
}

@article{Uehara:2024yyp,
    author = "Uehara, Koichiro and Escriv\`a, Albert and Harada, Tomohiro and Saito, Daiki and Yoo, Chul-Moon",
    title = "{Numerical simulation of type II primordial black hole formation}",
    eprint = "2401.06329",
    archivePrefix = "arXiv",
    primaryClass = "gr-qc",
    reportNumber = "RUP-24-1",
    doi = "10.1088/1475-7516/2025/01/003",
    journal = "JCAP",
    volume = "01",
    pages = "003",
    year = "2025"
}

@article{PhysRev.136.B571,
  title = {Relativistic Equations for Adiabatic, Spherically Symmetric Gravitational Collapse},
  author = {Misner, Charles W. and Sharp, David H.},
  journal = {Phys. Rev.},
  volume = {136},
  issue = {2B},
  pages = {B571--B576},
  numpages = {0},
  year = {1964},
  month = {Oct},
  publisher = {American Physical Society},
  doi = {10.1103/PhysRev.136.B571},
  url = {https://link.aps.org/doi/10.1103/PhysRev.136.B571}
}

@article{Shimada:2024eec,
    author = "Shimada, Masaaki and Escriv\'a, Albert and Saito, Daiki and Uehara, Koichiro and Yoo, Chul-Moon",
    title = "{Primordial black hole formation from type II fluctuations with primordial non-Gaussianity}",
    eprint = "2411.07648",
    archivePrefix = "arXiv",
    primaryClass = "gr-qc",
    doi = "10.1088/1475-7516/2025/02/018",
    journal = "JCAP",
    volume = "02",
    pages = "018",
    year = "2025"
}

@article{Inui:2024fgk,
    author = "Inui, Ryoto and Joana, Cristian and Motohashi, Hayato and Pi, Shi and Tada, Yuichiro and Yokoyama, Shuichiro",
    title = "{Primordial black holes and induced gravitational waves from logarithmic non-Gaussianity}",
    eprint = "2411.07647",
    archivePrefix = "arXiv",
    primaryClass = "astro-ph.CO",
    month = "11",
    year = "2024"
}

@article{Harada:2015yda,
    author = "Harada, Tomohiro and Yoo, Chul-Moon and Nakama, Tomohiro and Koga, Yasutaka",
    title = "{Cosmological long-wavelength solutions and primordial black hole formation}",
    eprint = "1503.03934",
    archivePrefix = "arXiv",
    primaryClass = "gr-qc",
    reportNumber = "RUP-15-5, RESCEU-4-15",
    doi = "10.1103/PhysRevD.91.084057",
    journal = "Phys. Rev. D",
    volume = "91",
    number = "8",
    pages = "084057",
    year = "2015"
}

@article{1964PhRv..136..571M,
       author = {{Misner}, Charles W. and {Sharp}, David H.},
        title = "{Relativistic Equations for Adiabatic, Spherically Symmetric Gravitational Collapse}",
      journal = {Physical Review},
         year = 1964,
        month = oct,
       volume = {136},
       number = {2B},
        pages = {571-576},
          doi = {10.1103/PhysRev.136.B571},
       adsurl = {https://ui.adsabs.harvard.edu/abs/1964PhRv..136..571M}
}

@article{Lyth:2004gb,
    author = "Lyth, David H. and Malik, Karim A. and Sasaki, Misao",
    title = "{A General proof of the conservation of the curvature perturbation}",
    eprint = "astro-ph/0411220",
    archivePrefix = "arXiv",
    reportNumber = "YITP-04-67",
    doi = "10.1088/1475-7516/2005/05/004",
    journal = "JCAP",
    volume = "05",
    pages = "004",
    year = "2005"
}

@article{Tanaka:2007gh,
    author = "Tanaka, Yoshiharu and Sasaki, Misao",
    title = "{Gradient expansion approach to nonlinear superhorizon perturbations. II. A Single scalar field}",
    eprint = "0706.0678",
    archivePrefix = "arXiv",
    primaryClass = "gr-qc",
    reportNumber = "YITP-07-31",
    doi = "10.1143/PTP.118.455",
    journal = "Prog. Theor. Phys.",
    volume = "118",
    pages = "455--473",
    year = "2007"
}

@article{Shibata:1999zs,
    author = "Shibata, Masaru and Sasaki, Misao",
    title = "{Black hole formation in the Friedmann universe: Formulation and computation in numerical relativity}",
    eprint = "gr-qc/9905064",
    archivePrefix = "arXiv",
    reportNumber = "OU-TAP-93",
    doi = "10.1103/PhysRevD.60.084002",
    journal = "Phys. Rev. D",
    volume = "60",
    pages = "084002",
    year = "1999"
}

@book{doi:10.1137/1.9780898719598,
author = {Trefethen, Lloyd N.},
title = {Spectral Methods in MATLAB},
publisher = {Society for Industrial and Applied Mathematics},
year = {2000},
doi = {10.1137/1.9780898719598},
address = {},
edition   = {},
URL = {https://epubs.siam.org/doi/abs/10.1137/1.9780898719598},
eprint = {https://epubs.siam.org/doi/pdf/10.1137/1.9780898719598}
}

@ARTICLE{1967IBMJ...11..215C,
       author = {{Courant}, R. and {Friedrichs}, K. and {Lewy}, H.},
        title = "{On the Partial Difference Equations of Mathematical Physics}",
      journal = {IBM Journal of Research and Development},
         year = 1967,
        month = mar,
       volume = {11},
        pages = {215-234},
          doi = {10.1147/rd.112.0215},
       adsurl = {https://ui.adsabs.harvard.edu/abs/1967IBMJ...11..215C}
}

@article{Escriva:2020tak,
    author = "Escriv\`a, Albert and Germani, Cristiano and Sheth, Ravi K.",
    title = "{Analytical thresholds for black hole formation in general cosmological backgrounds}",
    eprint = "2007.05564",
    archivePrefix = "arXiv",
    primaryClass = "gr-qc",
    reportNumber = "ICCUB-20-016",
    doi = "10.1088/1475-7516/2021/01/030",
    journal = "JCAP",
    volume = "01",
    pages = "030",
    year = "2021"
}

@article{Polnarev:2006aa,
    author = "Polnarev, Alexander G. and Musco, Ilia",
    title = "{Curvature profiles as initial conditions for primordial black hole formation}",
    eprint = "gr-qc/0605122",
    archivePrefix = "arXiv",
    doi = "10.1088/0264-9381/24/6/003",
    journal = "Class. Quant. Grav.",
    volume = "24",
    pages = "1405--1432",
    year = "2007"
}

@article{Musco:2018rwt,
    author = "Musco, Ilia",
    title = "{Threshold for primordial black holes: Dependence on the shape of the cosmological perturbations}",
    eprint = "1809.02127",
    archivePrefix = "arXiv",
    primaryClass = "gr-qc",
    doi = "10.1103/PhysRevD.100.123524",
    journal = "Phys. Rev. D",
    volume = "100",
    number = "12",
    pages = "123524",
    year = "2019"
}

@article{PhysRevD.49.6467,
  title = {General laws of black-hole dynamics},
  author = {Hayward, Sean A.},
  journal = {Phys. Rev. D},
  volume = {49},
  issue = {12},
  pages = {6467--6474},
  numpages = {0},
  year = {1994},
  month = {Jun},
  publisher = {American Physical Society},
  doi = {10.1103/PhysRevD.49.6467},
  url = {https://link.aps.org/doi/10.1103/PhysRevD.49.6467}
}

@article{PhysRevD.53.1938,
  title = {Gravitational energy in spherical symmetry},
  author = {Hayward, Sean A.},
  journal = {Phys. Rev. D},
  volume = {53},
  issue = {4},
  pages = {1938--1949},
  numpages = {0},
  year = {1996},
  month = {Feb},
  publisher = {American Physical Society},
  doi = {10.1103/PhysRevD.53.1938},
  url = {https://link.aps.org/doi/10.1103/PhysRevD.53.1938}
}

@article{Khlopov:1980mg,
    author = "Khlopov, M. Yu. and Polnarev, A. G.",
    title = "{PRIMORDIAL BLACK HOLES AS A COSMOLOGICAL TEST OF GRAND UNIFICATION}",
    doi = "10.1016/0370-2693(80)90624-3",
    journal = "Phys. Lett. B",
    volume = "97",
    pages = "383--387",
    year = "1980"
}

@article{excision,
  title = {Towards a singularity-proof scheme in numerical relativity},
  author = {Seidel, Edward and Suen, Wai-Mo},
  journal = {Phys. Rev. Lett.},
  volume = {69},
  issue = {13},
  pages = {1845--1848},
  numpages = {0},
  year = {1992},
  month = {Sep},
  publisher = {American Physical Society},
  doi = {10.1103/PhysRevLett.69.1845},
  url = {https://link.aps.org/doi/10.1103/PhysRevLett.69.1845}
}

@book{bookNR,
    author    = "T. W. Baumgarte and S. L. Shapiro",
    title     = "Numerical
relativity: solving Einstein's equations on the computer",
    publisher = "Cambridge
University Press",
    year      = "2010",
    
}

@ARTICLE{acreation1,
       author = {{Zel'dovich}, Ya. B. and {Novikov}, I.~D.},
        title = "{The Hypothesis of Cores Retarded during Expansion and the Hot Cosmological Model}",
      journal = {\sovast},
         year = "1967",
        month = "Feb",
       volume = {10},
        pages = {602},
       adsurl = {https://ui.adsabs.harvard.edu/abs/1967SvA....10..602Z}
}

@article{acreation2,
  title = {Primordial black holes in braneworld cosmologies: Accretion after formation},
  author = {Guedens, Raf and Clancy, Dominic and Liddle, Andrew R.},
  journal = {Phys. Rev. D},
  volume = {66},
  issue = {8},
  pages = {083509},
  numpages = {6},
  year = {2002},
  month = {Oct},
  publisher = {American Physical Society},
  doi = {10.1103/PhysRevD.66.083509},
  url = {https://link.aps.org/doi/10.1103/PhysRevD.66.083509}
}

@Article{acreation3,
author="NAYAK, B.
and SINGH, L. P.",
title="Accretion, primordial black holes and standard cosmology",
journal="Pramana",
year="2011",
month="Jan",
day="01",
volume="76",
number="1",
pages="173--181",
issn="0973-7111",
doi="10.1007/s12043-011-0002-x",
url="https://doi.org/10.1007/s12043-011-0002-x"
}

@article{PhysRevD.53.655,
  title = {Dynamics of topological defects and inflation},
  author = {Sakai, Nobuyuki and Shinkai, Hisa-aki and Tachizawa, Takashi and Maeda, Kei-ichi},
  journal = {Phys. Rev. D},
  volume = {53},
  issue = {2},
  pages = {655--661},
  numpages = {0},
  year = {1996},
  month = {Jan},
  publisher = {American Physical Society},
  doi = {10.1103/PhysRevD.53.655},
  url = {https://link.aps.org/doi/10.1103/PhysRevD.53.655}
}

@article{Deng:2017uwc,
    author = "Deng, Heling and Vilenkin, Alexander",
    title = "{Primordial black hole formation by vacuum bubbles}",
    eprint = "1710.02865",
    archivePrefix = "arXiv",
    primaryClass = "gr-qc",
    doi = "10.1088/1475-7516/2017/12/044",
    journal = "JCAP",
    volume = "12",
    pages = "044",
    year = "2017"
}

@article{Deng:2020mds,
    author = "Deng, Heling",
    title = "{Primordial black hole formation by vacuum bubbles. Part II}",
    eprint = "2006.11907",
    archivePrefix = "arXiv",
    primaryClass = "astro-ph.CO",
    doi = "10.1088/1475-7516/2020/09/023",
    journal = "JCAP",
    volume = "09",
    pages = "023",
    year = "2020"
}

@article{Cho:1997rb,
    author = "Cho, Inyong and Vilenkin, Alexander",
    title = "{Space-time structure of an inflating global monopole}",
    eprint = "gr-qc/9708005",
    archivePrefix = "arXiv",
    doi = "10.1103/PhysRevD.56.7621",
    journal = "Phys. Rev. D",
    volume = "56",
    pages = "7621--7626",
    year = "1997"
}

@article{Deng:2016vzb,
    author = "Deng, Heling and Garriga, Jaume and Vilenkin, Alexander",
    title = "{Primordial black hole and wormhole formation by domain walls}",
    eprint = "1612.03753",
    archivePrefix = "arXiv",
    primaryClass = "gr-qc",
    doi = "10.1088/1475-7516/2017/04/050",
    journal = "JCAP",
    volume = "04",
    pages = "050",
    year = "2017"
}

@article{Carr:2010wk,
    author = "Carr, B. J. and Harada, Tomohiro and Maeda, Hideki",
    title = "{Can a primordial black hole or wormhole grow as fast as the universe?}",
    eprint = "1003.3324",
    archivePrefix = "arXiv",
    primaryClass = "gr-qc",
    reportNumber = "RESCEU-8-10, CECS-PH-10-02",
    doi = "10.1088/0264-9381/27/18/183101",
    journal = "Class. Quant. Grav.",
    volume = "27",
    pages = "183101",
    year = "2010"
}

@article{PhysRevLett.70.9,
  title = {Universality and scaling in gravitational collapse of a massless scalar field},
  author = {Choptuik, Matthew W.},
  journal = {Phys. Rev. Lett.},
  volume = {70},
  issue = {1},
  pages = {9--12},
  numpages = {0},
  year = {1993},
  month = {Jan},
  publisher = {American Physical Society},
  doi = {10.1103/PhysRevLett.70.9},
  url = {https://link.aps.org/doi/10.1103/PhysRevLett.70.9}
}

@article{Koike:1995jm,
    author = "Koike, Tatsuhiko and Hara, Takashi and Adachi, Satoshi",
    title = "{Critical behavior in gravitational collapse of radiation fluid: A Renormalization group (linear perturbation) analysis}",
    eprint = "gr-qc/9503007",
    archivePrefix = "arXiv",
    reportNumber = "TIT-HEP-284, COSMO-53, TIT-HEP-284-COSMO-53",
    doi = "10.1103/PhysRevLett.74.5170",
    journal = "Phys. Rev. Lett.",
    volume = "74",
    pages = "5170--5173",
    year = "1995"
}

@article{Niemeyer:1999ak,
    author = "Niemeyer, Jens C. and Jedamzik, K.",
    title = "{Dynamics of primordial black hole formation}",
    eprint = "astro-ph/9901292",
    archivePrefix = "arXiv",
    doi = "10.1103/PhysRevD.59.124013",
    journal = "Phys. Rev. D",
    volume = "59",
    pages = "124013",
    year = "1999"
}

@article{Evans:1994pj,
    author = "Evans, Charles R. and Coleman, Jason S.",
    title = "{Observation of critical phenomena and selfsimilarity in the gravitational collapse of radiation fluid}",
    eprint = "gr-qc/9402041",
    archivePrefix = "arXiv",
    reportNumber = "TAR-039-UNC",
    doi = "10.1103/PhysRevLett.72.1782",
    journal = "Phys. Rev. Lett.",
    volume = "72",
    pages = "1782--1785",
    year = "1994"
}

@article{Young:2019yug,
    author = "Young, Sam and Musco, Ilia and Byrnes, Christian T.",
    title = "{Primordial black hole formation and abundance: contribution from the non-linear relation between the density and curvature perturbation}",
    eprint = "1904.00984",
    archivePrefix = "arXiv",
    primaryClass = "astro-ph.CO",
    doi = "10.1088/1475-7516/2019/11/012",
    journal = "JCAP",
    volume = "11",
    pages = "012",
    year = "2019"
}

@article{Faraoni:2016xgy,
    author = "Faraoni, Valerio and Ellis, George F. R. and Firouzjaee, Javad T. and Helou, Alexis and Musco, Ilia",
    title = "{Foliation dependence of black hole apparent horizons in spherical symmetry}",
    eprint = "1610.05822",
    archivePrefix = "arXiv",
    primaryClass = "gr-qc",
    doi = "10.1103/PhysRevD.95.024008",
    journal = "Phys. Rev. D",
    volume = "95",
    number = "2",
    pages = "024008",
    year = "2017"
}

@article{Helou:2016xyu,
    author = "Helou, Alexis and Musco, Ilia and Miller, John C.",
    title = "{Causal Nature and Dynamics of Trapping Horizons in Black Hole Collapse}",
    eprint = "1601.05109",
    archivePrefix = "arXiv",
    primaryClass = "gr-qc",
    doi = "10.1088/1361-6382/aa6d8f",
    journal = "Class. Quant. Grav.",
    volume = "34",
    number = "13",
    pages = "135012",
    year = "2017"
}

@article{Musco:2004ak,
    author = "Musco, Ilia and Miller, John C. and Rezzolla, Luciano",
    title = "{Computations of primordial black hole formation}",
    eprint = "gr-qc/0412063",
    archivePrefix = "arXiv",
    doi = "10.1088/0264-9381/22/7/013",
    journal = "Class. Quant. Grav.",
    volume = "22",
    pages = "1405--1424",
    year = "2005"
}

@article{Bloomfield:2015ila,
    author = "Bloomfield, Jolyon and Bulhosa, Daniel and Face, Stephen",
    title = "{Formalism for Primordial Black Hole Formation in Spherical Symmetry}",
    eprint = "1504.02071",
    archivePrefix = "arXiv",
    primaryClass = "gr-qc",
    reportNumber = "MIT-CTP-4662",
    month = "4",
    year = "2015"
}

@article{Nakama:2013ica,
    author = "Nakama, Tomohiro and Harada, Tomohiro and Polnarev, A. G. and Yokoyama, Jun'ichi",
    title = "{Identifying the most crucial parameters of the initial curvature profile for primordial black hole formation}",
    eprint = "1310.3007",
    archivePrefix = "arXiv",
    primaryClass = "gr-qc",
    reportNumber = "RESCEU-45-13, RUP-13-11",
    doi = "10.1088/1475-7516/2014/01/037",
    journal = "JCAP",
    volume = "01",
    pages = "037",
    year = "2014"
}

@article{Harada:2024trx,
    author = "Harada, Tomohiro and Iizuka, Hayami and Koga, Yasutaka and Yoo, Chul-Moon",
    title = "{Geometrical origin for the compaction function for primordial black hole formation}",
    eprint = "2409.05544",
    archivePrefix = "arXiv",
    primaryClass = "gr-qc",
    reportNumber = "RUP-24-17, YITP-24-111",
    doi = "10.1103/PhysRevD.111.023537",
    journal = "Phys. Rev. D",
    volume = "111",
    number = "2",
    pages = "023537",
    year = "2025"
}

@article{Harada:2023ffo,
    author = "Harada, Tomohiro and Yoo, Chul-Moon and Koga, Yasutaka",
    title = "{Revisiting compaction functions for primordial black hole formation}",
    eprint = "2304.13284",
    archivePrefix = "arXiv",
    primaryClass = "gr-qc",
    reportNumber = "RUP-23-10",
    doi = "10.1103/PhysRevD.108.043515",
    journal = "Phys. Rev. D",
    volume = "108",
    number = "4",
    pages = "043515",
    year = "2023"
}

@article{Yamamoto:2008js,
    author = "Yamamoto, Tetsuro and Shibata, Masaru and Taniguchi, Keisuke",
    title = "{Simulating coalescing compact binaries by a new code SACRA}",
    eprint = "0806.4007",
    archivePrefix = "arXiv",
    primaryClass = "gr-qc",
    doi = "10.1103/PhysRevD.78.064054",
    journal = "Phys. Rev. D",
    volume = "78",
    pages = "064054",
    year = "2008"
}

@article{Escriva:2019phb,
    author = "Escriv\`a, Albert and Germani, Cristiano and Sheth, Ravi K.",
    title = "{Universal threshold for primordial black hole formation}",
    eprint = "1907.13311",
    archivePrefix = "arXiv",
    primaryClass = "gr-qc",
    reportNumber = "ICC-19-013",
    doi = "10.1103/PhysRevD.101.044022",
    journal = "Phys. Rev. D",
    volume = "101",
    number = "4",
    pages = "044022",
    year = "2020"
}

@book{Baumgarte_Shapiro_2021, place={Cambridge}, title={Numerical Relativity: Starting from Scratch}, publisher={Cambridge University Press}, author={Baumgarte, Thomas W. and Shapiro, Stuart L.}, year={2021}}

@BOOK{2016nure.book.....S,
       author = {{Shibata}, Masaru},
        title = "{Numerical Relativity}",
         year = 2016,
          doi = {10.1142/9692},
       adsurl = {https://ui.adsabs.harvard.edu/abs/2016nure.book.....S}
}

@article{Polnarev:2012bi,
    author = "Polnarev, A. G. and Nakama, Tomohiro and Yokoyama, Jun'ichi",
    title = "{Self-consistent initial conditions for primordial black hole formation}",
    eprint = "1204.6601",
    archivePrefix = "arXiv",
    primaryClass = "gr-qc",
    reportNumber = "RESCEU-10-12",
    doi = "10.1088/1475-7516/2012/09/027",
    journal = "JCAP",
    volume = "09",
    pages = "027",
    year = "2012"
}

@ARTICLE{1966ApJ...143..452H,
       author = {{Hernandez}, Jr., Walter C. and {Misner}, Charles W.},
        title = "{Observer Time as a Coordinate in Relativistic Spherical Hydrodynamics}",
      journal = {\apj},
         year = 1966,
        month = feb,
       volume = {143},
        pages = {452},
          doi = {10.1086/148525},
       adsurl = {https://ui.adsabs.harvard.edu/abs/1966ApJ...143..452H}
}

@article{Escriva:2021pmf,
    author = "Escriv\`a, Albert and Romano, Antonio Enea",
    title = "{Effects of the shape of curvature peaks on the size of primordial black holes}",
    eprint = "2103.03867",
    archivePrefix = "arXiv",
    primaryClass = "gr-qc",
    reportNumber = "ICCUB-21-003",
    doi = "10.1088/1475-7516/2021/05/066",
    journal = "JCAP",
    volume = "05",
    pages = "066",
    year = "2021"
}

@article{Escriva:2023nzn,
    author = "Escriva, Albert and Tada, Yuichiro and Yoo, Chul-Moon",
    title = "{Primordial black holes and induced gravitational waves from a smooth crossover beyond standard model theories}",
    eprint = "2311.17760",
    archivePrefix = "arXiv",
    primaryClass = "astro-ph.CO",
    doi = "10.1103/PhysRevD.110.063521",
    journal = "Phys. Rev. D",
    volume = "110",
    number = "6",
    pages = "063521",
    year = "2024"
}

@article{Yoo:2020lmg,
    author = "Yoo, Chul-Moon and Harada, Tomohiro and Okawa, Hirotada",
    title = "{Threshold of Primordial Black Hole Formation in Nonspherical Collapse}",
    eprint = "2004.01042",
    archivePrefix = "arXiv",
    primaryClass = "gr-qc",
    reportNumber = "RUP-20-12",
    doi = "10.1103/PhysRevD.102.043526",
    journal = "Phys. Rev. D",
    volume = "102",
    number = "4",
    pages = "043526",
    year = "2020",
    note = "[Erratum: Phys.Rev.D 107, 049901 (2023)]"
}

@article{preparation,
    author = "Escriv{\`a}, Albert",
    title = "{Threshold for PBH formation in the type-II region and its analytical estimation}",
    eprint = "2504.05814",
    archivePrefix = "arXiv",
    primaryClass = "astro-ph.CO",
    doi = "10.1103/mq67-bbvj",
    journal = "Phys. Rev. D",
    volume = "112",
    number = "10",
    pages = "103527",
    year = "2025"
}

@misc{codigo_albert,
	title        = {{Albert Escriva-GitHub}},
	author       = {},
	year         = 2025,
	note         = {[Online]},
	howpublished = {\url{https://github.com/albert-escriva}}
}

@article{Ning:2025ogq,
    author = "Ning, Zhuan and Zeng, Xiang-Xi and Yuwen, Zi-Yan and Wang, Shao-Jiang and Deng, Heling and Cai, Rong-Gen",
    title = "{Sound waves from primordial black hole formations}",
    eprint = "2504.12243",
    archivePrefix = "arXiv",
    primaryClass = "gr-qc",
    month = "4",
    year = "2025"
}

\end{document}